\documentclass[12pt]{article}
\usepackage{graphicx}

\usepackage{amsmath}
\usepackage{amsfonts}
\usepackage{amssymb}
\usepackage{mathbbol}
\usepackage{amsbsy}
\usepackage{bm}
\usepackage{color}
\usepackage{wrapfig}
\usepackage{float}

\def\gtorder{\mathrel{\raise.3ex\hbox{$>$}\mkern-14mu
             \lower0.6ex\hbox{$\sim$}}}
\def\ltorder{\mathrel{\raise.3ex\hbox{$<$}\mkern-14mu
             \lower0.6ex\hbox{$\sim$}}}

\begin{document}

\pagenumbering{gobble}
\clearpage
\thispagestyle{empty}

\title{Physics of the cosmic microwave background anisotropy$^{\ast}$}

\author{Martin Bucher\\
\small
Laboratoire APC, Universit\'e Paris 7/CNRS\\
\small
B\^atiment Condorcet, Case 7020\\
\small
75205 Paris Cedex 13, France\\[3pt]
\small
bucher@apc.univ-paris7.fr
\small
and\\ 
\small
Astrophysics and Cosmology Research Unit\\
\small
School of Mathematics, Statistics and Computer Science \\
\small
University of KwaZulu-Natal\\
\small
Durban 4041, South Africa\\
}

\footnotetext[1]{This article is to be published also in the book ``One Hundred Years of
General Relativity: From Genesis and Empirical Foundations to Gravitational Waves,
Cosmology and Quantum Gravity,'' edited by Wei-Tou Ni (World Scientific, Singapore,
2015) as well as in Int. J. Mod. Phys. D (in press).}

\maketitle

\begin{abstract}
\noindent 
Observations of the cosmic microwave background (CMB), especially
of its frequency spectrum and its anisotropies, both in
temperature and in polarization, have played a key role in the
development of modern cosmology and our understanding of the
very early universe. We review the underlying physics of the CMB
and how the primordial temperature and polarization anisotropies
were imprinted. Possibilities for distinguishing competing
cosmological models are emphasized. The current status of CMB
experiments and experimental techniques with an emphasis toward
future observations, particularly in polarization, is reviewed.
The physics of foreground emissions, especially of polarized dust,
is discussed in detail, since this area is likely to become
crucial for measurements of the $B$ modes of the CMB polarization
at ever greater sensitivity.
\end{abstract}

\newpage
\setcounter{page}{1}
\pagenumbering{roman}
\tableofcontents

\newpage
\setcounter{page}{1}
\pagenumbering{arabic}

\section[Observing the microwave sky: a short history and observational overview]
{Observing the Microwave Sky: A Short History and Observational Overview}
\label{ObsMicroSky}

\begin{figure}
\begin{center}
\includegraphics[width=16cm]{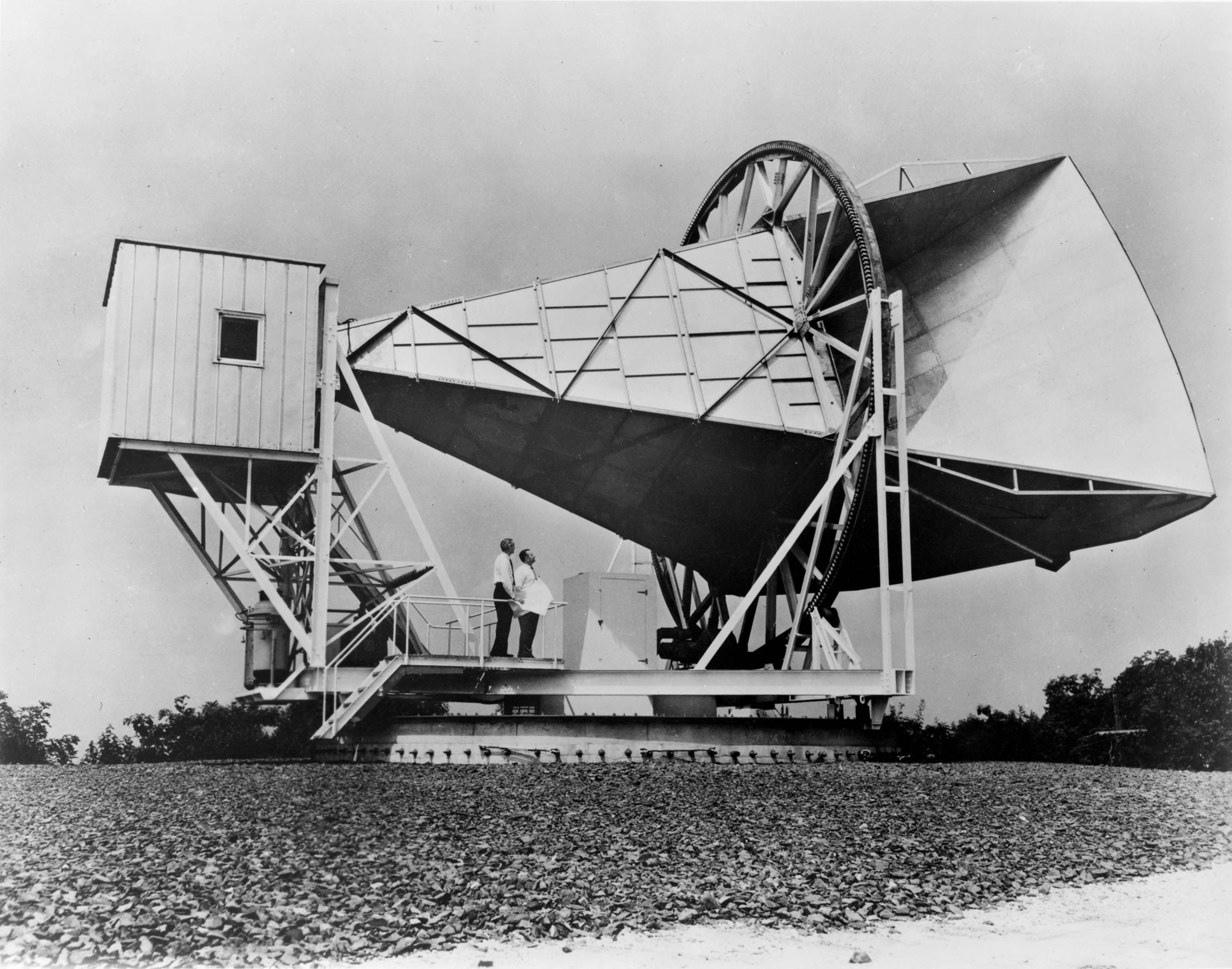}
\end{center}
\caption{\bf Horn antenna used in 1964 by Penzias and Wilson to discover the CMB. ({\it
Credit: NASA image})
\label{penziasWilsonAntennaPhoto}}
\end{figure}

In their 1965 landmark paper A.~Penzias and
R.~Wilson \cite{PenziasAndWilson1965}, who were investigating the
origin of radio interference at what at the time were considered
high frequencies, reported a 3.5\,K signal from the sky at 4\,GHz
that was ``isotropic, unpolarized, and free from seasonal
variations'' within the limits of their observations. Their
apparatus was a 20 foot horn directed at the zenith coupled to a
maser amplifier and a radiometer (see
Fig.~\ref{penziasWilsonAntennaPhoto}). The maser amplifier and
radiometer were switched between the sky and a liquid helium
cooled reference source used for comparison. Alternative
explanations such as ground pick-up from the sidelobes of their
antenna were ruled out in their analysis, and they noted that
known radio sources would contribute negligibly at this frequency
because their apparent temperature falls rapidly with frequency.

In a companion paper published in the same issue of the
Astrophysical Journal, Dicke {\it
et~al.} \cite{DickePeeblesRollWilkinson} proposed the explanation
that the isotropic sky signal seen by Penzias and Wilson was in
fact emanating from a hot big bang, as had been suggested in the
1948 paper of Alpher {\it et~al.} \cite{AlpherBetheGamov}
suggesting the presence of the photon blackbody component having a
temperature of approximately a few K. Their prediction was based
on considering the conditions required for successful
nucleosynthesis in an expanding universe---that is, to create an
appreciable fraction of primordial helium from the neutrons that
are decaying as the universe is expanding.

\begin{figure}
\begin{center}
\includegraphics[width=12cm]{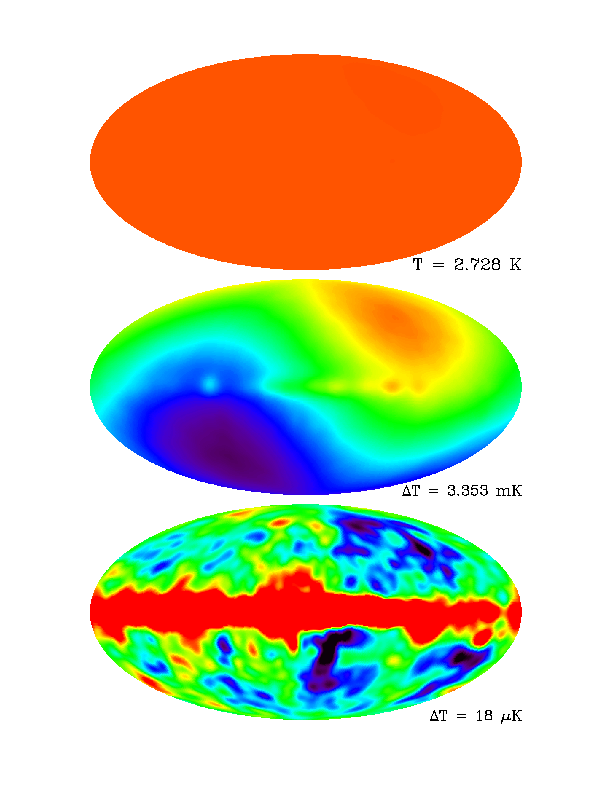}
\end{center}
\caption{
{\bf The microwave sky as seen by the COBE DMR (differential microwave
radiometer) instrument.}
The top panel shows the microwave sky as
seen on a linear temperature scale including zero. No anisotropies
are visible in this image, because the CMB monopole at 2.725\,K
dominates. In the middle panel, the monopole component has been
subtracted. Apart from some slight contamination from the galaxy
near the equator (corresponding to the plane of our Galaxy), one
sees only a nearly perfect dipole pattern, owing to our peculiar
motion with respect to the rest frame defined by the CMB. In the
bottom panel, both the monopole and dipole components have been
removed. Except for the galactic contamination around the equator,
one sees the cosmic microwave background anisotropy along with
some noise. ({\it Credit: NASA/COBE Science Team})
\label{cobe-fig-overview}}
\end{figure}

\begin{figure}
\begin{center}
\includegraphics[width=10cm]{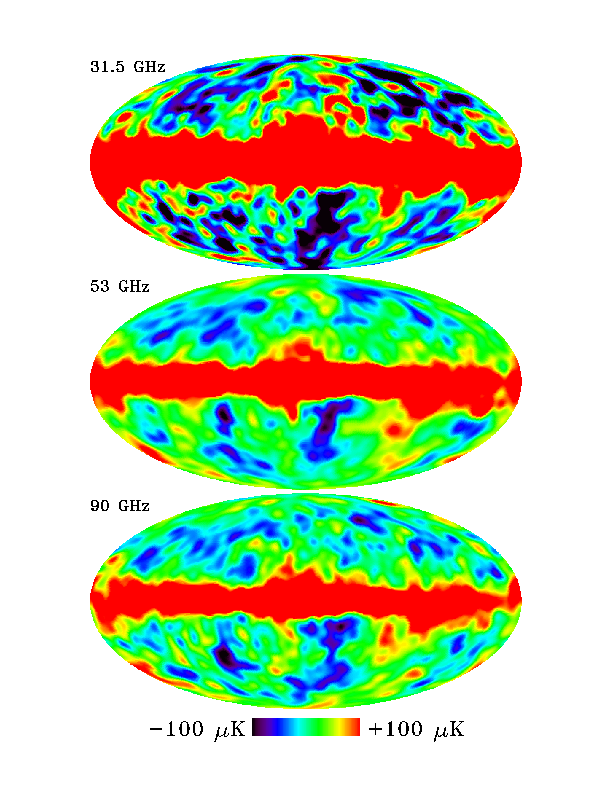}
\end{center}
\caption{{\bf COBE DMR individual frequency maps (with monopole and dipole
components removed).} Taking data at several frequencies is key to proving the
primordial origin of the signal and removing galactic contaminants. Here are the
three frequency maps from the COBE DMR observations. ({\it Credit: NASA/COBE
Science Team})
\label{cobe-freq_maps}}
\end{figure}

The importance of this discovery became almost immediately apparent, and others set out
to better characterize this excess emission, which later would become known as the Cosmic
Microwave Background (CMB), or sometimes in the older literature the Cosmic Microwave
Background Radiation (CMBR). The two principal questions were: (i) To what extent is this
background isotropic? (ii) How close is the spectrum to a perfect Planckian blackbody
spectrum?

\begin{figure}
\begin{center}
\includegraphics[width=9cm]{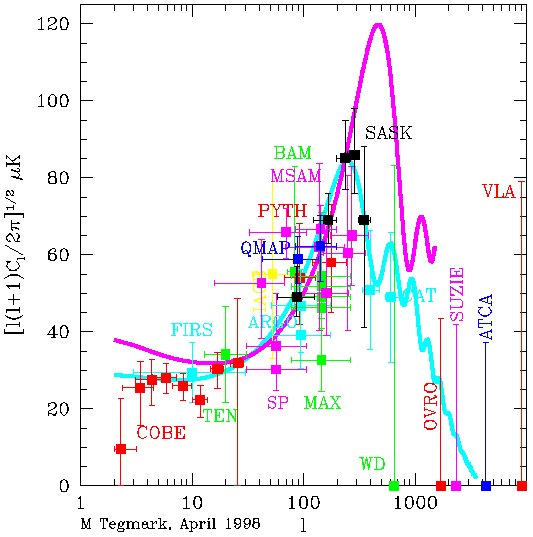}
\end{center}
\caption{{\bf State of CMB observations in 1998.} After the COBE DMR
detection at very large angular scales at the so-called
Sachs-Wolfe plateau, numerous groups sought to discover the
acoustic peaks, or to find their absence. This compilation from
1998 shows the state of play at that time. While an unmistakable
rise toward the first acoustic peak is apparent, it is not so
apparent what happens toward higher $\ell.$ Two theories broadly
compatible with the data at that time are plotted, one with
$\Omega _k=0$ and another with $\Omega _k\approx 0.7,$ which
predicts a higher amplitude of the primordial perturbations and an
acoustic peak shifted to smaller angular scales, as explained in
Sec.~\ref{AngularDiameterSection}. 
({\it Credit: Max Tegmark})
\label{stateOfPlay1998}} 
\end{figure}

\def\thefootnote{\alph{footnote}}

The main obstacle to answering these questions was the lack of adequate instrumentation,
and this was the main reason why the first detection of the CMB anisotropy had to wait
until 1992, when the COBE team announced their observation of a statistically significant
anisotropy of primordial origin after the dipole due to our motion with respect to the
CMB had been subtracted \cite{COBEdmr} (see Figs.~2 and~3). The COBE satellite, in a
low-Earth orbit, carried three instruments: the differential microwave radiometer
(DMR) \cite{DMR-instrument}, the far infrared absolute spectrophotometer
(FIRAS) \cite{firas-instrument-paper}, and the diffuse infrared background experiment
(DIRBE). By today's standards, the measurement of the cosmic microwave background
anisotropy was crude. The angular resolution was low---the width of the beam was
$7^\circ$ (FWHM), and the sky map used for the analysis was smoothed to $10^\circ$ to
suppress beam artefacts. The contribution from instrument noise was large by contemporary
standards. The COBE noise was 43,~16, and 22\,mK\,Hz$^{-1/2}$ for the 31, 53, and 90\,GHz
channels, respectively \cite{cobe-noise}.\footnote{By comparison, for WMAP the
detector sensitivities were $0.7, 0.7, 0.9, 1.1$, and $1.5$\,mK\,$\cdot$\,Hz$^{-1/2}$ for
the K, Ka, Q, V, and W bands, respectively \cite{jarosik}.} Nevertheless, COBE did
provide a convincing first detection of the CMB anisotropy, and most importantly
established the overall level of the primordial cosmological density perturbations, which
played a crucial role in determining the viability of the cosmological models in vogue at
that time, which were much more numerous and varied than today.

The COBE detection was followed by an intense effort to characterize the CMB anisotropy
at greater sensitivity and on smaller angular scales. There were numerous experiments
from the ground at locations where the column density of water in the atmosphere is
particularly low, such as Saskatoon, the Atacama Desert in Chile, the South Pole, and the
Canary Islands, as well as from stratospheric balloons. Figure~\ref{stateOfPlay1998}
shows the state of play about four years after COBE, with two competing theoretical
models plotted together with the data points available at that time. In
Sec.~\ref{Section:BasicCMBphysics} we shall present the physics of the CMB
systematically, but jumping ahead a bit, we give here a few words of explanation for
understanding this plot. In most theoretical models, the CMB is generated by an isotropic
Gaussian stochastic process, in which case all the available information concerning the
underlying theoretical model can be extracted by measuring the angular power spectrum of
the CMB anisotropies. Because of isotropy, one may expand the map in spherical harmonics
to extract its angular power spectrum, defined as
\begin{eqnarray}
C_\ell ^{TT,{\rm obs}}=\frac{1}{(2\ell +1)}\sum _{m=-\ell }^{+\ell } | a_{\ell
m}^T| ^2.
 \end{eqnarray}
It is customary to plot the quantity $\ell (\ell +1)C_\ell /(2\pi
),$ which would be constant for a scale invariant pattern on the
sky.\footnote{Strictly speaking, scale invariance on the sky can
be defined only asymptotically in the $\ell \to \infty $ limit,
because the curvature of the celestial sphere breaks scale
invariance. The use of $\ell (\ell +1)$ rather than $\ell ^2$ here
is an historical convention.} This is the quantity plotted in
Fig.~\ref{stateOfPlay1998}, where the question posed at the time
was whether there is a rise to a first acoustic peak followed by
several decaying secondary acoustic peaks, as indicated by the
solid theoretical curves. This structure is a prediction of simple
inflationary models---or more precisely, of models where only
the adiabatic growing mode is excited with an approximately scale
invariant primordial spectrum. In this plot, one sees fairly
convincing evidence for a rise in the angular power, but the
continuation of the curve is unclear.

\begin{figure}
\begin{center}
\includegraphics[width=6.5cm]{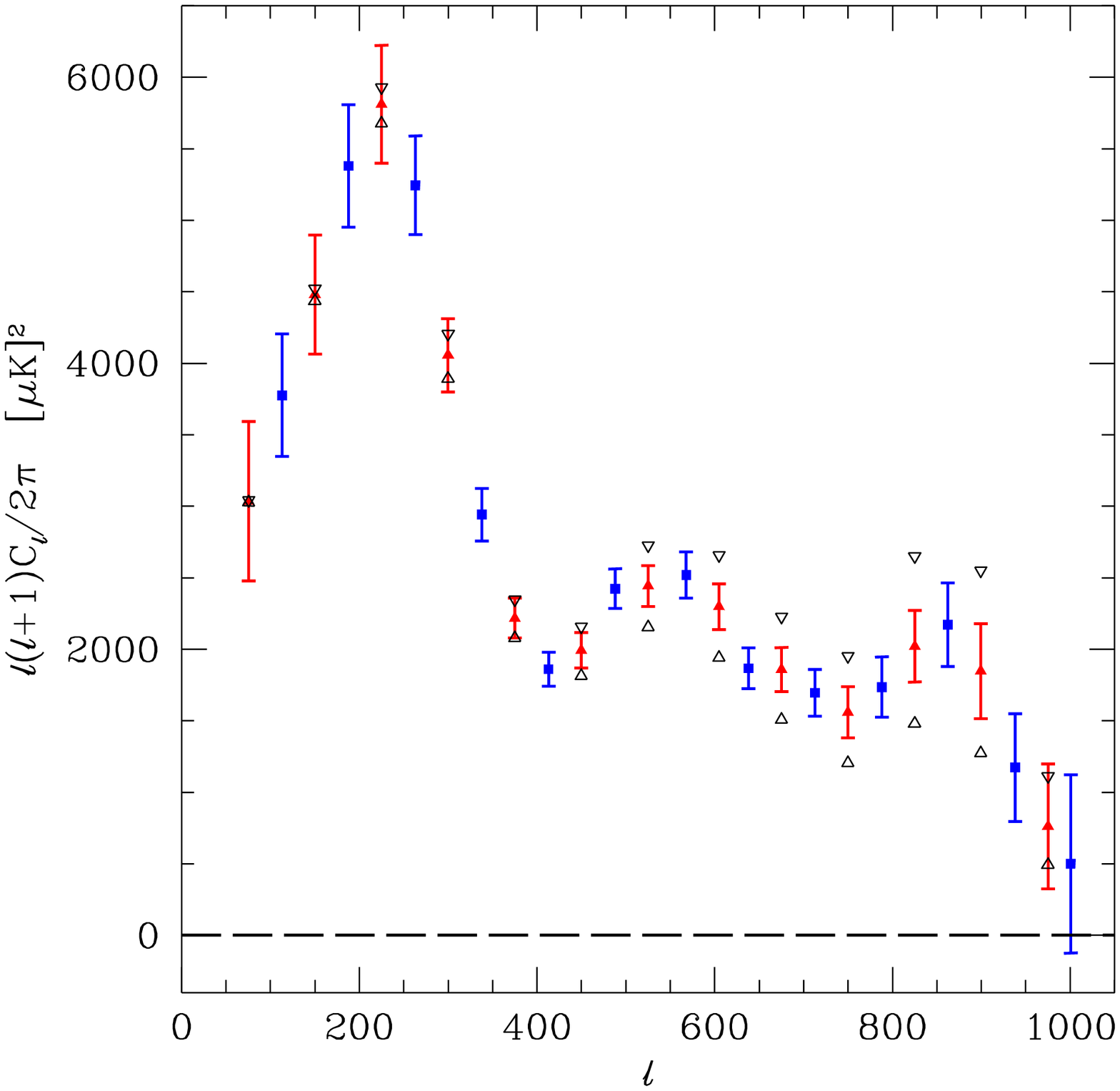}
\includegraphics[width=6.5cm]{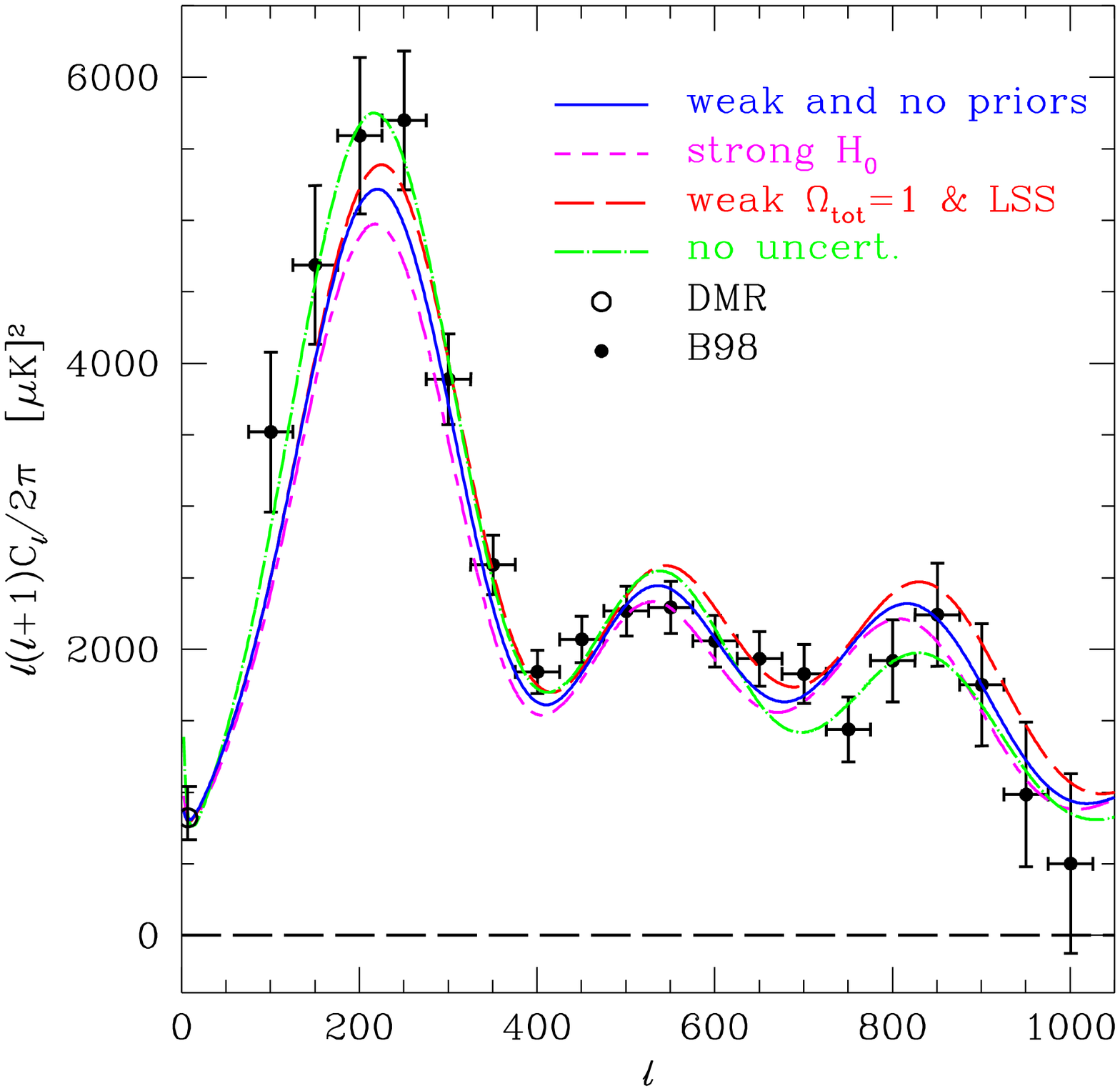}
\end{center}
\caption{{\bf Boomerang observation of acoustic oscillations.} The $C^{TT}_\ell $
power spectrum as measured by Boomerang is shown on the left without a fit to a
theoretical model and on the right with the theoretical predictions for a
spatially flat cosmological model with an exactly scale invariant primordial power
spectrum for the adiabatic growing mode. 
{\it (Credit: Boomerang Collaboration})
\label{BoomerangData}}
\end{figure}

\begin{figure}
\begin{center}
\includegraphics[width=12cm]{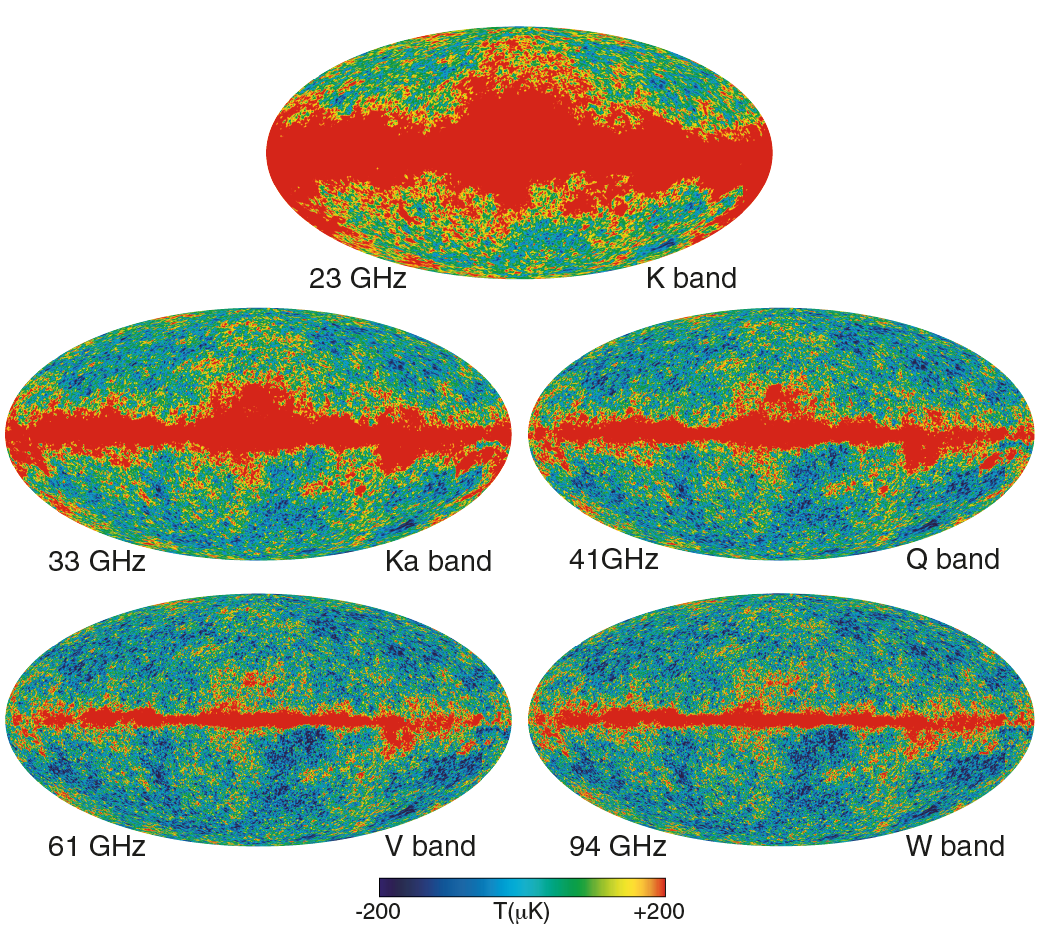}
\end{center}
\caption{{\bf WMAP single temperature frequency maps.} WMAP observed in
five frequency bands. ({\it Credit: NASA/WMAP Science Team})
\label{WMAP-Monofrequency-T}}
\end{figure}

Several experiments contributed to providing the first clear detection of the acoustic
oscillations, namely TOCO \cite{millerToco}, MAXIMA\cite{maxima}, and
Boomerang \cite{boomerang-netterfield,boomerang}. Figure~\ref{BoomerangData} shows the
power spectrum as measured by one of these experiments, namely Boomerang, where a series
of well defined acoustic oscillations is clearly visible.

In the meantime, parallel efforts were underway in
Europe and in the United States to prepare for another CMB space
mission to follow on COBE at higher sensitivity and angular
resolution. The COBE beam, which was $7^\circ $ FWHM, did not use
a telescope but rather microwave feed horns pointed directly at
the sky. While not allowing for a high angular resolution, the
feed horns had the advantage of producing well defined beams with
rapidly falling sidelobes. The US NASA WMAP satellite, launched in
2001, delivered its one-year data release in 2003 (including TT
and TE) 
\cite{wmap2003a}--\cite{wmap2003i},
and
its first polarization data (including also EE) in
2006
\cite{wmap2007a}--\cite{wmap2007d}
(see
Figs.~6 and~7). WMAP continued taking data for nine years, and
released \hbox{installments} of papers based on the five-, seven-,
and nine-year data releases, in which the results were further
refined benefitting from longer integration time (which nominally
would shrink error bars in proportion to $1/\sqrt{t_{{\rm obs}}}$)
as well as improved instrument
modeling \cite{wmap2009b}--\cite{wmap2013b}.
WMAP used horns pointed at a $1.4\,\textrm{m} \times
1.6\,\textrm{m}$ off-axis Gregorian mirror to obtain diffraction
limited beams (although the mirror was under-illuminated somewhat
to reduce far sidelobes). The 20 detectors were based on coherent
amplification using HEMTs (high electron mobility 
transistors)---the state of the art in coherent low-noise amplification at the
time. The coherent detection technology used by WMAP had the
advantage that the electronics could be passively cooled. The
competing incoherent bolometric detection technology permits
superior sensitivity, which operates almost at the quantum noise
limit of the incident photons, but requires cooling to
approximately 100\,mK. The cryogenic system required was judged to
present substantial risk for a space mission at the time, and this
was likely one of the reasons why WMAP was selected over one of
its competitor missions, which was similar to the European Planck
HFI proposal.

\begin{figure}
\begin{center}
\includegraphics[width=12cm]{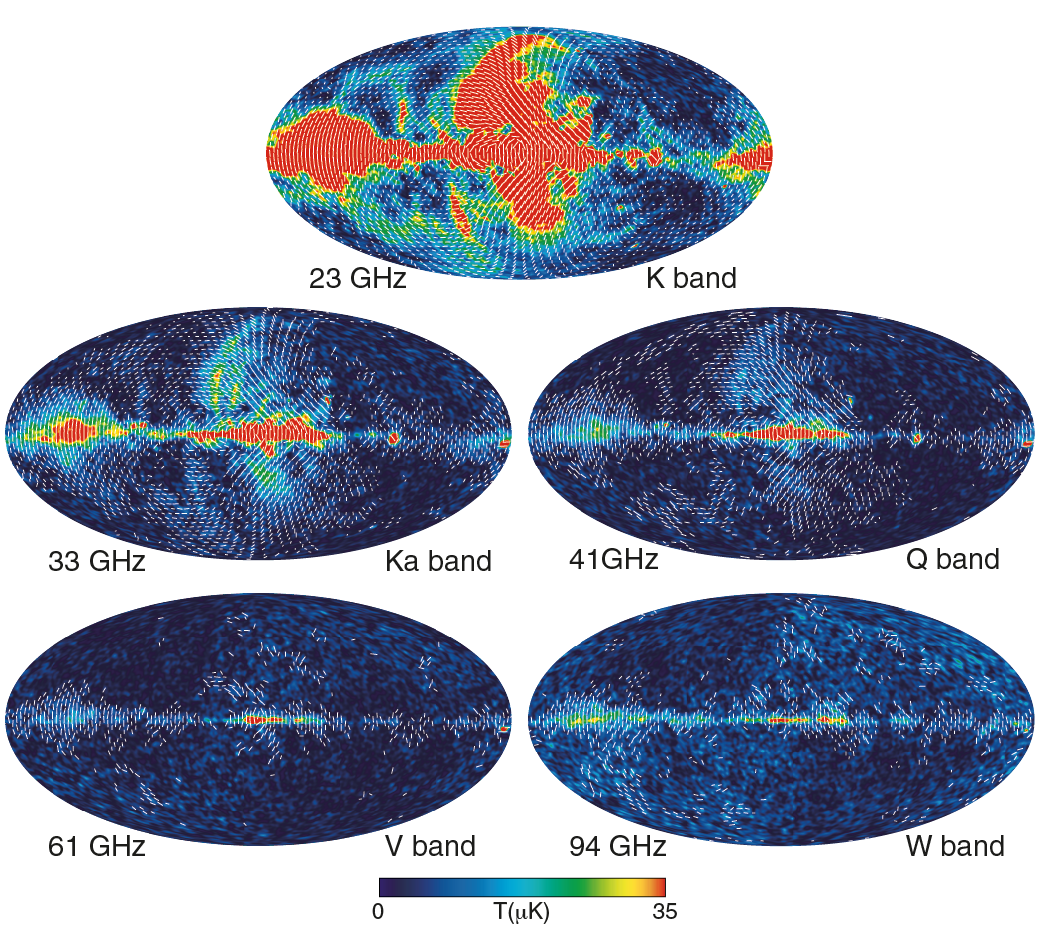}
\end{center}
\caption{{\bf WMAP single frequency polarization maps.} The lines (which may be thought
of as double-headed vectors) show the amplitude and orientation of the linear
polarization of the measured CMB anisotropy in the indicated bands. The amplitude
$P=\sqrt{Q^2+U^2}$ is also shown using the indicated color scale.
(\textit{Credit: NASA/WMAP Science Team})
\label{WMAP-Monofrequency-P}}
\end{figure}

\clearpage
\newpage 

The successor to WMAP was the European Space Agency (ESA) Planck
satellite, which was launched in May
2009 \cite{planck-cib-ps}--\cite{planck-hfi-cosmic-rays}.
Planck consisted of two instruments: (i) a low-frequency instrument with
three channels (at $30, 44$ and 70\,GHz) using a \hbox{coherent}
detection technology, and (ii) a high-frequency instrument using
cryogenically-cooled bolometric detectors observing in six bands
(100, 143, 217, 353, 545 and 857\,GHz). All bands except the
highest two bands were polarization sensitive and had a
diffraction limited angular resolution, starting at 33 arcmin for
the 30\,GHz channel and going down to 5.5 arcmin for the 217\,GHz
channel. (The top three channels have an angular resolution of
approximately 5'.) Planck HFI took data from August 2009 to
January 2012, when the coolant for its high-frequency instrument
ran out. The Planck Collaboration reported its first results for
cosmology in March 2013 based on its temperature anisotropy data.
The first results for cosmology using the polarization data
collected by Planck are expected in late 2014.

\begin{figure}
\begin{center}
\includegraphics[width=12cm]{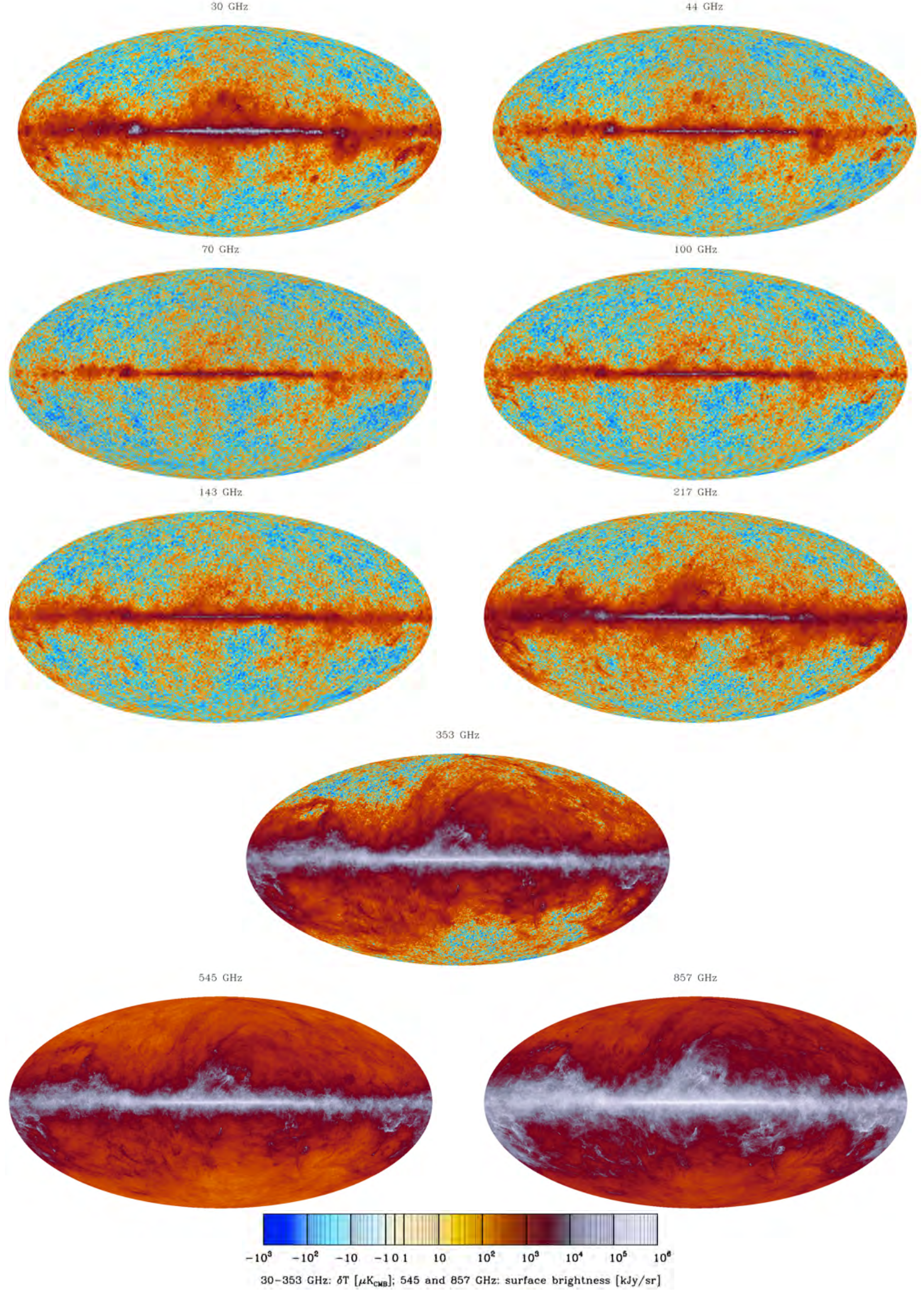}
\end{center}
\caption{{\bf Planck single frequency temperature maps.} The ESA Planck
satellite surveyed the sky in nine broad [i.e. $(\Delta \nu )/\nu
\approx 0.3$] frequency bands, centered at $30, 44, 70, 100, 143,
217, 353, 545$, and 857\,GHz, shown in galactic coordinates. The
units are CMB thermodynamic temperature [see
Eq.~(\ref{DefThermoTemp})]. The nonlinear scale avoids saturation
in regions of high galactic emission. (\textit{Credit:
ESA/Planck Collaboration})
\label{Planck-Monofrequency}}
\end{figure}

\begin{figure}
\begin{center}
\includegraphics[width=12cm]{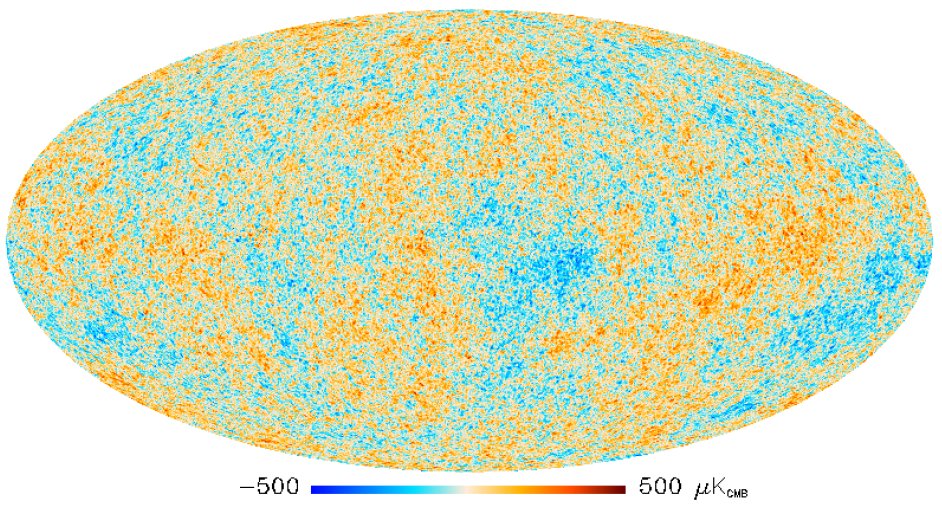}
\end{center}
\caption{{\bf Planck internal linear combination map.} A linear combination of the Planck
single frequency maps (shown in Fig.~\ref{Planck-Monofrequency}) is taken. This linear
combination is optimized to filter out any unwanted contaminants based on their differing
frequency dependence. The success of this procedure to remove the galactic contaminants
is evident from the disappearance of excess power along the equator corresponding to the
galactic plane. (For more details, see Ref.~161.) (\textit{Credit: ESA/Planck
Collaboration})
\label{Planck-ILC}}
\end{figure}

Figure~\ref{Planck-Monofrequency} shows the Planck full-sky maps of the intensity
of the microwave emission in nine frequency bands ranging from 30\,GHz to
857\,GHz. Figure~\ref{Planck-ILC} shows a cleaned full-sky map where a linear
combination of the single band maps has been taken in order to isolate the
primordial cosmic microwave background signal. While the fluctuations in the
cleaned map in Fig.~\ref{Planck-ILC} do not appear to single out any particular
direction in the sky and appear consistent with an isotropic Gaussian random
process, the maps in Fig.~\ref{Planck-Monofrequency} show a clear excess in the
galactic plane. These full-sky maps use a Molleweide projection in galactic
coordinates, with the galactic center at the center of the projection. Even though
the galactic contamination depends largely on the angle to the galactic plane,
with a lesser tendency to increase toward the galactic center, considerable
structure can be seen in the galactic emission over a broad range of angular
scales. The central bands include the least amount of galactic contamination,
which visibly is much larger for the lowest and the highest frequencies shown.

\begin{figure}
\begin{center}
\includegraphics[width=12cm]{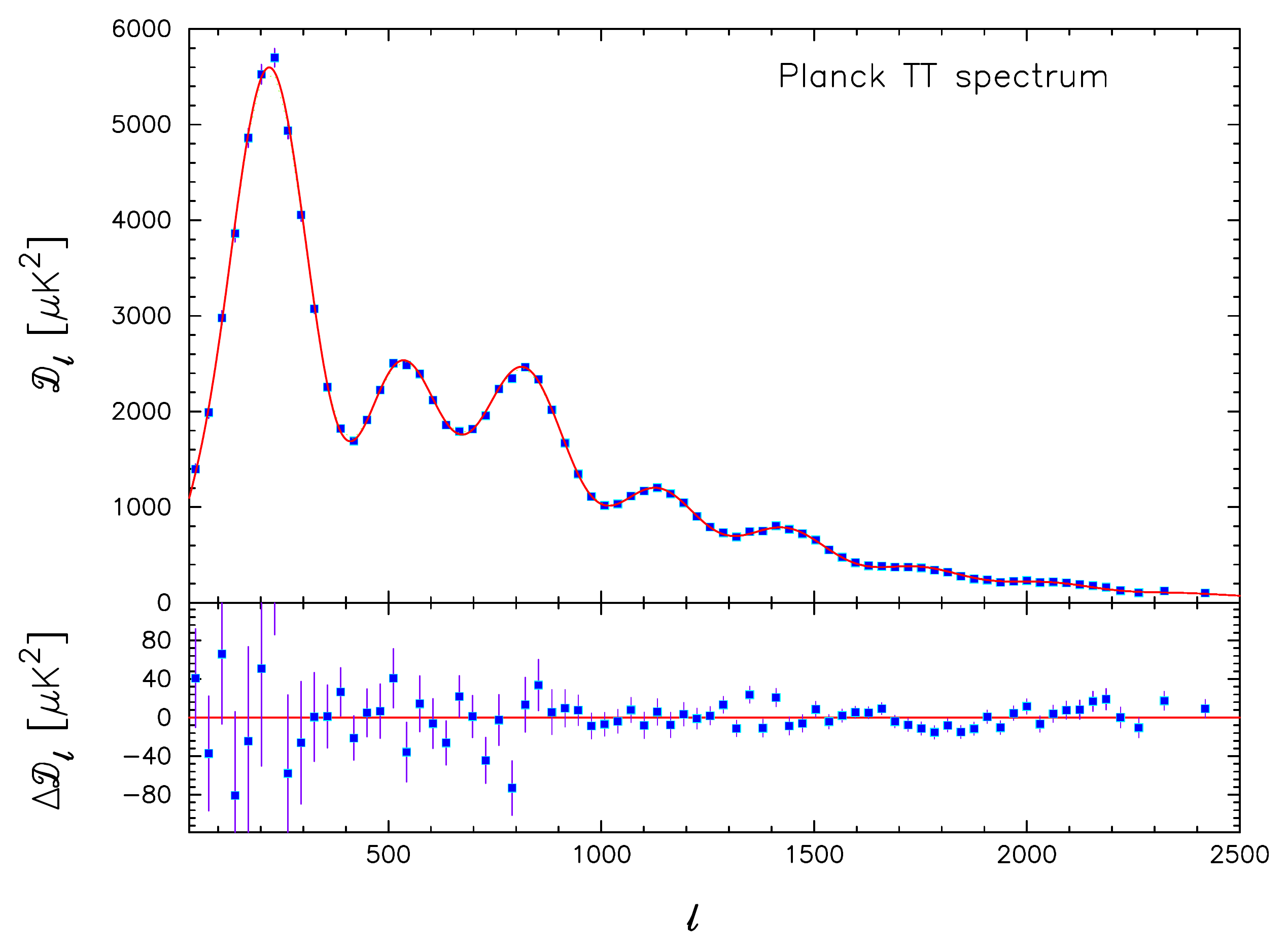}
\end{center}
\caption{{\bf CMB angular power spectrum as measured by Planck.} The binned $C^{TT}_\ell $
power spectrum as given in the Planck 2013 release is plotted with error bars that
combine uncertainties from cosmic variance (dominant at low $\ell $) and instrumental
noise (dominant at high $\ell $). The solid curve indicates the theoretical predictions
of the six-parameter concordance model with the $1\sigma $ cosmic variance for the
adopted binning scheme. The lower panel shows the residuals. For more details, see
Ref.~\protect\cite{planck-parameters}. The exquisite fit, seen here up to about $\ell
\approx 2200,$ into the damping tail, has been shown to extend to smaller angular scales
by ACT \protect\cite{act-cosmo-params} and SPT \protect\cite{keisler}.
(\textit{Credit: ESA/Planck Collaboration})
\label{PlanckPowerSpectrum}}
\end{figure}

The CMB dipole amplitude is $\Delta T=3.365\pm 0.027$\,mK and
directed toward $(l,b)=(264.4^\circ \pm 0.3^\circ, 48.4\pm
0.5^\circ )$ in galactic coordinates \cite{cmb-dipole}. The maps
have been processed to remove the 2.725\,K CMB monopole component
as well as the smaller CMB dipole having a peak-to-peak amplitude
of approximately 6.73\,mK, so that only the spherical harmonic
multipoles with $\ell \ge 2$ are included. CMB angular power is
typically expressed in terms of ${\cal D}_\ell =\ell (\ell
+1)C_\ell /2\pi,$ which in the flat sky approximation corresponds
to rms power (in $\mu K^2$) per unit logarithmic interval in the
spatial frequency. A scale invariant temperature spectrum on the
celestial sphere would correspond to $\ell ^2C_\ell$ being
constant. For comparison we give here a few ballpark numbers
characterizing the strength of the primordial CMB temperature
anisotropy. Figure~\ref{PlanckPowerSpectrum} plots the CMB power
spectrum as observed by Planck together with the predictions of
the fit to a six-parameter theoretical model, which we will
discuss further below. The magnitude at low $\ell,$ before the
rise to the first acoustic, or Doppler, peak at $\ell \approx
220,$ is ${\cal D}_\ell \approx 10^3\,\mu K^2,$ which would
correspond to an rms temperature in the neighborhood of $30\,\mu
K.$ The rise to the first acoustic peak increases the power by
about a factor of six, and then after approximately five
oscillations, damping effects take over, making the oscillations
less apparent and causing the spectrum to suffer a
quasi-exponential decay.

Although the detectors (to the extent that their response is ideal
or linear) measure the intensity expressed as the ``spectral
radiance'' or ``specific intensity'' $I_\nu $ (having units of
$\textrm{erg}~\textrm{s}^{-1}~\textrm{cm}^{-2}~\textrm{Hz}^{-1}~\textrm{str}^{-1}$)
averaged over a frequency band defined by the detector, it is
convenient to re-express these intensities in terms of an
effective temperature. There are two types of effective
temperature: the Rayleigh-Jeans (\hbox{R-J}) temperature and the
thermodynamic temperature. It is important to keep in mind the
distinction, especially at high frequencies [in this context $\nu
\gtorder \nu _{{\rm CMB}}=h^{-1}(k_{\rm B}T_{{\rm CMB}})=57$\,GHz]
where the factor relating the two becomes large.

In the R-J limit, where $(h\nu /k_{\rm B}T)\ll 1,$ the blackbody spectral
radiance may be approximated as $I_\nu =B_\nu(T)\approx 2(\nu /c)^2(k_{\rm B}T),$
as one would obtain classically from the density of states of the radiation field
and assigning an energy $(k_{\rm B}T)$ to each harmonic oscillator degree of
freedom. If we invert, assuming the R-J limit above (regardless of whether this
limit is valid), we obtain the following definition for the R-J temperature
 \begin{eqnarray}
T_{R-J}(\nu )=\frac{1}{2}\frac{1}{k_B}\left(\frac{c}{\nu
}\right)^2 I_\nu.
 \end{eqnarray}
The ``thermodynamic'' temperature corresponding to a specific intensity $I_\nu$ at a
certain frequency, on the other hand, is obtained by inverting the unapproximated
blackbody expression $I_\nu =B_\nu (T)=(2h\nu ^3/c^2)[\exp (h\nu /k_{\rm B}T)-1]^{-1}.$
For the CMB, where the variations about the average CMB temperature are small (i.e.,
$\Delta T/T\approx 10^{-5}$), linearized perturbation theory is valid and one can invert
this equation, obtaining
 \begin{eqnarray}
\delta T_{{\rm CMB}}= \frac{(e^x-1)^2}{x^2e^x}~ \delta
T_{R-J},\label{DefThermoTemp}
\end{eqnarray}
where $x=h\nu /k_{\rm B}T.$ This factor is approximately one for $x\ltorder 1,$
but rises exponentially for $x\gtorder 1.$

It should be remembered that for adding optically thin emissions, it is the
intensities, and not the thermodynamic temperatures, that add. Moreover,
foreground emission power laws know nothing about $T_{{\rm CMB}}$ and thus are
naturally expressed using either the specific intensity, or equivalently the R-J
temperature.

Other important milestones of CMB observation include the DASI discovery of the
polarization of the CMB in 2002 \cite{dasi} and the observations at small angular scales
carried out by the ACT and SPT teams, which at the time of writing constitute the best
measurements of the microwave anisotropies on large patches of the sky at high angular
resolution, measuring the power spectrum up to $\ell \approx 10$,000. The DASI experiment
was carried out using several feed horns pointed directly at the sky. Correlations were
taken of the signals from pairs of horns, thus measuring the sky signal
interferometrically. The 6\,m diameter Atacama Cosmology Telescope (ACT) situated in the
Atacama desert in Chile \cite{act-instrument,act-cosmo-params} and the 10\,m SPT (South
Pole Telescope) located at the Amundsen-Scott South Pole Station in Antarctica
\cite{keisler,spt-instrument,spt-secondary} probe the microwave sky on very small angular
scales, beyond the angular resolution of Planck. CMB observations are almost always
diffraction limited, so that the angular resolution is inversely proportional to the
telescope diameter. Ground based instruments, despite all their handicaps (i.e.,
atmospheric interference, lack of stability over long timescales, ground pickup from far
sidelobes) will always outperform space based experiments in angular resolution. There
are dozens of other suborbital experiments not mentioned here but which paved the road
for contemporary CMB observation. These experiments were important not only because of
their observations, but also because of their role in technology development and in the
development of new data analysis techniques. More information on these experiments can be
found in other earlier reviews.\footnote{See
http:/$\!$/lambda.gsfc.nasa.gov/links/experimental\_sites.cfm for an extensive
compilation of CMB experiments with links to their websites. The book
\cite{partridgeBook} provides an insightful account emphasizing the early history of CMB
observations with contributions from many of the major participants.}

\section[Brief thermal history of the universe]
{Brief Thermal History of the Universe}

The big bang model of the universe is an unfinished story. Certain
aspects of the big bang model are well established
observationally, while other aspects are more provisional and
represent our present best bet speculation. In the account below,
we endeavor to distinguish what is relatively certain and what is
more speculative.

If we assume: (i) the correctness of general relativity, (ii) that the universe is
homogeneous and isotropic on large scales (at least up to the size of that part of
the universe presently observable to us), and (iii) that for calculating the
behavior of the universe, small-scale anisotropies can be averaged over, we obtain
the Friedmann-Lama\^itre-Robertson Walker (FLRW) solutions to the Einstein field
equations, which we now describe. The metric for this family of solutions takes
the following form:
\begin{eqnarray}
ds^2=-dt^2+a^2(t)\gamma _{ij} dx^i dx^j, \label{FRWGen}
\end{eqnarray}
where $(i,j=1,2,3)$ and the line element $d\ell ^2=\gamma _{ij}
dx^i dx^j$ describes a maximally symmetric three-dimensional space
that may be Euclidean (flat), hyperbolic, or
spherical.\footnote{More generally, the three-dimensional line
element
\begin{eqnarray*}
ds^2= \gamma _{ij}dx^idx^j= \frac{{dx_1}^2+{dx_2}^2+{dx_3}^2}
{\left(1+\left(\frac{k}{4}\right)({x_1}^2+{x_2}^2+{x_3}^2)\right)^2},
\end{eqnarray*}
where $k=+1, 0,$ or $-1$ corresponds to spherical, flat (Euclidean), or hyperbolic three-dimensional
geometry, respectively. Apart from an overall change of scale, these are the only geometries satisfying
the hypotheses of spatial homogeneity and isotropy. By substituting $\tan (\chi /2)=r/2$ or $\tanh (\chi
/2)=r/2$ for the cases $k=+1$ or $k=-1,$ we may obtain the more familiar representations 
$ds^2=d\chi ^2+\sin \chi ^2 d{\Omega _{(2)}}^2$ 
or 
$ds^2=d\chi ^2+\sinh \chi ^2 d{\Omega _{(2)}}^2,$ 
respectively,
where $d{\Omega _{(2)}}^2=d\theta ^2+\sin \theta ^2d\phi ^2.$} 
For the flat case, which is the most important and most often discussed, Eq.~(\ref{FRWGen}) takes the form
\begin{eqnarray}
ds^2=-dt^2+ a^2(t)[dx_{1}^2 +{dx_2}^2 +{dx_3}^2].
\end{eqnarray}
Here $a(t)$ represents the {\it scale factor} of the universe.
Under the assumption of homogeneity and isotropy, the
stress-energy tensor (expressed as a mixed tensor with one
contravariant and one covariant index) must take the form
 \begin{eqnarray}
{T_{\mu }}^\nu = \left(\begin{array}{@{}c@{\quad}c@{\quad}c@{\quad}c@{}}
 \rho (t)& 0 & 0 & 0 \\
 0 & p(t)& 0 & 0 \\
 0 & 0 & p(t) & 0 \\
 0 & 0 & 0 & p(t)\\
\end{array}\right)\!.
 \end{eqnarray}

The Einstein field equations $G_{\mu \nu }=R_{\mu \nu }-(1/2)g_{\mu
\nu }R=(8\pi G)T_{\mu \nu }$ may be reduced to the two equations
\begin{eqnarray}
H^2(t)\equiv \frac{\dot a^2(t)}{a^2(t)}= \frac{8\pi G_N}{3}\rho (t)-\frac{k}{a^2(t)}
\end{eqnarray}
and
\begin{eqnarray}
\frac{\ddot a(t)}{a(t)}= -\frac{4\pi G}{3}(\rho +3p).
\end{eqnarray}
(For more details, see for example the discussion in the books by Weinberg
\cite{weinbergBook} or Wald \cite{waldBook}.)

The above discussion, which included only general relativity, is incomplete
because no details concerning the dynamics of $\rho (t)$ and $p(t)$ have been
given. The only constraint imposed by general relativity is stress-energy
conservation, which in the most general case is expressed as ${T_{\mu \nu }}^{;\nu
}=0,$ but in the special case above takes the form
\begin{eqnarray}
\frac{d\rho (t)}{dt}=-3H(t)[\rho (t)+p(t)]. \label{StressEnergyCons}
\end{eqnarray}

The relationship between $\rho $ and $p$ is known as the equation of state, and for the
special case of a perfect fluid $p$ depends only on $\rho $ and on no other variables.
Two special cases are of particular interest: a nonrelativistic fluid, for which $p/\rho
=0$, and an ultrarelativistic fluid, for which $p/\rho =1/3.$ Until the mid-90s it was
commonly believed that except at the very beginning of the universe, a two-component
fluid consisting of a radiation component and matter component suffices to account for
the stress-energy filling the universe. In such a model
\begin{eqnarray}
\rho (t)= \rho _{m,0} \left( \frac{a_0}{a(t)}\right) ^3 +\rho _{r,0} \left(
\frac{a_0}{a(t)}\right) ^4.
\end{eqnarray}
Here the subscript $0$ refers to the present time. For the radiation component, there is
obviously the contribution from the CMB photons, whose discovery was recounted above in
Sec.~\ref{ObsMicroSky}. If we assume that the universe started as a plasma at some very
large redshift with a certain baryon number density, most conveniently parametrized as a
baryon-to-photon or baryon-to-entropy ratio, we would conclude that there was statistical
equilibrium early on between the various species, and from this assumption we can
calculate the neutrino density today. It turns out that the neutrinos are colder than the
photons because the neutrinos fell out of statistical equilibrium before the electrons
and positrons annihilated. The $e^+e^-$ annihilation had the effect of heating the
photons relative to the neutrinos, because all the entropy that was in the electrons and
positrons was dumped into the photons. Consequently, assuming a minimal neutrino sector,
we can calculate the number of effective bosonic degrees of freedom due to the cosmic
neutrino background, which according to big bang theory must be present but has never
been detected. Measuring the number of baryons in the universe today is more difficult,
and is the subject of an extensive literature. The realization of a need for some
additional nonbaryonic ``dark'' matter dates back to Fritz Zwicky's study of the Coma
cluster in 1933 and is a long story that we do not have time to go into. Today ``cold
dark matter'' is part of the concordance model, where the name simply means that whatever
the details of this extra component may be (the lightest supersymmetric partner of the
ordinary particles in the Standard electroweak model, the axion, or yet something else),
for analyzing structure formation in the universe, we can treat this component as
nonrelativistic (cold) particles that can be idealized as interacting only through
gravitation.

In the mid-1990s it was realized that a model where the
stress-energy included only matter and radiation components could
not account for the observations. A crucial observation for many
researchers was the measurement of the apparent luminosities of
Type Ia supernovae as a function of redshift
\cite{supernovaA,supernovaB}, although at that time there were
already several other discordant observations indicating that the
universe with only matter and radiation could not account for the
observations.\footnote{See for example some of the contributions
in Ref.~\cite{CriticalDialogues} and references therein.} To
reconcile the observations with theory, another component having a
large negative pressure, such as would arise from a cosmological
constant, was required.

A cosmological constant makes a contribution to the stress-energy $T_{\mu \nu }$
proportional to the metric tensor $g_{\mu \nu }$ meaning that $p=-\rho $ exactly. This
relation inserted into the right-hand side of Eq.~(\ref{StressEnergyCons}) implies that
any expansion (or contraction) does not change the contribution to the density from the
cosmological constant.

With a cosmological constant, the density of the universe scales in the following
manner:
\begin{eqnarray}
\rho (t)= \rho _{\Lambda,0} \left( \frac{a_0}{a(t)}\right) ^0 +\rho _{m,0} \left(
\frac{a_0}{a(t)}\right) ^3 +\rho _{r,0} \left( \frac{a_0}{a(t)}\right) ^4.
\end{eqnarray}
The cosmological constant is most important at late times, when its density
rapidly comes to dominate over the contribution from nonrelativistic matter
(consisting according to our best current understanding of baryons and a so-called
cold dark matter component). It is customary to express the contributions of each
component as a fraction of the critical density $\rho _{{\rm crit}}=3H^2/8\pi G,$
which would equal the total density if the universe were exactly spatially flat.
For the $i$th component, we define $\Omega _i=(\rho _i/\rho _{{\rm crit}}).$

From the present data, we do not know whether the new component (or components)
with a large negative pressure is a cosmological constant, although current
observations indicate that $w=p/\rho $ for this component must lie somewhere near
$w=-1.$ A key question of many of the major initiatives of contemporary
observational cosmology is to measure $w$ (and its evolution in time) in order to
detect a deviation of the behavior of the dark energy from that predicted by a
cosmological constant.

The earliest probe of the hot big bang model arises from comparing
the primordial light element abundances as inferred from
observations to the theory of primordial nucleosynthesis. In a hot
big bang scenario, there is only one free \hbox{parameter} the
baryon-to-photon ratio $\eta _B,$ or equivalently, the
baryon-to-entropy ratio $\eta _S,$ which is related to the former
by a constant factor. As discussed above, in the hot big bang
theory at early times, in the limit as $a(t)\to 0,$ we have $\rho
\to \infty ,$ and $T\to \infty.$ For calculating the primordial
light element abundances, we do not need to extrapolate all the
way back to the initial singularity, where these quantities
diverge. Rather it suffices to begin the calculation at a
temperature sufficiently low so that all the baryon number of the
universe is concentrated in nucleons (rather than in free quarks),
but high enough so that there are only protons and neutrons rather
than bound nuclei consisting of several nucleons, and so that the
reaction
\begin{eqnarray}
p+e^- \rightleftharpoons n + \nu _e, \label{reacs}
\end{eqnarray}
(as well as related reactions) proceeds at a rate faster than the expansion rate
of the universe $H.$ Under this assumption, the ratio of protons to neutrons is
determined by an equilibrium condition expressed in terms of chemical potentials
\begin{eqnarray}
\mu (p)+\mu (e^-)= \mu (n) + \mu (\nu _e)+\Delta M,
\end{eqnarray}
where
\begin{eqnarray}
\Delta M=M_N-M_P-M_e =939.56\,{\rm MeV}-938.27\,{\rm MeV}-511\,{\rm KeV}=0.78\,{\rm
MeV}.
\end{eqnarray}
At first, when $T\gg (\Delta M),$ the protons and neutrons have almost precisely equal
abundances, but then, as the universe cools down and expands, the neutrons become less
abundant than the protons owing to the neutron's slightly greater mass. Then the
reactions (\ref{reacs}) freeze out---that is, their rate becomes small compared to
$H$---and the neutron-proton ratio becomes frozen in. Subsequently protons and
neutrons fuse to form deuterium, almost all of which combines to form ${}^4{\rm He},$
either directly from a pair of deuterons or through first forming ${}^3{\rm He},$ which
subsequently fuses with a free neutron. Almost all the neutrons that do not decay end up
in the form ${}^{4}{\rm He},$ because of its large binding energy relative to other low A
nuclei, but trace amounts of other light elements are also produced. Even though other
heavier nuclei have a larger binding energy per nucleon and their production would be
favored by equilibrium considerations, the rates for their production are too small,
principally because of the large Coulomb barrier that must be overcome.

In primordial big bang nucleosynthesis (BBN) the only free parameter is~$\eta_B.$ Since
nucleosynthesis takes place far into the epoch of radiation domination, the density as a
function of temperature is determined by the equation $\rho (T)=({\pi ^2}/{30})N_{{\rm
rel}}T^4$ where $N_{{\rm rel}}$ is the effective number of bosonic degrees of freedom.
The expansion rate then is important because it determines for example how many neutrons
are able to be integrated into nuclei heavier than hydrogen before decaying. Based on the
particles known to us that are ultrarelativistic during nucleosynthesis, the photon,
electron, positron, neutrinos, and antineutrinos, we think that we know what value
$N_{{\rm rel}}$ should have. But nucleosynthesis can be used to test for the presence of
extra relativistic degrees of freedom. We shall see later that the CMB can also be used
to constrain $N_{{\rm rel}}.$ For a review with more details about primordial
nucleosynthesis, see Ref.~\cite{nucleo-review}.

\section[Cosmological perturbation theory: describing a nearly perfect 
universe using general relativity]%
{Cosmological Perturbation Theory: Describing a Nearly Perfect Universe Using General Relativity}

The broad brush account of the history of the
universe presented in the previous section tells an average story, where spatial
homogeneity and isotropy have been assumed. As we shall see, early on, at large
$z,$ this story is not far from the truth, especially on large scales. But a
universe that is exactly homogeneous and isotropic would be quite unlike our own,
in that there would be no clustering of matter, observed very concretely in the
form of galaxies, clusters of galaxies, and so forth. In this section, we present
a brief sketch of the theory of cosmological perturbations. For excellent reviews
with a thorough discussion of the theory of cosmological perturbations in the
framework of general relativity, 
see Refs.~\cite{kodamaSasaki} and~\cite{MukhanovFeldmanBrandenberger}.

For the ``scalar'' perturbations,\footnote{In the theory of cosmological
perturbations, the terms ``scalar,'' ``vector,'' and ``tensor'' are given a
special meaning. A tensor of any rank that can be constructed using derivatives
acting on a scalar function and the Kronecker delta is regarded as a ``scalar''. A
quantity that is not a ``scalar'' but can be constructed in the same way from a
3-vector field is regarded as ``vector,'' and a ``tensor'' is a second-rank tensor
with its ``scalar'' and ``vector'' components removed. Under these definitions,
the linearized perturbation equations reduce to a block diagonal form in which the
``scalar,'' ``vector,'' and ``tensor'' sectors do not talk to each other.} we may
write the line element for the metric with its linearized perturbations $\Phi
(\mathbf{x}, \eta )$ and $\Psi (\mathbf{x}, \eta )$ in the following form:
 \begin{eqnarray}
ds^2=a^2(\eta ) [ -(1+2\Phi )d\eta ^2 +(1-2\Psi )d\mathbf{x}^2].
\label{NewtonianMetricPerturbation}
\end{eqnarray}
The functions $\Psi (\mathbf{x}, \eta )$ and $\Phi (\mathbf{x}, \eta )$ are the Newtonian
gravitational potentials. At low velocities $\Psi (\mathbf{x}, \eta )$ is most relevant,
but $\Phi (\mathbf{x}, \eta )$ can probed obsevationally using light or some other type
of ultrarelativistic particles. When there are no anisotropic stresses (i.e., all the
partial pressures of the stress-energy tensor are equal), $\Phi =\Psi.$

For a universe filled with a perfect fluid with an equation of state $p=p(\rho ),$
the evolution of the Newtonian gravitational potential $\Phi (\mathbf{x}, \eta )$
is governed by the equation (see Ref.~\cite{MukhanovFeldmanBrandenberger} for a
derivation and more details)
 \begin{eqnarray}
\Phi ^{\prime \prime } + 3 {\cal H} (1 +{c_s}^2) \Phi ^{\prime } -{c_s}^2\nabla
^2\Phi + [ 2{\cal H}^\prime + (1 +3{c_s}^2) {\cal H}^2 ] \Phi =0,
 \end{eqnarray}
where ${c_s}^2=dp/d\rho,$ ${\cal H}=a'/a,$ and ${}'=d/d\eta.$ Here
$\eta $ is the conformal time, related to the more physical proper
time by the relation $dt=a(\eta )d\eta .$

If the spatial derivative term (i.e., $\nabla ^2\Phi $) is
neglected, which is an approximation valid on superhorizon scales,
then the derived quantity \cite{bardeenGIF,lyth-zeta}
 \begin{eqnarray}
\zeta =\Phi +\frac{2}{3}\frac{(\Phi +{\cal H}^{-1}\Phi ^\prime) }{(1+w)},
 \end{eqnarray}
(where $w=p/\rho$) is conserved on superhorizon scales. This approximate invariant, which
can be related to the spatial curvature of the surfaces of constant density, is
invaluable for relating perturbations at the end of inflation (or some other very early
noninflationary epoch) to the later times of interest in this chapter. This property is
particularly useful given our ignorance of the intervening epoch, the period of
``reheating'' or ``entropy generation,'' the details of which are unknown and highly
model dependent.

To understand the qualitative behavior of the above equation, it is important to focus on the role of the
Hubble parameter $H(t)=\dot a(t)/a(t),$ which at time $t$ defines the dividing line between superhorizon
modes [for which $| \mathbf{k}_{{\rm phys}}| \ltorder H(t)$] and subhorizon modes [for which $|
\mathbf{k}_{{\rm phys}}| \gtorder H(t)$].\footnote{Here ``horizon'' means ``apparent horizon,'' which
depends only on the instantaneous expansion rate, unlike the ``causal horizon,'' which depends on the
entire previous expansion history.} This distinction relates to the relevance of the spatial gradient
terms in the evolution equations for the cosmological perturbations. This same distinction may be
re-expressed in terms of the conformal time $\eta $ and the co-moving wavenumber $\mathbf{k},$ where
${\cal H}=a'/a$ takes the place of $H=\dot a/a,$ with superhorizon meaning $| \mathbf{k}| \ltorder {\cal
H}(t)$ and subhorizon meaning $| \mathbf{k}| \gtorder {\cal H}(t).$ Here dots denote derivatives with
respect to proper time whereas the primes denote derivatives with respect to conformal time. It turns out
that for calculations, working with conformal time and co-moving wavenumbers is much more convenient than
working with physical variables.

What happens to the perturbations as the universe expands depends crucially on whether
the co-moving horizon size is expanding or contracting. During inflation, or more
generally when $w<-1/3,$ the co-moving horizon size is contracting. This means that the
dynamics of Fourier modes, which initially hardly felt the expansion of the universe,
become more and more affected by the expansion of the universe. Terms proportional to $H$
and $H^2$ in the evolution equations are becoming increasingly relevant whereas the
spatial gradient terms are becoming increasingly irrelevant as the mode ``exits'' the
horizon, finally to become completely ``frozen in'' on superhorizon scales. During
inflation a mode starts far within the horizon and crosses the horizon at some moment
during inflation. At the end of inflation, all the modes of interest lie frozen in, far
outside the horizon.

During inflation, $w=p/\rho $ was slightly more positive than
$-1.$ Formally inflation ends when $w$ crosses the milestone
$w=-1/3,$ which is when the co-moving horizon stops shrinking and
starts to expand. This moment roughly corresponds to the onset of
``reheating'' or ``entropy generation,'' when the vacuum energy of
the inflaton field gets converted into radiation, or equivalently
ultrarelativistic particles, which afterward presumably constitute
the dominant contribution to the stress-energy ${T_\mu }^\nu,$ so
that $w\approx 1/3.$ During the radiation epoch following
``entropy generation,'' modes successively re-enter the horizon
again becoming dynamical. The degrees of freedom describing the
modes are not the same as previously during inflation. This is at
present our best-bet story of what likely happened in the very
early universe. In this chapter, we shall be interested in the
later history of the universe, when modes are re-entering the
horizon.

We shall not enter into the details of how the primordial perturbations are generated
from quantum fluctuations of the vacuum during inflation, instead \hbox{referring} the
reader to the chapter by K.~Sato and J.~Yokoyama on {\it Cosmic Inflation} for this early
part of the story. Nor shall we discuss alternatives to inflation. The original papers in
which the scalar perturbations generated from inflation were first \hbox{calculated}
include Mukhanov \cite{Mukhanov:1981xt,Mukhanov:1982nu,Mukhanov:1985rz}, Hawking
\cite{Hawking:1982cz}, Guth \cite{Guth:1982ec}, Starobinsky \cite{Starobinsky:1982ee},
and Bardeen {\it et~al.} \cite{Bardeen:1983qw}. For early discussion of the generation of
gravity waves during inflation and their subsequent imprint on the CMB, see
Refs.~\cite{RR1,RR2,RR3,abbottWise}, and \cite{RR4}. A more
pedagogical account may be found for example in the books by Liddle and Lyth
\cite{LiddleLythBook} and by Peacock \cite{PeacockBook}, as well as in several review
articles, including for example \cite{inflationReviews}.

We emphasize the simplest type of cosmological perturbations, known as primordial
``adiabatic'' perturbations. The fact that the perturbations are primordial means
that at the late times of interest to us, only the growing adiabatic mode is
present, because whatever the amplitude of the decaying mode may initially have
been, sufficient time has passed for its amplitude to decay away and thus become
irrelevant. The term ``adiabatic'' deserves some explanation because its customary
meaning in thermodynamics does not precisely correspond to how it is used in the
context of cosmological perturbations. In a universe with only adiabatic
perturbations, at early times all the components initially shared a common
equation of state and common velocity field, and moreover on surfaces of constant
total density, the partial densities of the components are also spatially
constant. This is the state of affairs initially, on superhorizon scales, but
subsequently after the modes enter the horizon, the different components can and
do separate.

Adiabatic modes are the simplest possibility for the primordial cosmological
perturbations, but they are not the only possibility. Isocurvature perturbations where
the equation of state varies spatially are also possible, as discussed in detail for
example in Refs.~\cite{bondEfstathiou}, \cite{genCosmoPert}
and~\cite{MKamAKos1999} references therein. For isocurvature perturbations, the ratios
of components vary on a constant density surface, on which the spatial curvature is
constant as well. However the ratios between the components are allowed to vary. As the
universe expands, these variations in the equation of state eventually also generate
curvature perturbations. It might be argued that the isocurvature modes should
generically be expected to have been excited with some nonvanishing amplitude in
multi-field models of inflation. For recent constraints on isocurvature perturbations
from the CMB, see Planck Inflation 2013 \cite{planck-inflation} and the extensive list of
references therein.

\section{Characterizing the primordial power spectrum}

In the previous section, we defined the dimensionless invariant $\zeta (\textbf{x}),$
conserved on superhorizon scales, which can be used to characterize the primordial
cosmological perturbations at some convenient moment in the early universe when all the
relevant scales of interest lie far outside the horizon. Because this quantity is
conserved on superhorizon scales, the precise moment chosen is arbitrary within a certain
range. The existence of a quantity conserved on superhorizon scales allows us to make
precise predictions despite our nearly complete ignorance of the details of reheating.

Assuming a spatially flat universe, we may expand $\zeta (\textbf{x})$ into
Fourier modes, so that
\begin{eqnarray}
\zeta (\textbf{x})=\int \frac{d^3k}{(2\pi )^3} \zeta (\textbf{k}) \exp[i
\textbf{k} \cdot \textbf{x}].
\end{eqnarray}
If spatial homogeneity and isotropy are assumed, the correlation function of the
Fourier coefficients must take the form
\begin{eqnarray}
\langle \zeta (\textbf{k}) \zeta (\textbf{k}')\rangle= (2\pi )^3 k^{-3} {\cal
P}(k) \delta ^3( \textbf{k}-\textbf{k}'),
\end{eqnarray}
which also serves as the definition of the function ${\cal P}(k)$ known as the
primordial power spectrum. The $k^{-3}$ factor is present so that ${\cal P}(k)$ is
likewise dimensionless. An exactly {\it scale invariant} power spectrum
corresponds to ${\cal P}(k)\sim k^0.$ The meaning of scale invariance can be
understood by writing the mean square fluctuation at a point as the following
integral over wavenumber:
\begin{eqnarray}
\langle \zeta ^2(\textbf{x}=0) \rangle =\int _{-\infty }^{+\infty }d[\ln
(k)]{\cal P}(k).
\end{eqnarray}
Changes of scale correspond to rigid translations in $\ln (k),$
and the only choice for ${\cal P}(k)$ that does not single out a
particular scale is ${\cal P}(k)=\textrm{(constant)}.$

\section[Recombination, the blackbody spectrum, and spectral distortions]%
{Recombination, the Blackbody Spectrum, and Spectral Distortions}
\label{SpectralDistortions}

It is sometimes said that the CMB is the best blackbody in the universe. This is
not true. Observationally the only way to test how close the CMB is from a perfect
blackbody is to construct a possibly better artificial blackbody and perform a
differential measurement between the emission from the sky and this artificial
blackbody as a function of frequency.


In our discussion of the CMB, we stressed that theory predicted a
frequency spectrum having a blackbody form to a very high
accuracy. This indeed was one of the striking predictions of the
big bang theory, and the lack of any measurable deviation from the
perfect blackbody spectral shape over a broad range of frequencies
was probably one of the most important observational facts that
led to the demise of the alternative steady state cosmological
model. To date, the best measurements of the frequency spectrum of
the CMB are still those made by the COBE FIRAS instrument
\cite{mather1990,mather94,COBEfiras}. FIRAS
\cite{firas-instrument-paper} made differential measurements
comparing the CMB frequency spectrum on the sky to an artificial
blackbody. Despite the many years that have gone by since COBE and
the importance of measuring the absolute spectrum of the CMB, no
better measurement has been carried out because improving on FIRAS
would require going to space in order to avoid the spectral
imprint of the Earth's atmosphere. However a space mission concept
called PIXIE has been proposed that would essentially redo FIRAS
but with over two orders of magnitude better sensitivity and with
polarization sensitivity included \cite{pixie}. While differences
in temperature between different directions in the sky,
particularly on small angular scales, can be measured from the
ground, albeit with great difficulty, the same is not possible for
measurements of the absolute spectrum.

From a theoretical point of view, even in the simplest minimal cosmological model with no
new physics, deviations from a perfect blackbody spectrum are predicted that could be
measured with improved observations of the absolute spectrum having a sensitivity
significantly beyond that of FIRAS. It is often believed that when we look at the CMB, we
are probing conditions in the universe at around last scattering---that is, around
$z\approx 1100$---and that the CMB photons are unaffected by what happened during
significantly earlier epochs. This is not true. Although when $z\gtorder 1100,$ photons
are frequently scattered by electrons randomizing their direction, causing them to move
diffusively, this frequent scattering is inefficient at equilibrating the kinetic
temperature of the electrons to the energy spectrum of the photons and vice versa owing
to the smallness of the dimensionless parameter $E_\gamma /m_e,$ where $E_\gamma $ is a
typical photon energy. This ratio gives the order of magnitude of the fractional energy
exchange due to electron recoil for a typical collision. Given that in this limit the
approach to a blackbody spectrum is a diffusive process, we would estimate that
$(E_\gamma /m_e)^2$ collisions correspond to one decay time of the deviation of the
photon energy spectrum from the equilibrium spectrum. Plugging in $E_\gamma \approx
k_{\rm B}T_{z=1100}\approx 0.3\,{\rm eV},$ we get $(E_\gamma /m_e)^2\approx 4\times
10^{10}$! Thomson scattering, however, cannot equilibrate the photon number density with
the available energy density, so with Thomson scattering alone, an initially out of
equilibrium photon energy distribution, for example as the result of some sort of energy
injection, would generically settle down to a photon phase space distribution having a
positive chemical potential. Processes such as Bremsstrahlung and its inverse or double
Compton scattering are required to make the chemical potential decay to zero, and these
processes freeze out at $z\approx 10^6$ \cite{illarionov}. Therefore any energy injection
between $z\approx 10^6$ and $z\approx 10^3$ will leave its mark on the absolute spectrum
of the CMB photons, either in the form of a so-called $\mu $ distortion, or for energy
injected at later times, in the form of a more complicated energy dependence.

A broad range of interesting early universe science can be explored by searching
for deviations from a perfect blackbody spectrum. Some of the spectral distortions
are nearly certain to be present, while others are of a more speculative nature.
On the one hand, $y$-distortions in the field (that is, away from galaxy clusters
where the $y$-distortion is compact and particularly large) constitute a nearly
certain signal, as do the spectral lines from cosmological recombination. But
sources like decaying dark matter \cite{energy-release} may or may not be present.
Another interesting source of energy injection arise from Silk damping
\cite{silkDamping,psSD} of the primordial power spectrum on small scales, beyond
the range of scales where the primordial power spectrum can be probed by other
means. If a space-based instrument is deployed with the requisite sensitivity, a
formidable challenge will be to distinguish these signals from each other, and
also from the more mundane galactic and extragalactic backgrounds. From this
discussion one might conclude, borrowing terminology from stellar astrophysics,
that the CMB ``photosphere'' extends back to around $z\approx 10^6.$

\section[Sachs-Wolfe formula and more exact anisotropy calculations]%
{Sachs-Wolfe Formula and More Exact Anisotropy Calculations}
\label{Section:BasicCMBphysics}

\begin{wrapfigure}{L}{0.6\textwidth }
\begin{center}
\includegraphics[width=9cm]{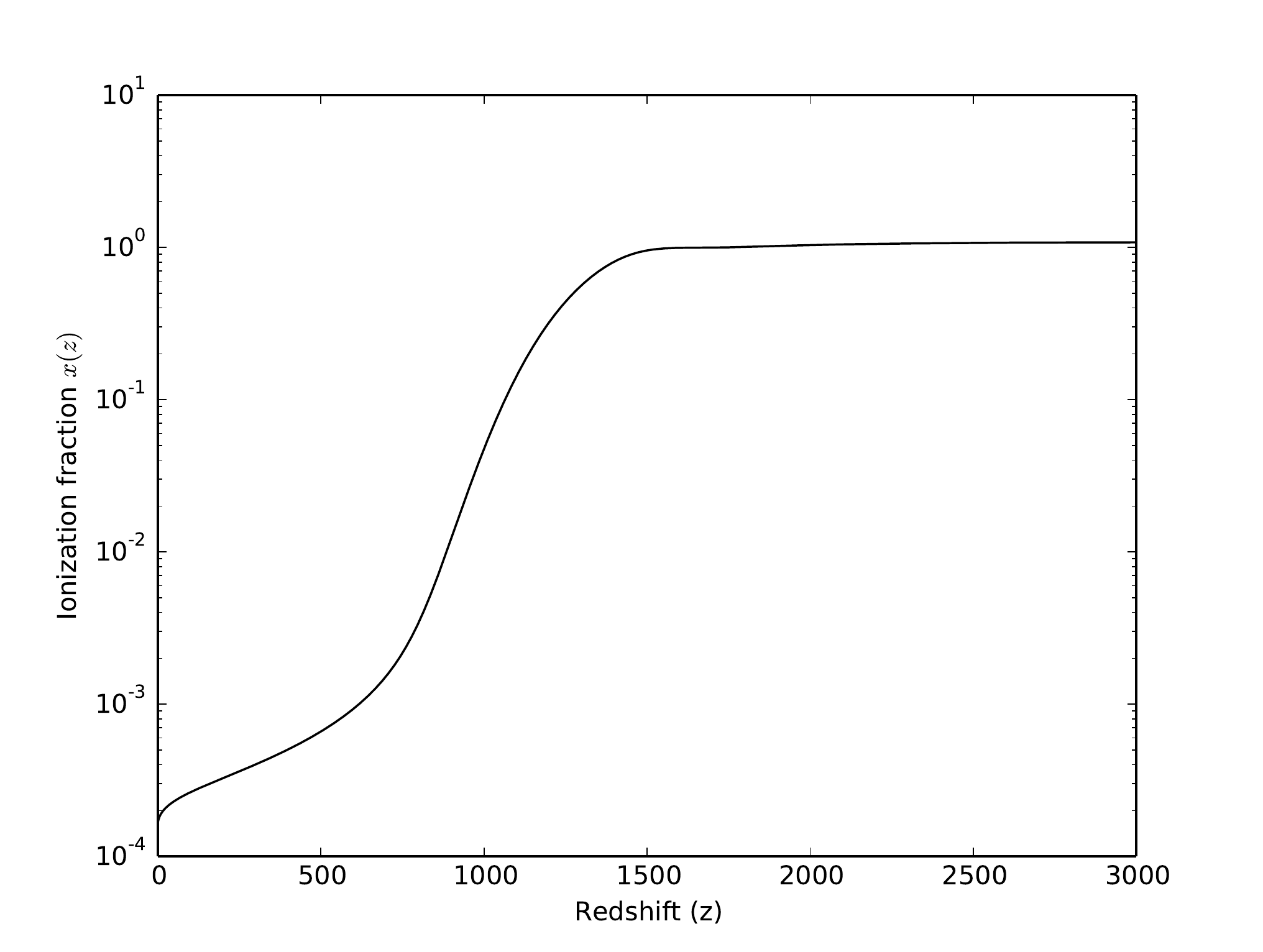}
\end{center}
\caption{{\bf Ionization fraction as function of redshift with effect of late-time ionizing
radiation ignored.} Calculated using the code {\it Cosmo Rec} by Jens Chluba (for details
see Ref.~\cite{WW1}).
\label{ionizationHistory}}
\end{wrapfigure}

At early times $(z\ltorder 1100)$ the universe is almost completely ionized and the
photons scatter frequently with free charged particles, primarily with electrons because
the proton--photon scattering cross-section is suppressed by a factor of $(m_e/m_p).$
During this period the plasma consisting of baryons, leptons, and photons may be regarded
as a single tightly-coupled fluid component, almost inviscid but with a viscosity
relevant on the shortest scales of interest for the calculation of the CMB temperature
and polarization anisotropies. Later, as the universe cools the electrons and protons
(and other nuclei) combine to form neutral atoms whose size greatly exceeds the thermal
wavelength of the photons, rendering the universe almost transparent. During this
process, known as ``recombination'' (despite the fact that the universe had never
previously been neutral), the photons cease to rescatter and free stream toward us
today \cite{peebles1968}. 
This last statement is only 90\% accurate, because the late-time
reionization starting at about $z\approx 6\hbox{--}7$ causes about 10\% of the CMB
photons to be rescattered. The effect of this rescattering on the CMB anisotropies will
be discussed in detail in Sec.~\ref{ReionizationSection}. It is during this transitional
epoch, situated around $z\approx 1100,$ that the CMB temperature and polarization
anisotropies were imprinted. Thus when we observe the microwave anisotropies on the sky
today, we are probing the physical conditions on a sphere approximately 47 billion light
years in radius today,\footnote{The radius of this sphere in terms of today's comoving
units is larger that the age of the universe (approximately 13.8 Gyr) converted into a
distance owing to the expansion of the universe.} or more precisely the intersection of
our past light cone at the surface of constant cosmic time at $z\approx 1100.$ This
sphere is known as the surface of last scatter or the last scattering surface.

The transition from a universe that is completely opaque to the
blackbody photons, in which the tight-coupling approximation holds
for the baryon-lepton plasma because the photon mean free path is
negligible, to a universe that is virtually transparent is not
instantaneous. This fact implies that the last scattering surface
(LSS) is not infinitely thin, rather having a finite width that
must be taken into account for calculating the small angle CMB
anisotropies, because this profile of finite thickness smears out
the small-scale three-dimensional inhomogeneities as they are
projected onto the two-dimensional celestial sphere.
Figure~\ref{ionizationHistory} shows a plot of the ionization
history of the universe (not taking into account the late-time
reionization at $z\gtorder 6$ mentioned above). (For a more
detailed discussion of the early ionization history of the
universe and the physics by which it is determined, see for
example Refs.~\cite{hs2008,sss1999,sss2000,sh2008} and
\cite{chlubaThomas}).

The first calculation of the CMB temperature anisotropies
predicted in a universe with linearized cosmological perturbations
was given by Sachs and Wolfe in their classic 1967 paper
\cite{SW1967} (see also Ref.~\cite{peeblesYu}). In their treatment, the LSS
surface is idealized as the surface of a three-dimensional
sphere---in other words, the transition from tight-coupling to
transparency is idealized to be instantaneous. Locally, on this
surface the photon-baryon fluid is subject to two kinds of
perturbations: (i) perturbations in density $\delta _{\gamma -b}$
and (ii) velocity perturbations $\mathbf{v}_b$. The former
translate into fluctuations in the photon blackbody temperature
$T_\gamma,$ in ``intrinsic'' temperature fluctuations at the last
scattering surface with $\delta T_\gamma/\bar T_\gamma $ and the
second translate into a Doppler shift of the CMB temperature. If
there were no metric perturbations, we would simply have
 \begin{eqnarray}
\frac{\delta T_f(\hat \Omega )}{\bar T_f}= \frac{\delta T_i(\hat \Omega )}{\bar
T_i}-\frac{1}{c}\delta \mathbf{v}_\gamma \cdot \hat \Omega,
 \end{eqnarray}
where $\hat \Omega$ is the unit outward normal on the last
scattering surface. But there are also additional terms due to the
metric perturbations, and careful calculations along the perturbed
geodesics yield the following modification to the above equation:
 \begin{eqnarray}
\frac{\delta T_f(\hat \Omega )}{\bar T_f}= \frac{\delta T_i(\hat \Omega )}{\bar
T_i}-\frac{1}{c}\delta \mathbf{v}_\gamma \cdot \hat \Omega + \Phi +\int _{\eta _i}^{\eta
_f}d\eta (\Phi' +\dot{\Psi}'), \label{SachsWolfeFormula}
 \end{eqnarray}
where the metric perturbations are parametrized as in
Eq.~(\ref{NewtonianMetricPerturbation}) using conformal Newtonian
gauge.\footnote{The Sachs-Wolfe formula as given here only
includes the ``scalar'' perturbations. The formula can be
generalized to include ``vector'' and ``tensor'' contributions,
for which the only contributions are from the Doppler and ISW
(integrated Sachs-Wolfe) terms.} This equation is known as the
Sachs-Wolfe formula. Because of the approximations involved, the
Sachs-Wolfe formula in not used for accurate calculations
(especially on small angular scales). However, it offers a good
approximation to the CMB anisotropies on large angular scales and
provides invaluable intuition that is lacking in the more precise
treatments.

\begin{wrapfigure}{L}{0.6\textwidth }
\begin{center}
\includegraphics[width=11cm]{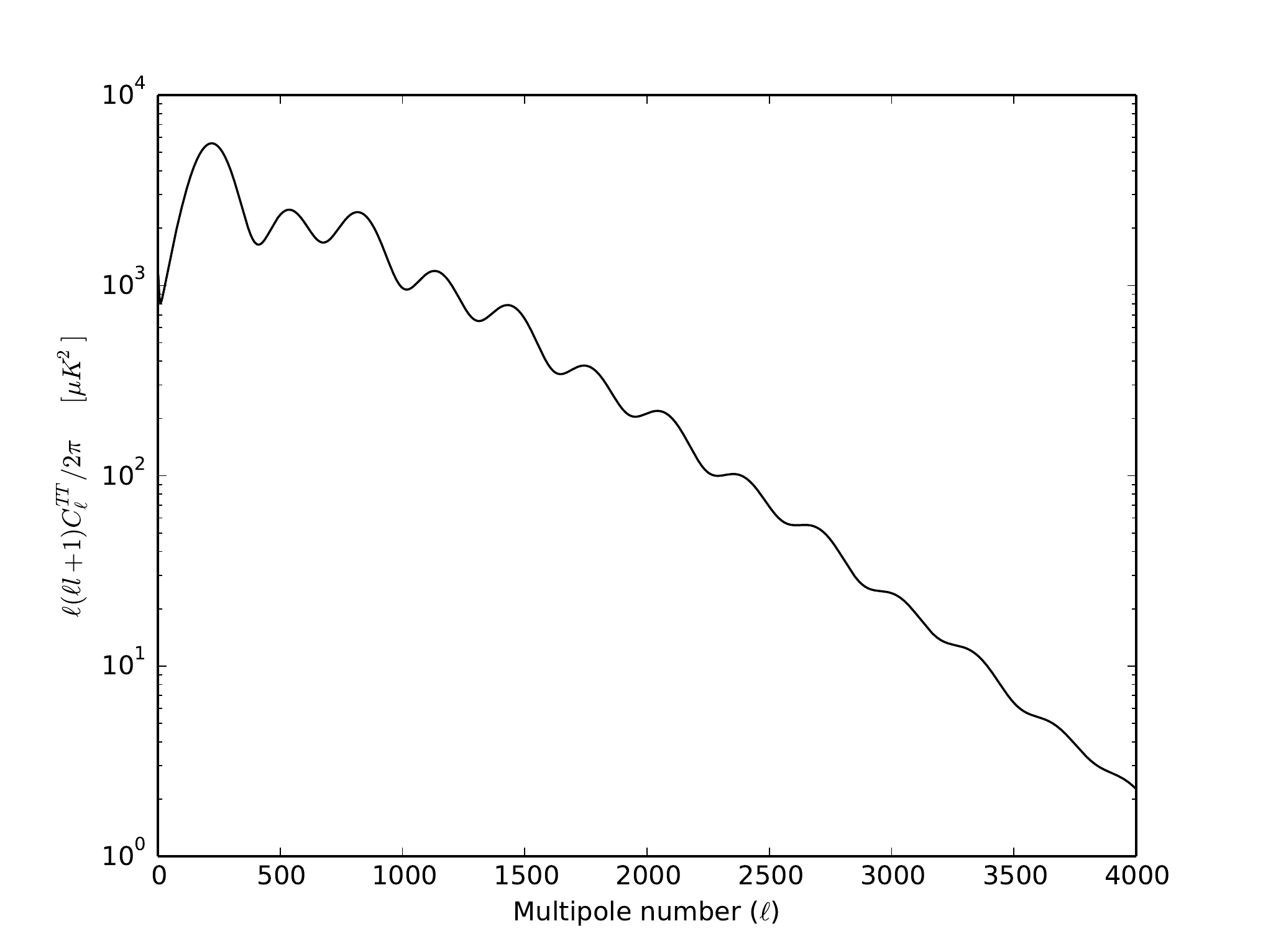}
\end{center}
\caption{{\bf Theoretical CMB spectrum.} CMB temperature power spectrum
predicted for a model with only the adiabatic growing mode
excited, as in the standard concordance cosmological model, is
shown. The theoretical model here assumes the best-fit
cosmological parameters taken from the Planck 2013 Results for
Cosmology \cite{planck-parameters}.
\label{CalculatedCMBSpectrum}}
\end{wrapfigure}

A few words about gauge dependence,\footnote{``Gauge dependence'' here
means invariance under general coordinate transformations, which consistent with the
linear approximation above may be truncated at linear order.} an issue that renders
linearized perturbation theory within the framework of general relativity somewhat messy.
While the final observed anisotropy ${\delta T_f(\hat \Omega )}/{\bar T_f}$ is the same
in all coordinate systems (except for the monopole and dipole terms), the attribution of
the total anisotropy among the various terms (i.e., intrinsic temperature fluctuation,
Doppler, gravitational redshift, integrated Sachs-Wolfe) depends on the choice of
coordinates. We may for example choose coordinates so that the photon temperature serves
as the time coordinate, making the first term disappear, or alternatively, we may make
the Doppler term disappear by making our co-moving observers move with the local photon
rest frame.

Students of cosmology are sometimes misled into
believing the solution to these ambiguities is to use conformal
Newtonian gauge, or equivalently the ``gauge invariant''
formalism, which is equivalent to transforming to conformal
Newtonian gauge. It is believed that this gauge choice is somehow
more ``physical'' because at least to linear order, the gauge
conditions lead to a unique choice of gauge. It is certainly nice
to have a gauge condition leading to a unique choice of
coordinates, but this uniqueness comes at a price. The Newtonian
gauge condition is ``nonlocal'' because of its reliance on the
decomposition of $h_{\mu \nu}$ into ``scalar,'' ``vector,'' and
``tensor'' components, which requires information extending all
the way out to spatially infinity. But no one has ever seen all
the way out to spatial infinity. For this reason, the Newtonian
potentials $\Psi$ and $\Phi$ are unphysical because they cannot be
measured. Their determination would require information extending
beyond our horizon.

We may further simplify Eq.~(\ref{SachsWolfeFormula}) assuming that only the
adiabatic growing mode has been excited. Using the fact that $\delta T_{i}$ and
$\mathbf{v}$ are not independent of $\Phi,$ we obtain the often cited result
\begin{eqnarray}
\frac{\delta T}{T}=-\frac{\Phi }{3} -\mathbf{v}\cdot \hat
{\boldsymbol{\Omega }} +\int d\eta (\Phi^\prime +\Psi ^\prime),
\end{eqnarray}
although the $-1/3$ factor is not quite right because a matter-dominated evolution
for $a(t)$ is assumed rather than a more careful treatment taking into account
matter-radiation equality occurring at around $z_{eq}\approx 3280.$ (For a
discussion see Ref.~\cite{whiteHuOneThird}.)

Using this result and ignoring the integrated Sachs-Wolfe term, we obtain
\begin{eqnarray}
\frac{\delta T}{T}(\hat {\boldsymbol{\Omega }})=-\frac{\Phi }{3}-\mathbf{v}\cdot
\hat {\boldsymbol{\Omega }}.
\end{eqnarray}
On large angular scales---that is, large compared to the angle subtended by the horizon
on the last scattering surface---the first term dominates. The modes labeled by
$\mathbf{k}$ obey an oscillatory system of coupled ODEs, and at the putative big bang
each mode starts with a definite sharp phase corresponding to the part of the cycle where
$\mathbf{v}=0.$ It is only at horizon crossing that the phase has evolved sufficiently
for the velocity term to contribute appreciably to the CMB anisotropy.
Figure~\ref{CalculatedCMBSpectrum} shows the calculated form of the CMB temperature power
spectrum. For $\ell \ltorder 100,$ $\Delta T/T\approx -\Phi /3$ provides an adequate
explanation for this leftmost plateau of the curve, but at higher $\ell $ a rise to a
first acoustic peak, situated at about $\ell \approx 220,$ is observed followed by a
series of decaying secondary oscillations.

\begin{wrapfigure}{L}{0.6\textwidth }
\begin{center}
\includegraphics[width=10cm]{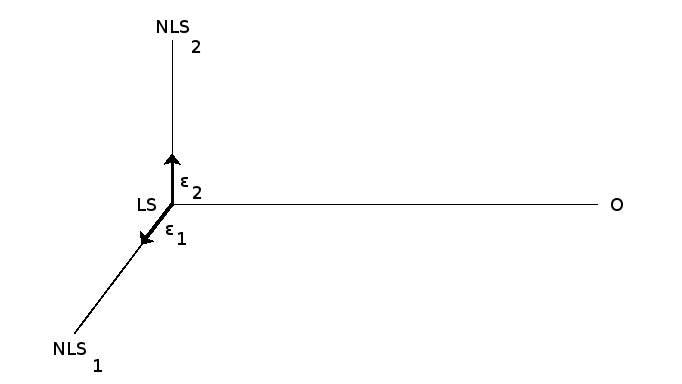}
\end{center}
\caption{{\bf Anisotropy of polarized Thomson scattering and origin of the CMB
polarization.} We show two examples of photon scattering from (NLS) (next-to-last
scattering) to (LS) (last scattering) and finally to the observer (at $O$.) The three
segments in the diagram have been chosen mutually at right angles in order to maximize
the effect, and for simplicity we assume that the radiation emanating from (NLS)$_1$ and
(NLS)$_2$ is unpolarized. All the radiation that scatters coming from (NLS)$_1$ to (LS)
and then towards $O$ is completely linearly polarized in the 
$\boldsymbol{\hat \epsilon }_2$ direction, because the other polarization 
is parallel to $\overline{({\rm LS})O}$
and thus cannot be scattered in that direction. Likewise, the scattered photons emanating
from (NLS)$_2$ are completely linearly polarized in the orthogonal direction for the same
reason. Consequently if there were only two sources as above, measuring the Stokes
parameter $Q$ at $O$ amounts to measuring the difference in intensity between these two
sources.
\label{PolarizationOrigin}}
\end{wrapfigure}

A more accurate integration of the evolution of the adiabatic mode up to last
scattering can be obtained in the fluid approximation. This approximation assumes
that the stress-energy content of the universe can be described by a fluid
description consisting of two components coupled to each other only by gravity
\cite{SeljakTwoFluid}. However more accurate calculations must go beyond the fluid
approximation using a Boltzmann formalism that includes higher order moments.

However, before describing the Boltzmann formalism, capable of describing this
intermediate regime for the photons, we note one final shortcoming of the
approximation with tight-coupling and instantaneous recombination---namely, the
treatment of polarization. We present here an approximate, heuristic treatment in
order to provide intuition.

The scattering of photons by electrons is polarization dependent, and this effect leads
to a polarization of the CMB anisotropy when the fact that recombination does not occur
instantaneously is taken into account. The Thomson scattering
\hbox{cross-section} of a photon off an electron is given by
 \begin{eqnarray}
\frac{d\sigma }{d\Omega }=\left( \frac{e^2}{mc^2} \right)^2(
\boldsymbol{\hat \epsilon}_i\cdot \boldsymbol{\hat \epsilon}_f)
^2,
\end{eqnarray}
where $\boldsymbol{\hat \epsilon}_i$ and $\boldsymbol{\hat \epsilon}_f$ are the
initial and final polarization vectors, respectively.

To see qualitatively how polarization of the CMB comes about, let us for the moment
assume that the photons coming from next-to-last scattering are unpo- larized and
calculate the polarization introduced at last scattering, as indicated in Fig.~13.
Measuring photons with one linear polarization selects photons that propagated from
next-to-last to last scattering at a small angle to the axis of linear polarization,
while photons with the other linear polarization tend to come from a direction from
next-to-last to last scattering with a small angle with the other axis. Consequently,
measuring the polarization amounts to measuring the temperature quadrupole as seen by the
electron of last scattering. If we make the simplifying assumption that the radiation
emanating from next-to-last scattering is unpolarized, we obtain the following expression
for the Stokes parameters from the linear polarization,
\begin{eqnarray}
\begin{pmatrix} Q\vspace*{5pt}\cr U
\end{pmatrix} _{l.s.}
&=& \int _{\rho =0}^{\rho =\rho _{\rm max}} d( -\!\exp [ -\tau (r_{l.s.}+\rho
)]) \int d\hat {\boldsymbol{\Omega }}~ \sin ^2\theta
\begin{pmatrix}
\cos \phi \cr \sin \phi
\end{pmatrix}\nonumber\\[3pt]
&& \times ~~T( \rho \sin \theta \cos \phi , \rho \sin \theta \sin \phi , r_{l.s.}+\rho \cos
\theta ) .
\end{eqnarray}
Here we assume that the line of sight is along the $z$ direction. The expression gives
two of the five components of quadrupole moment of the temperature as seen by the
electron at last scattering, and after integration along the line of sight would give the
total linear polarization with the polarization at next-to-last scattering neglected.

\section[What can we learn from the CMB temperature and polarization \break anisotropies?]%
{What Can We Learn From the CMB Temperature and Polarization Anisotropies?}

The previous Section showed how starting from Gaussian isotropic and homogeneous
initial conditions on superhorizon scales, the predicted CMB temperature and
polarization anisotropies are calculated, describing in detail all the relevant
physical processes at play. In this Section, we turn to the question of what we
can learn about the universe from these observations. We focus on how to exploit
the temperature and polarization two-point functions, which under the assumption
of Gaussianity would summarize all the available information characterizing the
underlying stochastic process. Gaussianity is a hypothesis to be tested using the
data, as discussed separately in Sec.~\ref{NonGaussianity}.


\subsection[Character of primordial perturbations: adiabatic growing mode versus field ordering]%
{Character of primordial perturbations: adiabatic growing mode versus field ordering}

\begin{figure}[t]
\begin{center}
\includegraphics[width=10cm]{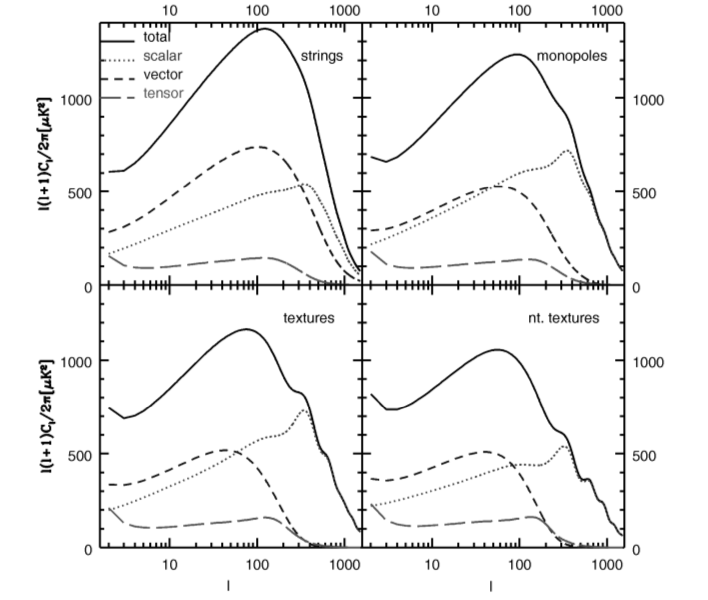}
\end{center}
\caption{{\bf Structure of acoustic oscillations: field ordering predictions.} These acoustic
oscillation predictions from four field-ordering models are to be contrasted with the
sharp, well-defined peaks in Fig.~\ref{CalculatedCMBSpectrum} as predicted by inflation.
(Reprinted with permission from Ref.~\protect\cite{PenSeljakTurok}.) ({\it
Credit: Pen, Seljak and Turok})\label{InflationVsFieldOrdering}}
\end{figure}

In the aftermath of the COBE/DMR announcement of the detection of the CMB temperature
anisotropy, two paradigms offered competing explanations for the origin of structure in
the early universe. On the one hand, there was cosmic inflation, which in its simplest
incarnations predicts homogeneous and isotropic Gaussian initial perturbations where only
the adiabatic growing mode is excited. On the other hand, there was also another class of
models in which the universe is postulated initially to have been perfectly homogeneous
and isotropic. However, subsequently a symmetry breaking phase transition takes place,
with the order parameter field taking uncorrelated values in causally disconnected
regions of spacetime. Then, as the universe expands and the co-moving size of the horizon
grows, the field aligns itself over domains of increasingly large co-moving size,
generally in a self-similar way. In many of these models, topological defects of varying
co-dimension---such as monopoles, cosmic strings, and domain walls---arise after the
phase transition, but textures where the spatial gradient 
energy is more diffusely
distributed are also possible.\footnote{For the physics of topological defects, see
Ref.~\cite{VMPreskill} and for a comprehensive account emphasizing the connection to
cosmology, see Ref.~\cite{ShellardVilenkinBook}.} In these models the contribution of
the field ordering sector to the total energy is always subdominant, and cosmological
perturbations are generated in a continuous manner extending to the present time. The
stress-energy $\Theta _{\mu \nu }$ from this sector sources perturbations in the metric
perturbation $h_{\mu \nu }$. These metric perturbations in turn generates perturbations
in the dominant components contributing to the stress-energy: the baryons, photons,
neutrinos, and cold dark matter. In these models, the cosmological perturbations are not
primordial, but rather are continuously generated so that the perturbations on the scale
$\mathbf{k}$ are generated primarily as the mode $\mathbf{k}$ enters the apparent
horizon \cite{penSpergelTurok}.
This property implies that the decaying mode as well as the growing mode
of the adiabatic perturbations is excited, and this fact has a spectacular effect on the
shape of the predicted CMB power spectrum. While inflationary models, which excite only
the growing mode, predict a series of sharp, well-defined acoustic oscillations, field
ordering models 
predict broad, washed out oscillations or no oscillations at all in the
angular CMB spectrum \cite{AlbrechtCausality,TexturePeaks,PenSeljakTurok}.
Heuristically one can understand this
behavior by arguing that the positions of the peaks reflect the phase of the
oscillations. Therefore, if both the growing and decaying modes are present in exactly
equal proportion, there should be no oscillations in the angular spectrum. However,
precise predictions require difficult numerical simulations and depend on the precise
model for the topological defect or field ordering sector.
\hbox{Figure}~\ref{InflationVsFieldOrdering} shows the shape of the predicted CMB TT
power spectrum for some field ordering models, to be compared with the predictions for
minimal inflation shown in Fig.~\ref{CalculatedCMBSpectrum}.

After COBE, a big question was whether improved degree-scale CMB observations would
unveil the first Doppler peak followed by a succession of decaying secondary peaks at
smaller angular scales, as predicted by an approximately scale invariant primordial power
spectrum for the growing adiabatic mode, or whether some other shape would be observed,
perhaps the one predicted by a field ordering model.

We already presented (in Sec.~\ref{ObsMicroSky}) the experimental
results for the first observations of the acoustic oscillations
and their subsequent precise mapping (see in particular
Figs.~\ref{BoomerangData}, \ref{stateOfPlay1998} and
\ref{PlanckPowerSpectrum}.) The data clearly favor simple models
of inflation producing adiabatic growing mode perturbations with
an approximately scale \hbox{invariant} power spectrum and exclude
scenarios where topological defects serve as the \hbox{primary}
source for the initial density perturbations. However, models
including a small admixture of defect induced perturbations cannot
be ruled out \cite{planck-defects}.

\newcommand\norm[1]{\| #1\|}
\subsection{Boltzmann hierarchy evolution}

In the previous section, we derived the Sachs-Wolfe formula, which provides an intuitive
understanding of how the CMB anisotropies are imprinted. The Sachs-Wolfe treatment
provides an approximate calculation of the CMB temperature anisotropies, correctly
capturing their qualitative features. However to calculate predictions accurate at the
sub-percent level, as is required for confronting current observations to theoretical
models, the transition from the tight-coupling regime to transparency, or the
free-streaming regime, must be modeled more realistically. This requires going beyond the
fluid approximation, where the photon gas can be described by its first few moments.
Modeling this transition faithfully requires a formalism with an infinite number of
moments---that is, a formalism in which the fundamental dynamical variable is the
photon phase space distribution function $f( \mathbf{x}, t, \nu, \hat {\mathbf{n}}, \hat
{\boldsymbol{\epsilon }})$ where $\nu $ is the photon frequency, and the unit vectors
$\hat {\mathbf{n}}$ and $\hat {\boldsymbol{\epsilon }}$ denote the photon propagation
direction and electric field polarization, respectively. The perturbations about a
perfect blackbody spectrum are small, so truncating at linear order provides an adequate
approximation. Moreover, Thomson scattering is independent of frequency, and thus (to
linear order) does not alter the blackbody character of the spectrum. Consequently, a
description in terms of a perturbation of the blackbody temperature that depends on
spacetime position, propagation direction, and polarization direction can be used in
place of the full six-dimensional phase space density, thus reducing the number of
dimensions of the phase space by one.\footnote{Although written as a continuum variable.
$\hat{\boldsymbol{\epsilon}}$ is in reality a discrete variable corresponding to the
Stokes parameters $(E, Q$ and $U$).} For a calculation correct to second order, a
description in terms of a perturbed temperature would not suffice.

As before we consider a single ``scalar'' Fourier mode $\mathbf{k}$ of flat
three-dimensional space, so that an $\exp [i\mathbf{k}\,{\cdot}\, \mathbf{x}]$ spatial
dependence is implied. A complication in \hbox{describing} polarization arises in
specifying a basis, which preferably is accomplished using only scalar quantities. We
describe the polarization using the unit vectors $\hat {\mathbf{n}}$ and $\hat
{\mathbf{k}}$ to define the unit vectors
\begin{eqnarray}
\hat {\boldsymbol{\theta }}= \frac{ \hat {\mathbf{k}}- \hat {\mathbf{n}} (\hat
{\mathbf{n}}\cdot \hat {\mathbf{k}}) }{ \norm{ \hat {\mathbf{k}}- \hat
{\mathbf{n}} (\hat {\mathbf{n}}\cdot \hat {\mathbf{k}}) } }, \quad \hat
{\boldsymbol{\phi }}= \frac{\hat {\mathbf{n}}\times \hat {\mathbf{k}}}{ \norm{
\hat {\mathbf{n}}\times \hat {\mathbf{k}} } },
 \end{eqnarray}
which together with photon propagation direction $\hat {\mathbf{n}}$ form an
orthonormal basis. We thus describe the Stokes parameters of the radiation
propagating along $\hat {\mathbf{n}}$ as follows:
 \begin{eqnarray}
\langle {\cal E}_a(\hat {\mathbf{n}}, \nu ) {\cal E}_b(\hat
{\mathbf{n}}, \nu ) \rangle = I(\hat {\mathbf{n}}, \nu ; \hat
{\mathbf{k}}) \delta _{ab} + Q(\hat {\mathbf{n}}, \nu ; \hat
{\mathbf{k}})[ \hat {\boldsymbol{\theta }}\otimes \hat
{\boldsymbol{\theta }} - \hat {\boldsymbol{\phi }}\otimes \hat
{\boldsymbol{\phi}}] _{ab}.
\end{eqnarray}
For the ``scalar'' mode, the Stokes parameters $U$ and $V$ vanish in this basis.

The Thomson scattering cross-section does not vary with frequency,
a property that provides substantial simplification. Suppose that
at some initial time the photon distribution function has a
blackbody dependence on frequency, with a blackbody temperature
suffering only small linearized perturbations about an average
temperature, and that these temperature perturbations depend on
the spacetime position, propagation direction, and linear
polarization direction. The frequency independence of Thomson
scattering ensures that a photon distribution function initially
having this special form will retain this special form during its
subsequent time evolution. Initially, when tight-coupling is an
excellent approximation, any deviation in the frequency dependence
from a perfect blackbody spectrum decays rapidly. Moreover, any
initial polarization is erased by the frequent scattering.
Therefore a blackbody form described by a perturbation in the
spectral radiance of the form
\begin{eqnarray}
\delta I_\nu (\hat {\mathbf{n}}, \hat {\boldsymbol{\epsilon }}, \mathbf{x}, t) =
\bar T_\gamma \frac{\partial{\kern-1pt} B_\nu (T)}{\partial T} \Delta _{ab}(\hat
{\mathbf{n}}, \mathbf{x}, t) \hat \epsilon ^a \hat \epsilon ^b
 \end{eqnarray}
is well motivated. Here
\begin{eqnarray}
B_\nu (T)=\left( \frac{2h\nu ^3}{c^2}\right) \left( \exp\! \left[\frac{h\nu }{k_{\rm
B}T}\right]-1\right)^{-1}
\end{eqnarray}
is the spectral radiance for a blackbody at temperature $T$. The unperturbed
spectral radiance is $I_{\nu, 0}(\hat {\mathbf{n}}, \hat {\boldsymbol{\epsilon }},
\mathbf{x}, t)=B_\nu (T_\gamma (t))$, where $T_\gamma (t)$ is the photon
temperature at cosmic time $t$ in the unperturbed background expanding FLRW
spacetime. (In Sec.~\ref{SpectralDistortions}, however, we discuss some
interesting possible caveats to this assumption.) For the ``scalar'' mode of
wavenumber $\mathbf{k},$ we may decompose
\begin{eqnarray}
\Delta _{ab}(\hat {\mathbf{n}}, \mathbf{x}, t)= \Delta ^{({\rm scl})}_{T}(\hat
{\mathbf{n}}, \mathbf{x}, t)\delta _{ab} + \Delta ^{({\rm scl})}_{P}(\hat
{\mathbf{n}}, \mathbf{x}, t)( {\hat \theta }_a {\hat \theta }_b - {\hat \phi }_a
{\hat \phi }_b).
 \end{eqnarray}
We now turn to the evolution of the variables
$\Delta^{({\rm scl})}_{T}(\hat {\mathbf{n}}, t)$ and
$\Delta ^{({\rm scl})}_{P}(\hat{\mathbf{n}}, t).$
It is convenient, exploiting the symmetry under rotations
about $\hat{\mathbf{k}},$ to expand
\begin{eqnarray}
\begin{split}
\Delta ^{({\rm scl})}_{T}(\hat {\mathbf{n}}, t; \mathbf{k})&= \sum _{\ell
=0}^\infty \Delta ^{({\rm scl})}_{T, \ell }(t; \mathbf{k} )(-i)^\ell (2\ell
+1) P_\ell ( \hat {\mathbf{k}} \cdot \hat {\mathbf{n}}),\\[3pt]
\Delta ^{({\rm scl})}_{P}(\hat {\mathbf{n}}, t; \mathbf{k} )&= \sum _{\ell
=0}^\infty \Delta ^{({\rm scl})}_{P, \ell }(t; \mathbf{k}) (-i)^\ell (2\ell
+1) P_\ell ( \hat {\mathbf{k}} \cdot \hat {\mathbf{n}}).
\end{split}
\end{eqnarray}
The evolution of $\Delta ^{({\rm scl})}_{T, \ell }$ and $\Delta ^{({\rm scl})}_{P, \ell
}$ is governed by the following infinite system of equations \cite{MaBert}:
\begin{eqnarray}
\Delta _{T0} ^{({\rm scl})}&=& -k\Delta _{T1} ^{({\rm scl})}+\phi',\cr
\Delta _{T1}^{({\rm scl})} &=& \frac{k}{3}( \Delta _{T0} ^{({\rm scl})}-2\Delta
_{T2}^{({\rm scl})}+\psi ) +a(t)n_e(t)\sigma _T\left(
\frac{v_b}{3}-\Delta _{T1} ^{({\rm scl})}\right)\!,\cr
\Delta _{T2}^{({\rm scl})}&=& \frac{k}{5}( 2\Delta _{T1}^{({\rm scl})}-3\Delta
_{T3}^{({\rm scl})}) + a(t)n_e(t)\sigma _T \left( \frac{\Pi }{10}+\Delta _{T2}^{({\rm
scl})}\right)\!, \cr 
\Delta _{T\ell }^{({\rm scl})}&=&
\frac{k}{(2\ell +1)}\{ \ell \Delta _{T(\ell -1)}^{({\rm scl})} - (\ell +1) \Delta
_{T(\ell +1)}^{({\rm scl})}\} -
a(t)n_e(t)\sigma _T \Delta _{T \ell }^{({\rm scl})}, \quad \mbox{for }\ell \ge 3,\cr
\Delta _{P\ell }^{({\rm scl})}&=& \frac{k}{(2\ell +1)}\{ \ell \Delta _{P(\ell -1)}^{({\rm
scl})} - (\ell +1) \Delta _{P(\ell +1)}^{({\rm scl})}\}-
a(t)n_e(t)\sigma _T \Delta _{P \ell }^{({\rm scl})}\cr
&+&\,\frac{1}{2}a(t)n_e(t)\sigma _T \Pi \left(\delta _{\ell,0}+\frac{1}{5} \delta
_{\ell,2} \right)\!, \quad \mbox{for }\ell \ge 0, 
\label{PhotonBH}
 \end{eqnarray}
where
\begin{eqnarray}
\Pi =\Delta _{T0}^{({\rm scl})}+ \Delta _{P0}^{({\rm scl})}+ \Delta _{P2}^{({\rm scl})}.
\end{eqnarray}
These equations are known as the {\it Boltzmann hierarchy} for the photon phase space
distribution function. In the last line of Eq.~(33) $\delta _{ab}$ denotes the Kronecker
delta function. For practical numerical calculations, this infinite set of coupled
equations must be truncated at some $\ell _{\max}.$

In principle, ignoring questions of computational efficiency, we could truncate
the system of equations in (\ref{PhotonBH}) at some large $\ell _{\max},$
sufficiently above the maximum multipole number of interest today, and integrate
the coupled system from some sufficiently early initial time to the present time
$t_0$ to find the ``scalar'' temperature and polarization anisotropies today,
which would be given by the following integrals over plane wave modes:
\begin{eqnarray}
\begin{split}
\Delta ^{({\rm scl})}_{T}(\hat {\mathbf{n}}, t_0)&= \int d^3k \Delta ^{({\rm
scl})}_{T}(\hat {\mathbf{n}}, t_0; \mathbf{k}),\\[4pt]
\Delta ^{({\rm scl})}_{P}(\hat {\mathbf{n}}, t_0)&= \int d^3k \Delta ^{({\rm
scl})}_{P}(\hat {\mathbf{n}}, t_0; \mathbf{k}).
\end{split}
\end{eqnarray}
This was the approach used in the COSMICS code
\cite{cosmics}, one of the first codes providing accurate computations of the CMB angular
power spectrum. However in this approach much computational effort is expended at late
times when there is virtually no scattering.

A computationally simpler but mathematically completely equivalent approach known
as the line of sight formalism was introduced in Ref.~\cite{lineOfSight}. The
crucial idea is to express the anisotropies today (at $\mathbf{x}=0$ and $\eta
=\eta -0$) in terms of a line of sight integral, so that
\begin{eqnarray}
\Delta ^{(S)}_T(\hat {\mathbf{n}})&=& \int _0^{\eta _0}d\eta\exp [-\tau (\eta
)]\exp[ik(\eta -\eta _0)]\\
&&\times \left[ \left( -\frac{d\tau }{d\eta }\right) \left( \Delta
_{T0}+i\mu v_B+\frac{1}{2}P_2(\mu )\Pi \right) + (\phi ^\prime
-ik\mu \psi ) \right] \label{LOS-T}
\end{eqnarray}
and
\begin{eqnarray}
\Delta ^{(S)}_P(\hat {\mathbf{n}})&=& \int _0^{\eta _0}d\eta
\exp[-\tau (\eta )]\exp [ik(\eta -\eta _0)] \left(-\frac{d\tau
}{d\eta }\right) (1-P_2(\mu )) \Pi \label{LOS-P}
 \end{eqnarray}
where $\Pi = \Delta _{T2}^{(S)} +\Delta _{P2}^{(S)} + \Delta _{P0}^{(S)}.$ We may think
of $\exp [-\tau (\eta )] (-d\tau /d\eta )d\eta ={\cal V}(\eta )d\eta$ as a measure along
the line of sight weighted according to where last scattering takes place. We shall call
${\cal V}(\eta )$ the visibility function. In Eq.~(\ref{LOS-T}), the first term
represents the nonpolarized contribution at last scatter and the second term represents
the contribution from the integrated Sachs-Wolfe effect. In Eq.~(\ref{LOS-P}) for the
polarized anisotropy, the only contribution is from last scatter. It is convenient to
rewrite the above integrals using integration by parts so that all spatial derivatives
along the line of sight and occurrences of $\mu$ are eliminated, and in the process we
also omit the monopole surface term at the endpoint corresponding to the observer as this
contribution is not measurable. After this integration by parts, the integrals above take
the form
\begin{eqnarray}
\Delta ^{(S)}_{T,P}(\hat {\mathbf{n}})&=& \int _0^{\eta _0}d\eta \exp[ik\mu
(\eta_0-\eta)]{\cal S}^{(S)}_{T,P}((\eta _0-\eta )\hat {\mathbf{n}}, \eta),
 \end{eqnarray}
where the scalar source functions are
\begin{eqnarray}
\begin{split} 
{\cal S}^{(S)}_{T}&= \exp [-\tau (\eta )](\phi' + \psi' )
+{\cal V}\left( \Delta _{T0}+\psi +\frac{v'_b}{k}+\frac{3\Pi'}{4k^2}\right) \\[3pt]
&\quad+ {\cal V}' \left( \frac{v_b}{k}+\frac{3\Pi'}{4k^2}
\right) +{\cal V}'' \frac{3\Pi}{4k^2}, \\[3pt]
{\cal S}^{(S)}_{P}&=\frac{3}{4k^2}({\cal V} (k^2\Pi + \Pi '')+ 2{\cal V}' \Pi' + {\cal
V}'' \Pi),
\end{split}
\end{eqnarray}
and the primes denote derivatives with respect to conformal time $\eta.$ Now that the
spatial derivatives have been eliminated, it is no longer necessary to use the plane
waves $\exp [ik\mu (\eta_0-\eta )]$ as the eigenfunctions of the Laplacian operator with
eigenvalue $-k^2$. It is convenient instead to use a spherical wave expansion with
$j_\ell (k(\eta -\eta )$
\begin{eqnarray}
\Delta ^{(S)}_{\ell ; A} &=& \int _0^{\eta _0}d\eta j_\ell
\Big(k(\eta_0-\eta )\Big){\cal S}^{(S)}_{A}\Big((\eta _0-\eta
)\hat {\mathbf{n}}, \eta\Big),
\end{eqnarray}
where $A=T,P.$

The line of sight formulation results in a significant reduction of the
computational effort required for computing the high $\ell $ coefficients for two
reasons. First, the Boltzmann hierarchy can be truncated at low $\ell,$ because
only moments up to $\ell =2$ appear in the integral. Second, when the universe has
become transparent, there is no longer any contribution from the Thomson
scattering.

We express the initial state for the growing adiabatic mode at some early reference time
whose exact value must be chosen after the epoch of entropy generation following
inflation but before any of the relevant modes re-entered the horizon. We use the
variable $\zeta (\mathbf{x}),$ whose value is conserved on superhorizon scales, to
characterize the primordial perturbations in the adiabatic growing mode. We may expand,
either using plane wave Fourier modes or a spherical wave expansion, so that
\begin{eqnarray}
\zeta (\mathbf{x})=\int d^3k \zeta (\mathbf{k})\exp [i\mathbf{k}\cdot \mathbf{x}]
\end{eqnarray}
or
\begin{eqnarray}
\zeta (\mathbf{x})= \sum _{\ell =0}^\infty \sum _{m=-\ell
}^{+\ell } \int _0^\infty dk \zeta _{\ell, m}(k) j_\ell (kr) Y_{\ell m}(\theta,
\phi ).
\end{eqnarray}

We characterize the homogeneous and isotropic Gaussian statistical ensemble for $\zeta (\mathbf{x})$ by
means of a power spectrum ${\cal P}(k),$ which suffices to completely define the statistical process
generating $\zeta (\mathbf{x}).$ We define ${\cal P}(k)$ according to the following expectation values:
\begin{eqnarray}
\langle \zeta (\mathbf{k}) \zeta (\mathbf{k}') \rangle = (2\pi )^3\delta
^3(\mathbf{k}-\mathbf{k}') {\cal P}(k),
 \end{eqnarray}
or equivalently
\begin{eqnarray}
\langle \zeta _{\ell m}(k) \zeta _{\ell 'm'}(k') \rangle = \delta _{\ell, \ell '}
\delta _{m, m'} \delta (k-k') {\cal P}(k).
 \end{eqnarray}
It follows that
\begin{eqnarray}
a^{(S),T}_{\ell m}=\int _0^\infty dk\int _0^{\eta _0}d\eta j_\ell
\Big(k(\eta _0-\eta )\Big) {\cal S}^{(S)}_T(\eta, k) \zeta _{\ell
m}(k),
 \end{eqnarray}
so that
\begin{eqnarray}
C_\ell ^{(S),TT}= \langle [ a^{(S),T}_{\ell m}] ^* a^{(S),T}_{\ell m} \rangle
=\int _0^\infty dk {\cal P}(k) \Delta _\ell ^{(S),T}(k) \Delta _\ell
^{(S),T}(k),
 \end{eqnarray}
where
\begin{eqnarray}
\Delta _\ell ^{(S),T}(k)=\int _0^{\eta _0}d\eta j_\ell \Big(
k(\eta _0-\eta )\Big) {\cal S}^{(S)}_T\Big(\eta, k\Big).
\end{eqnarray}
The above approach was first implemented in the publicly available code CMBFAST
\cite{lineOfSight} and later in the code
CAMB \cite{lewisChallinorLasenby}.{\kern1pt}\footnote{For a download and information on CAMB, see
www.camb.info.} More details about the computational issues may be found in these
papers.

\subsection{Angular diameter distance}
\label{AngularDiameterSection}

The radius of the last scattering surface expressed in terms of co-moving units
today (for example in light-years or Mpc) is given by the integral
\begin{eqnarray}
x_{ls}={H_0}^{-1}\int _{a_{ls}}^1\frac{da}{a^2}\frac{1}{ \sqrt{ \Omega _{r}a^{-4}+ \Omega
_{m}a^{-3}+\Omega _{\Lambda }}}
 \end{eqnarray}
for a universe whose stress-energy content includes radiation,
matter, and a cosmological constant. Here the scale factor is
given by $a_{ls}=(z_{ls}+1)^{-1}.$ For other equations of state
the argument of the square root above is modified accordingly. A
physical scale $d$ at last scattering is converted to a co-moving
scale today $x$ according to $x=(z_{ls}+1)d,$ and under the
assumption of a flat spatial geometry, the angle in radians
subtended by an arc of length $d$ on the last scattering sphere is
$\theta =(d/d_{ls}).$

Except for the integrated Sachs-Wolfe effect (discussed below in
Sec.~\ref{IntegratedSachsWolfe}), which is very hard to measure because the $S/N$
is negligible but for the very first multipoles, the projection effect described
above is the only way in which late-time physics enters into determining the
angular power spectrum. Possible spatial curvature, parametrized through $\Omega
_k,$ the details of a possible quintessence field, or similar new physics that
alter the expansion history at late time enters into the CMB power spectra only
through a rescaling of the relation between angular scales, on the one hand, and
physical scales, on the other, around last scattering, and this rescaling can be
encapsulated into a single variable $d_{ls}.$

If the family of FLRW solutions extended to admit spatial
manifolds of constant positive or negative curvature, the
expression for the angular size must be modified to the following
for a negatively curved (hyperbolic) universe
\begin{eqnarray}
\theta = \frac{ \Omega _k^{1/2}d_{ls} } { \sinh [\Omega _k^{1/2}d_{ls}] }
\frac{d}{d_{ls}},
\end{eqnarray}
or in the case of positive curvature (i.e., a spherical universe)
\begin{eqnarray}
\theta = \frac{ (-\Omega _k)^{1/2}d_{ls} } { \sin [(-\Omega _k)^{1/2}d_{ls}] }
\frac{d}{d_{ls}}.
 \end{eqnarray}
In the former case, the additional prefactor giving the
contribution for the non-Euclidean character of the geometry has a
demagnifying effect and increases the angular diameter distance,
whereas for the spherical case the effect is the opposite.

In the early 1990s, many theorists working in early universe cosmology fervently
believed on grounds of simplicity in a cosmological model with $\Omega _k=0$
(supposedly an inexorable prediction of inflation) and no cosmological constant
(regarded as extremely finely tuned and ``unnatural'') containing only matter and
radiation so that $\Omega _r \,{+}\,\Omega _m=1$ exactly. This belief was maintained
despite a number of discordant astronomical observations. Many in the astronomical
community, however, rather adopted the attitude ``I believe only what I see," and
thus favored some kind of a low-density universe, either a negatively curved
universe where $\Omega _k=1-\Omega _m-\Omega _r$ or a universe with a nonzero
cosmological constant, where $\Omega _\Lambda =1-\Omega _m-\Omega
_r.$\footnote{For a snapshot of the debates in cosmology in the mid-1990s, see the
proceedings of the conference {\it Critical dialogues in cosmology} held in 1995
at Princeton University \cite{CriticalDialogues}.}

Although it was widely claimed that spatial flatness was a prediction of
inflation, the basic idea of how to produce a negatively curved universe within
the framework of inflation had already been proposed in 1982 by Gott and Statler
\cite{gottA,gottB}. However, since they had not calculated the perturbations
predicted in such a model, it was not clear that such a model would work. In 1995
Bucher {\it et~al.} \cite{bgt,btArbMass} and Yamamoto {\it et~al.} \cite{sasakiLD}
calculated the perturbations for single bubble open inflation, showing them to be
consistent with all the observations available at the time. Present constraints
from the CMB limit $|\Omega _k-1|$ to a few percent at most, the exact number
depending on the precise parametric model assumed
\cite{planck-parameters,planck-inflation}.

\subsection{Integrated Sachs-Wolfe effect}
\label{IntegratedSachsWolfe}

The discussion above claimed that the impact of late-time physics on the CMB power
spectrum could be completely encapsulated into a single parameter: the angular diameter
distance to last scattering $d_{ls}.$ This is only approximately true, because the
geometric argument above assumed that all the CMB anisotropy can be localized on the last
scattering surface and ignored the integrated Sachs-Wolfe term in
Eq.~(\ref{SachsWolfeFormula}), which gives the part of the linearized CMB anisotropy
imprinted after last scattering.

Intuitively, the integrated Sachs-Wolfe term can be understood as
follows. The Newtonian gravitational potential blueshifts the CMB
photons as they fall into potential wells and similarly redshifts
them as they again climb out. If the depth of the potential well
does not change with time, the two effects cancel. In this case,
there is no integrated \hbox{Sachs-Wolfe} contribution because
the integral can be reduced to the sum of two surface terms. But
if the depth of the potential well changes with time, in
particular if the overall scale of the potential is decaying, the
two effects no longer cancel and an integrated Sachs-Wolfe
contribution is imprinted.

In the linear theory, the gravitational potential at late times may be factorized
in the following way:
\begin{eqnarray}
\Phi (\mathbf{x}, t)= \Phi (\mathbf{x}) T(t),
\end{eqnarray}
and we may normalize $T(t)$ to one when the universe has become matter dominated,
but before the cosmological constant---or some other dark energy sector---has
started to alter the expansion history.

From the CMB anisotropies alone, it is hard to determine the
contribution of the integrated Sachs-Wolfe effect to the total
CMB anisotropy on large angular scales. However on can try to
isolate its contribution by measuring the cross-correlation of the
CMB temperature anisotropy with the large-scale structure using a
broad radial window function extending to large redshifts as was
first proposed in Ref.~\cite{reesSciama}, whose aim was to find
evidence for a nonzero cosmological constant.

The evolution equation for the linearized density contrast of a single
pressureless matter component is \cite{PeeblesLSSofUniverseBook}
\begin{eqnarray}
\ddot \delta _m(t)+2H(t)\dot \delta _m(t)-\frac{3}{2}H^2(t)\Omega _m(t)\delta
_m(t)=0, \label{MatterContrast}
 \end{eqnarray}
where the dots denote derivatives with respect to proper time and $H(t)=\dot
a(t)/a(t).$ Here we treat the baryons and the CDM (cold dark matter) as a single
component ignoring any relative velocity between them. This is not a bad
approximation at late times, particularly on very large scales.

\begin{wrapfigure}{L}{0.6\textwidth }
\begin{center}
\includegraphics[width=10cm]{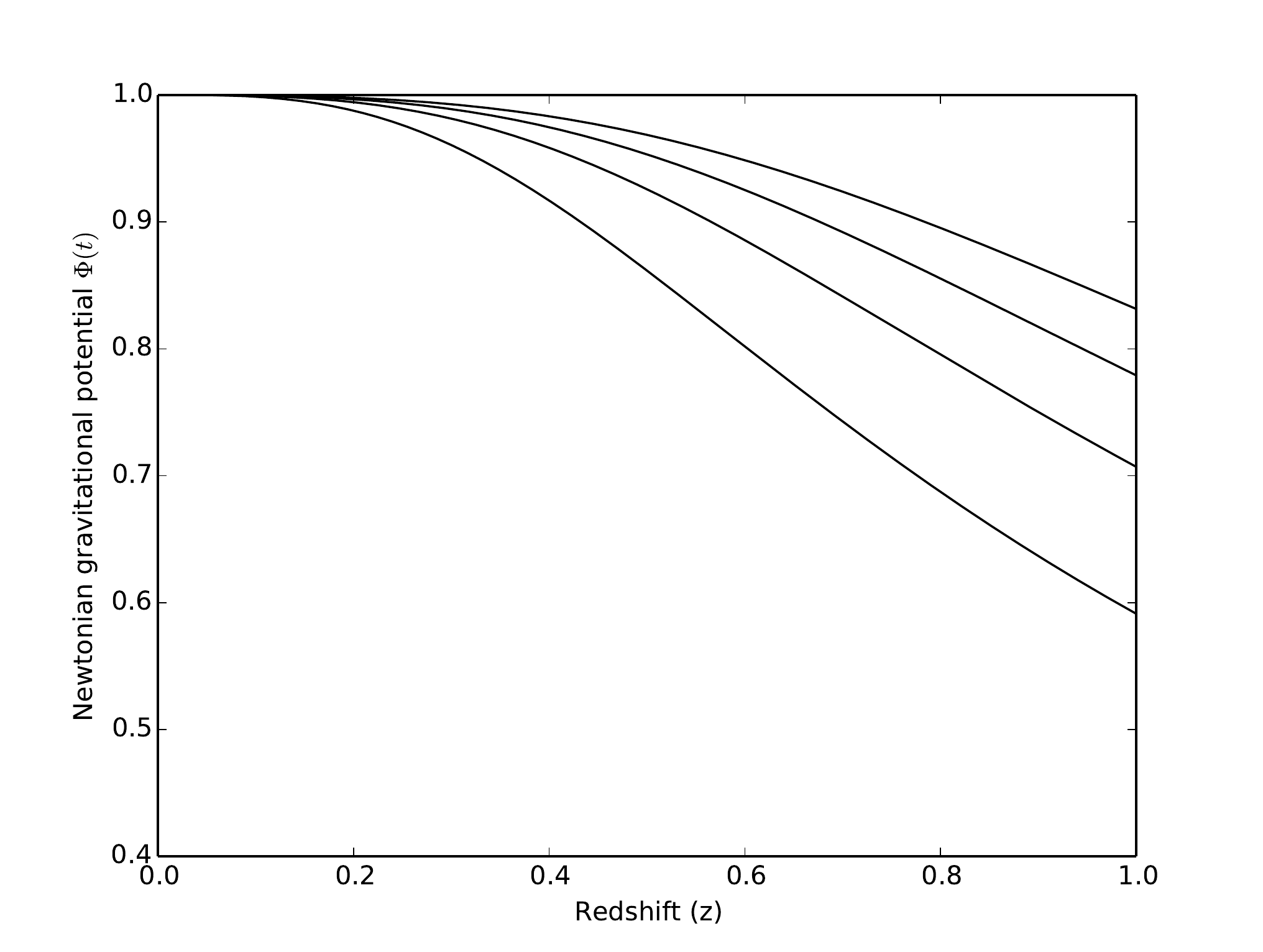}
\end{center}
\caption{{\bf Decay of gravitational potential for late-time integrated Sachs-Wolfe
effect.} We sketch the time dependence of the gravitational potential for a universe with
$\Omega _{\Lambda 0}=$ 0.9, 0.8, 0.7, and 0.6 (from top to bottom on the right) with
$a=1$ today. The qualitative behavior will be similar for other kinds of the dark
energy.
\label{LambdaPotentialEvolution}}
\end{wrapfigure}

For an Einstein-de Sitter universe (for which $\Omega _r$ is negligible and $\Omega
_\Lambda =0$), which is a good approximation for our universe at intermediate redshifts,
because for $z\ltorder z_{eq}\approx 3.4\times 10^4$ radiation is unimportant and after
the decoupling of the photons and baryons, it is valid to set the velocity of sound equal
to zero as above. Let $z_{m\Lambda }=1/(1+a_{m\Lambda })$ be the redshift where the
nonrelativistic matter and the putative cosmological constant contribute equally to the
mean density of the universe, so that $\Omega _m={a_{m\Lambda }}^3/(1+{a_{m\Lambda }}^3)$
and $\Omega _\Lambda =1/(1+{a_{m\Lambda }}^3).$ For $\Omega _\Lambda =0.68,$ for example,
we would have $z_{m\Lambda }=0.3$ and $a_{m\Lambda }=0.78.$ For the Einstein-de Sitter
case, $a(t)=t^{2/3}$ and the growing and decaying solution to Eq.~(\ref{MatterContrast})
are $t^{2/3}$ and $t^{-1},$ respectively. $(1/a^2)\nabla ^2\phi =(3/2)H^2\Omega _m\delta
_m,$ and for the growing mode $\Phi (\mathbf{x})$ is independent of time, implying that
there is no integrated \hbox{Sachs-Wolfe} contribution at late times. However, as the
cosmological constant begins to take over and $\Omega _\Lambda $ starts to rise toward
one, $\Phi $ starts to decay, as sketched in Fig.~\ref{LambdaPotentialEvolution}.

As explained in detail in Ref.~\cite{reesSciama}, the maximum signal-to-noise (S/N)
that can be extracted from the ISW correlation, even with an ideal measurement of the
large-scale structure anisotropy within our past light cone, is modest because of the
``noise'' from the primary CMB anisotropies emanating from the last scattering surface.
This cosmic variance noise cuts off the available S/N because on small angular scales,
the relative contribution from ISW plummets. The S/N is centered around $\ell \approx 20$
and integrates to about $S/N\approx 6.5$ (assuming $\Omega _{\Lambda 0}=0.7).$ For an
early detection of a correlation, see Ref.~\cite{boughnCrittenden}, and for the most
recent work in this area see Ref.~\cite{planck-isw} and references therein.

\subsection{Reionization}
\label{ReionizationSection}

According to our best current understanding of the ionization
history of the universe, the early ionization fraction around
recombination is accurately modeled assuming homogeneity and using
atomic rate and radiative transfer equations. As a rough estimate,
one may assume the reionization fraction predicted by equilibrium
thermodynamics given by the Saha equation, in which there is
exponential sensitivity to changes in temperature. After $x\approx
\frac{1}{2},$ which occurs at about $z\approx 1270,$ the
ionization fraction $x(z)$ rapidly decays to almost zero, except
for trace amounts of ionized hydrogen (denoted as ${\it HII}$ in
the astrophysical literature) and of free electrons, due to the
inefficiency of recombination in the presence of trace
concentrations. If it were not for inhomogeneities and structure
formation, this nearly perfect neutral phase would persist to the
present day. However, the initially small primordial
inhomogeneities grow and subsequently evolve nonlinearly. The
first highly nonlinear gravitationally collapsed regions give rise
to the first generation of stars and quasars, which serve as a
source of UV radiation that reionizes the neutral hydrogen almost
completely, except in a minority of regions of extremely high gas
density, where recombination occurs sufficiently frequently to
counteract the ionizing flux of UV photons.

The first observational evidence for such nearly complete reionization came from
observations of the spectra of quasars at large redshift in the interval
$10.4\,{\rm eV}/(1+z_{{\rm quasar}})<E_\gamma <10.4\,{\rm eV}$---that is,
blueward of the redshifted Lyman $\alpha $ line \cite{GunnPeterson}. If the
majority of the hydrogen along the line of sight to the quasar were neutral,
resonant Ly$\alpha $ $(2p\to 1s)$ scattering would deflect the photons in this
frequency range out of the line of sight, and this part of the spectrum would be
devoid of photons. The fact that the spectrum is not completely blocked in this
range indicates that the universe along the line of sight was reionized. Almost
complete absorption by neutral hydrogen---that is, the so-called Gunn-Peterson
trough---was not observed in quasar spectra until 2001, when spectroscopic
follow-up of several high-redshift quasars discovered by the Sloan digital sky
survey (SDSS) detected such a trough from $z\approx 5.7$ to $z\approx 6.3,$ where
the continuum emission blueward of the Ly$\alpha $ appears to have been completely
absorbed by neutral hydrogen \cite{becker}.

The big question here is exactly when did the universe first become reionized. Or said
another way: When did the first generation of stars and quasars form and become capable
of generating enough ionizing radiation to convert the neutral gas to its present ionized
state?

This is a vast subject of very current research, sometimes described as the
exploration of the ``Dark Ages" of the universe. While we believe that we know a
lot about the conditions around the time when the CMB anisotropies were imprinted,
and we also know a lot about the recent universe extending to moderate redshift,
little is known with any degree of certainty concerning this intermediate epoch.
Good reviews of this subject include Ref.~\cite{BarkanaLoeb}. Future
observations of the 21 cm hyperfine transition of atomic hydrogen (HI) by the next
generation of radio telescopes, in particular the Square Kilometer Array
(SKA),\footnote{https:/$\!$/www.skatelescope.org/.} promise to provide
three-dimensional maps of the neutral hydrogen in our past light cone.

Here we limit ourselves to discussing the impact of reionization on the CMB
anisotropies. The most simplistic model---or caricature---of reionization
postulates that the ionization fraction $x(z)$ changes instantaneously from $x=0$
for $z>z_{{\rm reion}}$ to $x=1$ for $z<z_{{\rm reion}}.$ Under this assumption,
the optical depth for rescattering by the reionized electrons is
\begin{eqnarray}
\tau =\int_0^{z_{{\rm reion}}}dz \sigma _{T}n_e(z)\frac{d\ell }{dz}.
 \end{eqnarray}
The Thomson scattering optical depth from redshift $z$ to the present day is given
by the integral
\begin{eqnarray}
\begin{split}
\tau (z) &=& \frac{c\sigma _T}{H_0} \int _{a(z)}^1 \frac{da}{a}
\frac{n_e(a)}{(\Omega _ra^{-4}+\Omega _ma^{-3}+\Omega _\Lambda )^{1/2}}\\
&=& \frac{c \sigma _T n_{e0} }{H_0} \int _{a(z)}^1 \frac{da}{a^4}
\frac{x_e(a)}{(\Omega _ra^{-4}+\Omega _ma^{-3}+\Omega _\Lambda )^{1/2}},
\end{split}
\end{eqnarray}
where $a(z)=1/(z+1),$ $\sigma _T$ is the Thomson scattering
cross-section, $H_0$ is the present value of the Hubble constant
(in units of inverse length), and $n_{e0}$ is the electron density
today, assuming $x_e=1.$ The functions $n_{e}(a)$ and $x_{e}(a)$
are the~electron density\footnote{It is customary to define $x_e$
as $n_e/n_B,$ so that $x_e$ slightly exceeds one when the helium
is completely ionized as well.} and the ionization fraction at
redshift $z=1/a-1,$ respectively.

On small angular scales, the effect of reionization on the
predicted CMB temperature and polarization anisotropy spectra is
straightforward to calculate. Of the CMB photons emanating from
the LSS, a fraction $\exp (-\tau )$ does not undergo rescattering
by the reionized electrons and their fractional temperature
perturbations are preserved. However, for the remaining fraction
$(1-\exp (-\tau ))\approx \tau $ that is rescattered, the
small-angle CMB anisotropies are completely erased. The reionized
electrons nearly completely wash out small-scale detail much like
a window pane of ground glass. The net effect is that the
small-scale CMB anisotropies are attenuated, by a factor of $\exp
(-\tau )$ in amplitude and by a factor of $\exp (-2\tau )$ in the
power spectrum. On larger angular scales, the attenuation of the
anisotropy from the rescattered photon component is incomplete,
and a more detailed calculation is needed to determine the
detailed form of the $\ell $-dependent attenuation factor. The
order of magnitude of the angular scale where the attenuation
starts to become incomplete is given by the ratio of the co-moving
distance travelled from next-to-last to last scattering by a
rescattered photon, $d_{{\rm reion}},$ to the radius of the LSS
$d_{{\rm LS}}.$ We thus obtain an angular scale $\theta _{{\rm
reion}}\approx d_{{\rm reion}}/d_{{\rm LS}}$ for this transition.

For determining the cosmological model from the $C_{{\rm TT}}$ power spectrum
alone, the reionization optical depth is highly degenerate with the overall
amplitude of the cosmological perturbations $A.$ This is because except on very
large scales, where cosmic variance introduces substantial uncertainty, it is the
combination $A\exp (-2\tau )$ that determines the observed amplitude of the CMB
angular power spectrum.

\begin{wrapfigure}{L}{0.6\textwidth }
\begin{center}
\includegraphics[width=10cm]{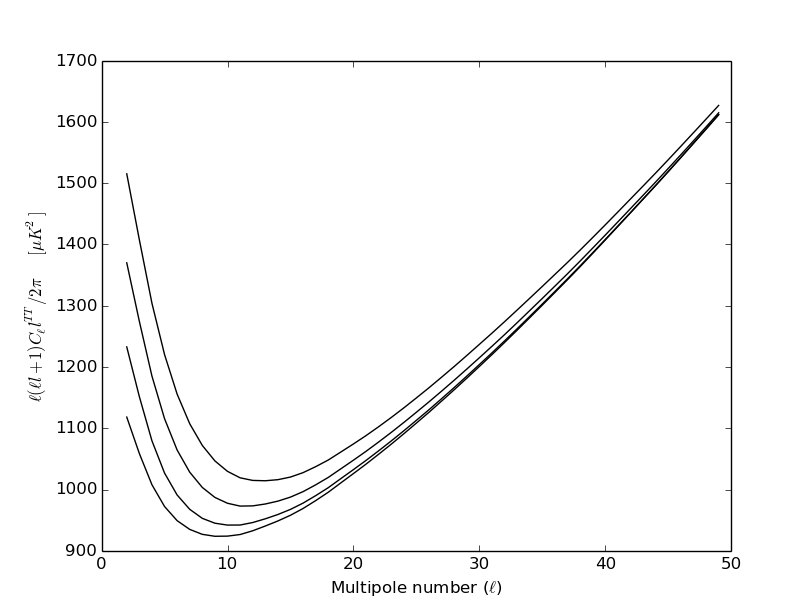}
\end{center}
\caption{{\bf CMB power spectra for four
values of the reionization optical depth.} We plot the indicated power spectra for $\tau
=0.00,$ $0.04,$ $0.08$ and $0.12$ assuming the same concordance values for the other
cosmological parameters. We multiply the spectra by $\exp (2\tau )$ in order to remove
the trivial dependence on $\tau $ at small angular scales and thus highlight what occurs
at low $\ell.$
\label{ReionFigure}}
\end{wrapfigure}

The polarization power spectra, meaning $C^{{\rm EE}}$ and also
$C^{{\rm TE}},$ help to break this degeneracy by providing a
rather model independent measurement of $\tau _{{\rm reion}}.$
Recall our order of magnitude estimate for the polarization
introduced by the finite thickness of the LSS
 \begin{eqnarray}
P\sim d^2\frac{\partial ^2T}{\partial x^2},
\end{eqnarray}
where $d$ and $x$ are expressed in co-moving units and $d$ is the mean co-moving distance that
the photon travels from next-to-last scattering to last scattering. The second derivative is that of the
temperature anisotropy evaluated at the position of a typical electron of last scatter. Let us compare
the polarization imprinted during recombination to that imprinted later for the fraction of photons
$(1\,{-}\,\exp (-\tau ))\approx \tau $ rescattered by the free electrons from reionization, for the
moment assuming that the second derivative factors are comparable for the two cases. However, $d_{{\rm
rec}}\ll d_{{\rm reion}};$ therefore, we may expect that $P_{{\rm reion}}/P_{{\rm rec}}\approx \tau
(d_{{\rm rec}}/d_{{\rm reion}})^2\gg 1.$ This polarization from reionization, however, is concentrated
almost exclusively in the first few multipoles, because directions separated by a small angle see the
same quadrupole anisotropy from last scattering from the vantage point of a typical free electron after
reionization. Figure~\ref{ReionFigure} shows the scalar power spectra for four values of $\tau $ keeping
the other cosmological parameters fixed. We shall see that the same amplification applies to the tensor
modes, or primordial gravitational waves, presumably generated during inflation.

We now turn to observations allowing us to fix the value of $\tau $ using the low
$\ell, C^{{\rm EE}}$ spectra and $C^{{\rm TE}}$ spectra based on the qualitative
theoretical arguments above but made more precise by calculating $\tau $ within the
framework of the previously described six-parameter concordance model. While the first
observations of the polarization of the CMB were reported by the DASI
experiment \cite{dasi}, it was the WMAP large-angle polarization results, first the TE
correlations reported in the one-year release, later followed by the EE correlation
results from the third-year results, that provided the first determination of $\tau_{{\rm
reion}}$ using the CMB, and subsequently refined in later updates to the WMAP results
with more integration time. As part of their first-year results \cite{wmap2003f}, the
WMAP team reported a value for the reionization optical depth of $\tau =0.17\pm 0.04$ at
$68\% $ confidence level, which would correspond to $11<z_r<30$ at 95\% confidence where
a step function profile for the ionization fraction is assumed. This determination was
based on exploiting only the measurement of the TE cross-correlation power spectrum,
because for this first release the analysis of the polarized sky maps was not
sufficiently advanced to include a reliable EE auto-correlation power spectrum
measurement. The reported value of $\tau$ was larger than expected. In their third-year
release \cite{wmap2007c}, when WMAP presented a full analysis of the polarization with
$EE$ included, the determination of $\tau$ shifted downward, namely to $\tau =0.10\pm
0.03$ (using only EE data) and $\tau =0.09\pm 0.03$ with (TT, TE and EE all included).
The final nine-year WMAP release \cite{wmap2013b} cites a value of $\tau \approx 0.089
\,{\pm}\, 0.014$ using WMAP alone. Including other external data sets results in small
shifts about this value.

The Planck 2013 Results for Cosmology
\cite{planck-mission-paper,planck-parameters,planck-inflation} release did not
include an analysis of the Planck polarization data. Therefore the likelihoods
used to determine the cosmological parameters included the WMAP determination of
$\tau $ as {\it a prior}, and in some cases the WMAP low-$\ell $ polarization likelihood
was used in order to break the degeneracy between $\tau $ and $A$ described above
that arises when only temperature data is used.

It has been pointed out that reionization histories, which are completely
characterized by function $x_e(z),$ cannot be reduced to a single number. Some
work has been done to characterize what further information can be extracted from
the \hbox{low-$\ell $} CMB concerning the reionization history, as described in
Refs.~\cite{huHolder} and \cite{kaplinghat}, but the conclusion is that from
the CMB data at most a few nonzero numbers can be extracted at $S/N\gtorder 1$
given the large cosmic variance at low $\ell.$

In the above discussion we have assumed that reionization occurs homogeneously in
space, an approximation adequate for calculating its effect on the anisotropies at
large and intermediate angular scales. However, according to our best
understanding, the universe becomes reionized by the formation of bubbles, or
Str\"omgren spheres, surrounding the first sources of ionizing radiation, which
grow and become more numerous, and eventually percolate. While this basic picture
is highly plausible, the details of modeling exactly how this happens remain
speculative. The effect on the CMB is significant only at very small angular
scales, far into the damping tail of the primordial anisotropies. Power from the
low-$\ell $ anisotropies is transformed into power at very high $\ell.$ This can
be understood by considering the power in Fourier space resulting from the rather
sharp bubble walls, which in projection look almost like jump discontinuities.
Some simulations of this effect, along with other effects at large $\ell,$ can be
found for example in Ref.~\cite{zahn}.

\subsection{What we have not mentioned}

The preceding subsections provide a sampling of what can be
learned from the CMB TT and polarization power spectra with each
point discussed in some detail. For lack of space, we cannot cover
all the parameters and extensions of the simple six-parameter
concordance model that can be explored, and here we provide a few
words regarding what we have not covered giving some general
references for more details.

The structure of the acoustic oscillations and the damping tail (relative heights,
positions, etc.) of the CMB power spectra depends on the cosmological parameters, as
forecast and explained pedagogically in Refs.~\cite{peda-cosmo,jungman}, and most
recently applied to the Planck 2013 data for determining the cosmological parameters
\cite{planck-parameters}. Some of the constraints are obtained in the framework of the
standard six-parameter concordance model, but many extensions to this model can be
constrained, such as theories with varying $\alpha_{{\rm QED}},$ nonminimal numbers of
neutrinos, varying neutrino masses, isocurvature modes, nonstandard recombination, to
name just a few examples. Moreover the question can be addressed whether there exists
\hbox{statistically} significant evidence in favor of extending the basic six-parameter
model to a model having additional parameters. See Refs.~\cite{planck-parameters} and
\cite{planck-inflation} for an extensive set of references.

\section[Gravitational lensing of the CMB]
{Gravitational Lensing of the CMB}
\label{GravitationalLensingSection}

The previous sections presented a simplified view of the CMB where the CMB anisotropies
are imprinted on the so-called last scattering surface, situated around $z\approx 1100.$
It is assumed that we look back to this surface along straight line geodesics in the
unperturbed coordinate system. This is not a bad approximation, but it is not the whole
story. Inhomogeneities in the distribution of matter, particularly at late times when the
clustering of matter has become nonlinear, act to curve the photon, or geometric optics,
trajectories. These nonuniform deflections distort the appearance of the surface of last
scatter, much like a fun house mirror. Gravitational lensing of the CMB is a small effect
but large enough so that it has to be taken into account in order to compare models of
the primordial universe with the observations correctly. In one sense, gravitational
lensing may be viewed as a nuisance effect to be removed. But gravitational lensing of
the CMB may also be regarded as an invaluable tool for probing the inhomogeneities of the
matter distribution between us and the last scattering surface. Gravitational lensing is
presently a very active area of astronomy, and the CMB is only one of the many types of
objects whose gravitational lensing may be measured in order to probe the underlying mass
inhomogeneities in the universe. Gravitational lensing of the CMB may be contrasted with
competing weak lensing probes in that: (i) for the CMB, the sources (i.e., the objects
being lensed) are the most distant possible, situated at $z\approx 1100,$ (ii) because of
the above, CMB lensing is sensitive to clustering at larger redshifts, thus less
sensitive to nonlinear corrections, and (iii) unlike observations of the correlations of
the observed ellipticities of galaxies, where there are systematic errors due to
intrinsic alignments, for CMB lensing there is no such problem.

In the linear theory (which is a good approximation, especially
for a heuristic discussion), the deflection due to lensing may be
described by means of a lensing potential, defined as a function
of position $\hat \Omega $ on the celestial sphere. Let $\hat
\Omega_{{\rm unlens}}$ coordinatize the last scattering surface as
it would look in the absence of lensing, and define $\hat
\Omega_{{\rm lens}}$ to be the position on the last scattering
surface with lensing taken into account. We may express
 \begin{eqnarray}
\hat \Omega _{{\rm lens}}=\hat \Omega_{{\rm unlens}}+\nabla \Phi_{{\rm lens}}(\hat
\Omega ).
 \end{eqnarray}
It follows that
 \begin{eqnarray}
T_{{\rm lens}} = T_{{\rm unlens}}-(\nabla \Phi_{{\rm lens}})\cdot (\nabla T_{{\rm
unlens}}),
 \end{eqnarray}
and similar expressions may be derived for the polarization. As in weak lensing,
we may derive a projected Poisson equation for the lensing potential
 \begin{eqnarray}
\nabla ^2\Phi_{{\rm lens}}(\hat \Omega ) = \int _0^{x_{{\rm
source}}} dx\,
 W(x;x_{{\rm source}} ) \,\delta \rho (\hat \Omega,x),
 \end{eqnarray}
where $\nabla ^2$ is the Laplacian operator on the
sphere. More details about the lensing of the CMB and what can be learned from it may be
found in the review \cite{lensingReview} and references therein.

\begin{figure}[t]
\begin{center}
\includegraphics[width=15cm]{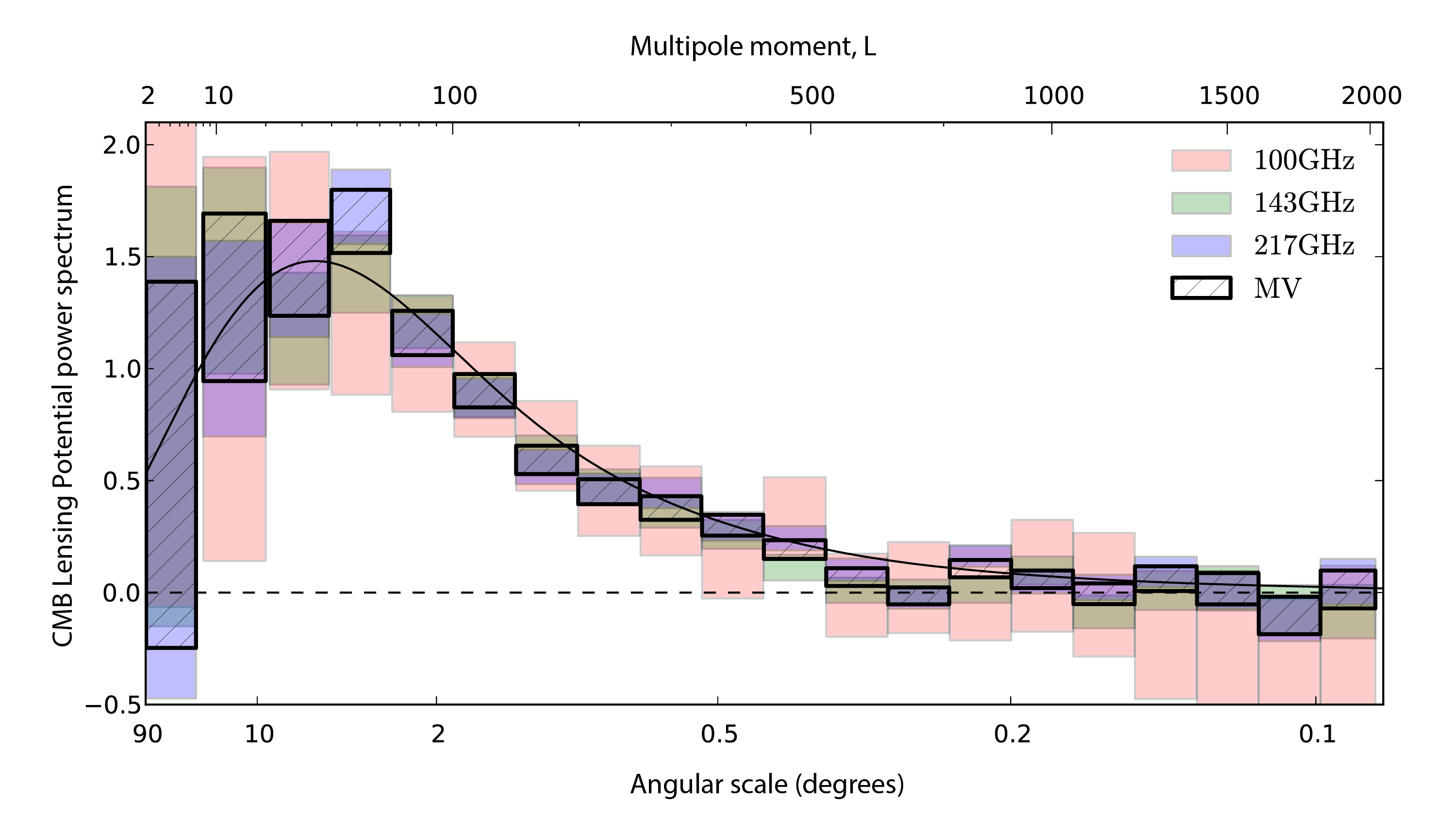}
\end{center}
\caption{{\bf Power spectrum of CMB gravitational lensing potential as measured by
Planck.} The angular power spectrum for the lensing potential as reported in the
Planck 2013 results is shown above. The overall statistical significance for a
lensing detection is greater than $25\sigma.$ The solid curve indicates the
expectation based on the concordance model whose six parameters have been fixed by
the Planck data combined with ancillary data. For more details, see
Ref.~\protect\cite{planck-lensing}. ({\it {\kern-1pt}Credit\/$:$ ESA/Planck
Collaboration})\label{planckGLspectrum}}
\end{figure}

The first discovery of gravitational lensing of the CMB involved detecting a statistically significant
nonzero cross-correlation between a lensing map reconstructed using the WMAP data and infrared galaxies
\cite{lensing-cross}. The use of a cross-correlation allows one to detect a small signal within a noisy
map. Figure~\ref{planckGLspectrum} shows the gravitational lensing spectrum as measured by Planck
\cite{planck-lensing}. Measurements of gravitational lensing have also been made by ACT
\cite{act-lensing} and SPT \cite{spt-lensing}, and the Polarbear experiment has recently reported the
discovery of B modes due to lensing \cite{polar-bear-bmode-lensing}. The lensing field described above
(which can be expressed either as a potential, as a deflection field, or as a dilatation and shear field)
can in principle be recovered from distortions in the small-scale power spectrum due to the lensing field
on larger scales. As a general rule, the higher the resolution of the survey, the better the
reconstruction. Hu and Okamoto \cite{okamoto,okamotoHu} proposed a reconstruction based on Fourier
modes, and later many other estimators have been proposed, which are all quite similar because they are
all exploiting the same signal. In the future, as observations at higher sensitivity and in particular
due to B modes at small scales become available, lensing promises to become a powerful probe of the
clustering of matter and halo structure. It is hoped that lensing will eventually be able to determine
the absolute neutrino masses, even if they are as low as allowed by present neutrino oscillation data
\cite{lesgourguesPastor}. Lensing detected through B modes is particularly advantageous because the fact
that the B modes should be zero in the absence of lensing means that cosmic variance does not intervene
and that only the instrument noise is an issue.\footnote{This conclusion hardly depends on $r$ whatever
its final value may turn out to be, because the angular scales best for probing lensing lie beyond the
recombination bump of the B modes.}

\section[CMB statistics]{CMB Statistics}
\label{NonGaussianity}

\subsection[Gaussianity, non-Gaussianity, and all that]%
{Gaussianity, non-Gaussianity, and all that}

For the simplest models analyzed at linear order, the primordial CMB signal is the
outcome of an isotropic Gaussian random process on the celestial sphere. While any
particular realization of this random process produces a sky temperature map that
is not isotropic, the hypothesis is that the underlying stochastic process is
isotropic. Concretely the hypotheses of isotropy and Gaussianity imply that the
probability of obtaining a sky map $T(\hat \Omega ),$ whose multipole expansion in
terms of spherical harmonic coefficients is given by
\begin{eqnarray}
T(\hat \Omega )= \sum _{\ell =0}^\infty \sum _{m=-\ell}^{+\ell } a_{\ell m}
 Y_{\ell m}(\hat \Omega ),
 \end{eqnarray}
is as follows:
 \begin{eqnarray}
P( \{ a_{\ell m}\}) = \prod _{\ell =0}^\infty \prod
_{m=-\ell}^{+\ell } \frac{1}{\sqrt{2\pi C_{\ell}^{th}}}
\exp\!\left[- \frac{1}{2} \frac{| a_{\ell m}| ^2 }{C_{\ell}^{th}}
\right]\!. \label{GaussIso}
 \end{eqnarray}
Here $\hat \Omega =(\theta, \phi )$ denotes a position on the
celestial sphere and the set of positive coefficients $\{
C_{\ell}^{th} \}$ represents the angular power spectrum of the
underlying stochastic process, or the theoretical power spectrum.

If we had postulated Gaussianity alone without the additional hypothesis of
isotropy, we would instead have obtained a more general probability distribution
function having the form
\begin{eqnarray}
P( \{ a_{\ell m}\}) = {\det }^{-1/2}( 2\pi C_{th}) \exp\!\left[
-\frac{1}{2} \sum _{\ell,m} \sum _{\ell ',m'} {a_{\ell m}}^*
{({C_{th}})^{-1}}_{\ell m, \ell 'm'} a_{\ell 'm'} \right]\!.
\label{genGauss}\qquad
\end{eqnarray}
Here $C_{th}$ is the covariance matrix for the spherical harmonic
multipole coefficients. This Ansatz is too general to be useful
because there are many more parameters than observables, and in
cosmology we can observe only a single sky---or said another
way, only a single realization of the stochastic process defined
in Eq.~(\ref{genGauss}). The assumption of isotropy is very
strong, greatly restricting the number of degrees of freedom by
setting all the off-diagonal elements to zero and requiring that
the diagonal elements depend only on $\ell.$

Happily, the present data, subject to some caveats (such as
gravitational lensing, discussed above in
Sec.~\ref{GravitationalLensingSection}), seem consistent with the
hypotheses of isotropy and Gaussianity, notwithstanding some
``anomalies'' at moderate statistical significance (discussed in
detail below in Sec.~\ref{AnomaliesSection}). These anomalies
could be a sign of something new, but given their limited
statistical significance, the argument that they are simply
statistical ``flukes'' cannot be rejected.

For completeness, we now indicate how the above formalism is
extended to include the polarization of the primordial CMB signal.
There are several formalisms for describing the polarization of
the CMB, for example, using spin-weighted spherical harmonics, or
using the Stokes parameters $I,$ $Q,$ $U,$ $V$ \cite{BornAndWolf}.
(The polarization of the CMB and its description are discussed in
Refs.~35, 96, and~108) We may define $T_{ab}(\hat \Omega )=
\langle {\mathcal{E}}_{a}(\hat \Omega )^* {\mathcal{E}}_{b}(\hat
\Omega ) \rangle _{\textrm{time}}$ where ${\mathcal{E}}_{a}(\hat
\Omega )$ is the electromagnetic field component propagating from
the direction $\hat \Omega $ and units are chosen so that for a
unit electric polarization vector ${\boldsymbol{\varepsilon }}$,
and $T_{ab}(\hat \Omega ) \varepsilon _a \varepsilon _b$
represents the CMB thermodynamic temperature when by means of a
linear polarizer only the component along
${\boldsymbol{\varepsilon }}$ is measured. We may decompose
\begin{eqnarray}
T_{ab}(\hat \Omega )= T(\hat \Omega ) \delta _{ab} +P_{ab}(\hat \Omega ),
\end{eqnarray}
where the polarization tensor (containing the Stokes parameters
$Q,U$ and $V$) is trace-free and Hermitian (and also symmetric in
the absence of circular polarization).

We must now choose a convenient basis for the polarization. If we considered a
single point on the celestial sphere, we could using Stokes parameters set
$Q=T_{11}-T_{22}$ and $U=2T_{12},$ but no natural choice for the directions
orthonormal unit vectors $\hat e_1$ and $\hat e_2$ in the tangent space at $\hat
\Omega $ can be singled out as preferred. Had we instead chosen
\begin{eqnarray}
 \left(\begin{array}{@{}c@{}}
\hat e_1^\prime \\[3pt]
\hat e_2^\prime
\end{array}\right)=
\left(\begin{array}{@{}c@{\quad}c@{}}
 \cos \theta & \sin \theta \\[3pt]
 -\!\sin \theta & \cos \theta
\end{array}\right)
 \left(\begin{array}{@{}c@{}}
\hat e_1 \\[3pt]
\hat e_2
\end{array}\right)\!,
\end{eqnarray}
we would instead have the Stokes parameters
\begin{eqnarray}
\left(\begin{array}{@{}c@{}}
 Q'\\[3pt]
 U'
\end{array}\right)=
\left(\begin{array}{@{}c@{\quad}c@{}}
 \cos 2\theta & \sin 2\theta \\[3pt]
 -\!\sin 2\theta & \cos 2\theta
\end{array}\right)
\left(\begin{array}{@{}c@{}}
 Q\\[3pt]
 U
\end{array}\right)\!.
\label{QUTrans}
\end{eqnarray}

\noindent The transformation law in Eq.~(\ref{QUTrans}) indicates that the traceless part
of the polarization is of spin-2. If the sky were flat with the geometry of $R^2$ instead
of $S^2$ and we had to include homogeneous modes as well, there would be no way out of
this quandary. But there is a celebrated result of differential geometry that a fur on a
sphere of even dimension cannot be combed without somewhere leaving a bald spot or
singularity.\footnote{This is a consequence of the ``hairy ball'' theorem
\cite{HairyBall}.} A traceless tensor field on the sphere cannot have an everywhere
vanishing covariant derivative.

We can generate a complete basis for the polarization on the celestial sphere by
means of derivative operators acting on the spherical harmonics. Using $Y_{\ell
m}(\hat \Omega )$ as a starting point, we may define an E mode basis vector as
follows:
\begin{eqnarray}
E_{ab; \ell m}(\hat \Omega ) =N_\ell ^{{\rm E,B}} \left( \nabla _a\nabla
_b-\frac{1}{2}\delta _{ab}\nabla ^2\right) Y_{\ell m}(\hat \Omega ),
 \end{eqnarray}
where $N_\ell ^{{\rm E,B}}$ is a normalization factor. Here
$a,b=1,2$ indicate components with respect to an arbitrary
orthonormal basis on $S^2.$ Without introducing more structure,
this is the only way to produce a second-rank traceless tensor on
the sphere with derivative operators acting on $Y_{\ell m}(\hat
\Omega )$ in a way that does not break isotropy, a feature that
ensures that $E_{ab; \ell m}(\hat \Omega )$ transforms under
rotations according to the quantum numbers $\ell m.$ However,
there is another polarization, rotated with respect to the E mode
by $45^\circ,$ and in order to write down a basis for this other
polarization, known as the B modes, it is necessary to introduce
an orientation on the sphere through the unit antisymmetric tensor
$\epsilon _{12}=1.$ This choice of orientation, or volume element
in the language of differential geometry, defines a handedness, or
direction in which the $45^\circ $ rotation is made.

\begin{figure}
\begin{center}
\includegraphics[width=16cm]{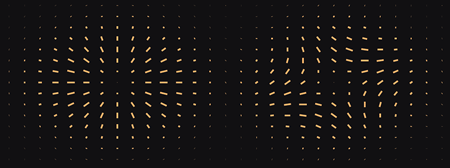}
\end{center}
\caption{{\bf E and B modes of the CMB polarization.} The polarization
pattern in the left panel can be expressed as the double
derivative acting on a potential with its trace removed (in this
case a Gaussian potential) and thus is an E mode. The B mode
pattern on the right, however, cannot be represented in this way,
but can be represented by a psuedo-scalar potential whose double
derivative with the trace removed followed by a $45^\circ $
rotation gives the indicated pattern.\label{EB-Figure}}
\end{figure}

We may define
\begin{eqnarray}
B_{ab}= \frac{1}{2} ( \delta _{aa'}+\epsilon _{aa'}) ( \delta _{bb'}+\epsilon
_{bb'}) E_{a'b'},
 \end{eqnarray}
or in terms of derivative operators
\begin{eqnarray}
B_{ab}= \frac{1}{2} N_\ell ^{E,B} ( \epsilon _{ac}\delta _{bd}+\epsilon
_{bc}\delta _{ad})\nabla _c\nabla _d Y_{\ell m}(\hat \Omega ).
 \end{eqnarray}
The difference between E modes and B modes is illustrated in Fig.~\ref{EB-Figure}.
We want to normalize so that
\begin{eqnarray}
\int d\hat \Omega E_{ab, \ell m} (\hat \Omega ) E_{ab, \ell '
m'}^*(\hat \Omega ) = \int d\hat \Omega B_{ab, \ell m} (\hat
\Omega ) B_{ab, \ell ' m'}^*(\hat \Omega ) = \delta _{\ell, \ell
'}\delta _{m, m'}
\end{eqnarray}

\noindent and
\begin{eqnarray}
\int d\hat \Omega E_{ab, \ell m} (\hat \Omega ) B_{ab, \ell '
m'}^*(\hat \Omega ) =0,
 \end{eqnarray}

\noindent so we require that \cite{stebbins}
\begin{eqnarray}
N_\ell ^{E,B}=\sqrt{\frac{2}{ (\ell+2) (\ell+1) \ell (\ell-1)}}.
 \end{eqnarray}
Although the above discussion emphasized the role of spherical harmonics, the
distinction between $E$ and $B$ modes does not rely on spherical harmonics, and
much confusion has resulted from not fully appreciating this point. The absence of
an $E$ mode component or a $B$ mode component may be formulated in a completely
local manner using covariant derivative operators acting on the polarization
tensor $P_{ab}(\hat \Omega ).$ If $\nabla _aP_{ab}=0,$ then the $E$ mode is
absent, and similarly if $\epsilon _{ab} \nabla _aP_{bc}=0,$ then the $B$ mode is
absent. Consequently, if we are willing to differentiate the data, we could detect
$B$ modes locally with absolutely no leakage. However, the above does not imply
that a unique decomposition of $P_{ab}$ into $E$ and $B$ on an incomplete sky is
possible because nonsingular vector fields (akin to harmonic functions) can be
found that satisfy both conditions, although no nontrivial such vector fields
exist defined on the full sky. Where global topology does enter is in eliminating
configurations that satisfy both the above differential constraints.

Having defined the $E$ and $B$ modes, we now indicate how the statistical distribution
defined in Eq.~(\ref{GaussIso}) is modified to include polarization. We modify our
notation somewhat to avoid unnecessary clutter. When polarization is included, our vector
$\{ a^{\rm T}_{\ell m}\}$ is generalized to become $\{ a^{\rm T}_{\ell m}, a^{\rm
E}_{\ell m}, a^{\rm B}_{\ell m} \}$, which we shall denote as the vector $(\textbf{t},
\textbf{e}, \textbf{b} ).$ For each $\ell $ we have a positive definite covariance matrix
of the form
 \begin{eqnarray}
\textbf{C}_\ell =
 \left(\begin{array}{@{}c@{\quad}c@{\quad}c@{}}
C^{{\rm TT}}_\ell & C^{{\rm TE}}_\ell & C^{{\rm TB}}_\ell \\[9pt]
C^{{\rm ET}}_\ell & C^{{\rm EE}}_\ell & C^{{\rm EB}}_\ell \\[9pt]
C^{{\rm BT}}_\ell & C^{{\rm BE}}_\ell & C^{{\rm BB}}_\ell\\
\end{array}\right)\!.\label{psMatrixA}
\end{eqnarray}
\noindent If we assume that the underlying physics imprinting the cosmological
perturbations is invariant under spatial inversion, then the above covariance matrix
simplifies to
 \begin{eqnarray}
\textbf{C}_\ell =
 \left(\begin{array}{@{}c@{\quad}c@{\quad}c@{}}
C^{{\rm TT}}_\ell & C^{{\rm TE}}_\ell & 0\\[8pt]
C^{{\rm ET}}_\ell & C^{{\rm EE}}_\ell & 0\\[8pt]
0& 0& C^{{\rm BB}}_\ell\\
\end{array}\right)\!,
\label{psMatrixB}
\end{eqnarray}
because under spatial inversion
\begin{eqnarray}
 C^{{\rm TB}}_\ell \to -C^{{\rm TB}}_\ell, \quad C^{{\rm EB}}_\ell
\to -C^{{\rm EB}}_\ell,
\end{eqnarray}
whereas the other power spectra coefficients preserve their sign.

If parity is a symmetry of the physical processes generating the primordial
perturbations and imprinting the CMB anisotropies, the expectation values $C^{{\rm
TB},th}_\ell $ and $C^{{\rm EB},th}_\ell $ must vanish, although for any
particular sky realization $C^{{\rm TB,obs}}_\ell $ and $C^{{\rm EB,obs}}_\ell $
will not vanish because of cosmic variance. (Recall that $C^{{\rm TB},th}_\ell $
and $C^{{\rm EB},th}_\ell $ is the average over a fictitious ensemble including an
infinite number of sky realizations.) Studying $C^{{\rm TB,obs}}_\ell$ and
$C^{{\rm EB,obs}}_\ell $ to search for statistically significant deviations from
zero is a way to search for parity violation in the very early universe. However,
in practice adjusting experimental parameters to minimize $C^{{\rm
TB,obs}}_\ell $ and
$C^{{\rm EB,obs}}_\ell $ is often used to calibrate detector angles in $B$ mode
polarization experiments.

The propagation of light, or electromagnetic radiation, at least
historically and conceptually, occupies a privileged place in the
theories of special and general relativity, because of the
constancy of the speed of light in vacuum and the fact that at
least in the eikonal approximation, light travels along null
geodesics, with its electric polarization vector propagated along
by parallel transport. It is by thinking of bundles of light rays
that we construct for ourselves a physical picture of the causal
structure of spacetime. A modification of the propagation of light
(for example, from dilaton and axion couplings \cite{niOne}, a
possible birefringence of
spacetime \cite{ferte,sorbo,keating,contaldi}, or alternative
theories of
electromagnetism \cite{goldhaberOne,goldhaberTwo,niTwo}) would not
necessarily undermine the foundations of relativity theory, but
would probably merely lead to a more complicated theory, possibly
only in the electromagnetic sector. The CMB may be considered an
extreme environment because of the exceptionally long travel time
of the photons, and one can search for new effects beyond the
known interactions with matter (e.g., Thomson and other
scattering, collective plasma effects such as dispersion and
Faraday rotation). Some such effects would lead to spectral
distortions, for example through missing photons. The rotation of
the polarization vector during this long journey would lead to
mixing of $E$ and $B$ modes, and since the primordial $E$ modes
are so much larger, leads to interesting constraints on
birefringence, which will greatly improve as more data comes in
from searches for primordial $B$ modes.

The statistical description in Eqs.~(\ref{psMatrixA}) and (\ref{psMatrixB})
suffice for confronting the predictions of theory with idealized measurements of
the microwave sky with complete sky coverage and no instrument noise. We must
further assume that there are no secondary anisotropies, nor galactic foregrounds,
both of which have been perfectly removed.

Even under these idealized assumptions, it is not possible to pin down the theory
completely because of a phenomenon known as cosmic variance. The problem is that while we
would like to characterize the covariance of the underlying theory $C^{{\rm AB},th}_\ell$
(where ${\rm A,B}={\rm T,E,B}$), we have only a single realization of the microwave sky,
and $C^{{\rm AB,obs}}_\ell$ is only an estimate of $C^{{\rm AB},th}_\ell$ with
fluctuations about its expectation value which is equal to $C^{{\rm AB},th}_\ell$. If we
consider just the temperature fluctuations, we may define the sample variance for the
$\ell$th multipole
 \begin{eqnarray}
C^{{\rm TT,obs}}_\ell =\frac{1}{(2\ell +1)}\sum _{m=-\ell }^{+\ell
}| a_{\ell m}| ^2.
\end{eqnarray}
The random variable $C^{{\rm TT,obs}}_\ell $ obeys a $\chi
^2$-distribution with $(2\ell +1)$ degrees of freedom and thus has
fractional variance of $\sqrt{2/(2\ell +1)}.$ If we consider a
microwave sky map bandlimited up to $\ell _{\max},$ there are
approximately ${\ell _{\max}}^2$ degrees of freedom, so for
example the overall amplitude of the cosmological perturbations
may be determined with a fractional accuracy of approximately
$1/\ell _{\max}.$

Most analyses of the CMB data (possibly combined with other data sets) assume a
model where the theoretical power spectrum depends on a number of cosmological
parameters, which we may abstractly denote as $ \alpha _1, \alpha _2, \ldots,
\alpha _D,$ also written more compactly as the vector $\boldsymbol{\alpha }.$ For
example, in the six-parameter minimal cosmological model claimed sufficient in the
2013 Planck Analysis \cite{planck-parameters} to explain the present data, the
parameters comprising this vector were $A_S,n_s,H_0,\omega _b,\omega _{{\rm CDM}}$
and $\tau.$

Whether one adopts a frequentist or Bayesian analysis to analyze the data, the
input is always the relative likelihood of the competing models given the
observations, or a ratio of the form:
 \begin{eqnarray}
\frac{ P(\mbox{observations}\,{|}\,\boldsymbol{\alpha
}_1)}{P(\mbox{observations}\,{|}\, \boldsymbol{\alpha }_2)}.
 \end{eqnarray}
In this ratio, the measure for the probability density of the outcome, represented
by a continuous variable, cancels.

In this abridged discussion, we shall follow a Bayesian analysis
where {\it a prior}
probability $P_{{\rm prior}}(\boldsymbol{\alpha })$ is assumed on the space of
models. In Bayesian statistics, Bayes' theorem is used to tell us how we should
rationally update our prior beliefs (or prejudices) in light of the data or
observations, to yield a posterior probability to characterize our updated beliefs
given by the formula
 \begin{eqnarray}
P_{{\rm posterior}}(\boldsymbol{\alpha })= \frac{P(\mbox{observations}\,{|}\,
\boldsymbol{\alpha }) P_{{\rm prior}}(\boldsymbol{\alpha }) }{ \int
d^D\boldsymbol{\alpha}' P(\mbox{observations}\,{|}\, \boldsymbol{\alpha }')
P_{{\rm prior}}(\boldsymbol{\alpha }')}.
\end{eqnarray}
There is no reason why the posterior distribution should lend itself to a simple
analytic form, and a common procedure is to explore the form of the posterior
distribution using Markov Chain Monte Carlo (MCMC) methods
\cite{cosmomc,DunkleyMCMC}, which are particularly well suited to explore
distributions of large dimensionality.

However, it often occurs that the likelihood and the posterior
distribution can be adequately represented as a Gaussian by
expanding about its maximum likelihood value. In this case, the
maximum of the log of the probability can be found using a
numerical optimization routine, and then by computing
second derivatives using numerical finite differences, the
Gaussian approximation to the posterior can be found. This
infinitesimal analysis is often referred to as a Fisher analysis.

\subsection{Non-Gaussian alternatives}

It is incredibly difficult to test the Gaussianity of the primordial microwave sky
without some guidance from theory as to what non-Gaussian alternatives are well
motivated. The space of non-Gaussian stochastic models is dauntingly vast, with
the space of Gaussian models occupying by any measure an almost negligible
fraction of the model space. If one tests enough models, one is guaranteed to
produce a result beyond any given threshold of significance. Consequently, one of
the questions lurking behind any claim of a detection of non-Gaussianity is how
many similar models did one test.

\section[Bispectral non-Gaussianity]{Bispectral Non-Gaussianity}
\label{bispectralNG}

Although it was recognized that inflation, in its simplest form
described by Einstein gravity and a scalar field minimally coupled
to gravity, would have some nonlinear corrections, early analyses
of the cosmological perturbations generated by inflation
linearized about a homogeneous solution and calculated the
perturbations in the framework of a linearized (free) field
theory. According to the lore at the time, the nonlinear
corrections to this approximation would be too small to be
observed. The first calculations of the leading nonlinear
corrections were given by Maldacena \cite{maldacena} and Aquaviva
{\it et~al.} \cite{aquaviva}, who calculated the bispectrum or
three-point correlations of the primordial cosmological
perturbations within the framework of single-field inflation.

Subsequent work indicated that in models of multi-field inflation,
bispectral non-Gaussianity of an amplitude much larger than for
the minimal single-field inflationary models can be obtained. The
predictions of many of these multi-field models were within the
range that would be detectable by the Planck satellite. For many
of these models, the non-Gaussianity, owing to the fact that it is
generated by dynamics on superhorizon scales where derivative
terms are unimportant, is well described by the ``local'' Ansatz
for bispectral non-Gaussianity, under which the non-Gaussian field
$\zeta _{{\it NL}}$ is generated from an underlying Gaussian field
$\zeta _L$ according to the rule
\begin{eqnarray}
\zeta _{{\it NL}}(\mathbf{x}) = \zeta _{L} (\mathbf{x})
 + \big[\zeta_L (\mathbf{x})\big]^2,
 \label{LocalFNLAnsatz}
\end{eqnarray}
where $f_{\it NL}$ is a dimensionless parameter that quantifies
the degree of non-Gaussianity.

Here $f_{\it NL}$ is independent of wavenumber, but a dependence on wavenumber can also
be contemplated, so that the above relation is generalized to the following (now working
in wavenumber space):
\begin{eqnarray}
\zeta _{\it NL}(\mathbf{k})= \int d^3k_1 \int d^3k_2 f_{\it NL}(k;k_1;k_2)\delta ^3(
\mathbf{k} -\mathbf{k}_1 -\mathbf{k}_2 ) \zeta _{L}(\mathbf{k}_1) \zeta
_{L}(\mathbf{k}_2). \label{GeneralFNLAnsatz}
 \end{eqnarray}
Here as a consequence of the hypothesis of isotropy, we have
expressed $f_{\it NL}$ as a function depending only on the lengths
of the vectors. Any two triangles (where the three vectors close)
that may be mapped into each other by an isometry are assigned the
same $f_{\it NL}.$

For bispectral non-Gaussianity of the {\it local} form, as defined
in Eq.~(\ref{LocalFNLAnsatz}), we briefly sketch how the order of
magnitude of $(S/N)^2$ for a detection of $f_{\it NL}$ from a CMB
map extending up to $\ell _{\max}$ may be estimated. We simplify
the estimate by employing the flat sky approximation, allowing us
to replace the unintuitive discrete sums and Wigner $3j$ symbols
with more the transparent continuum integrals and a
two-dimensional $\delta$-function. Approximating $C_\ell \approx
C_0\ell ^{-2}$ (in other words, we ignore acoustic oscillations,
damping tails, and all that) and assuming that the CMB is scale
invariant on the sky, as would occur if one could extrapolate the
simplified Sachs-Wolfe formula $(\Delta T)/T=-\Phi /3$ to
arbitrarily small scales), we may write
\begin{eqnarray}
\begin{split}
\left( \frac{S}{N} \right) ^2&= O(1) \int \frac{d^2\boldsymbol{\ell }}{(2\pi )^2}
\int \frac{d^2\boldsymbol{\ell }_2}{(2\pi )^2} \int \frac{d^2\boldsymbol{\ell
}_3}{(2\pi )^2} (2\pi )^2 \delta ^2 ( \boldsymbol{\ell }_1+ \boldsymbol{\ell
}_2+ \boldsymbol{\ell }_3)\\
&\times \,\frac{ {f_{\it NL}}^2 {C_0}^4 (
 {\ell _1}^{-2} {\ell _2}^{-2}
+{\ell _2}^{-2} {\ell _3}^{-2} +{\ell _3}^{-2} {\ell _1}^{-2}) ^2 }{ {C_0}^3 {\ell
_1}^{-2} {\ell _2}^{-2} {\ell _3}^{-2} } \\
&= O(1){(f_{\it NL})}^2 C_0 {\ell _{\max}}^2 \ln \left( \frac{\ell _{\max}}{\ell
_{\min}} \right)\!.
\end{split}
 \end{eqnarray}
The quantity in the first line simply approximates the sum over distinguishable
triangles, the combinatorial factors being absorbed in the $O(1)$ factor, and the second
line gives the signal-to-noise squared of an individual triangle, as calculated
diagrammatically. More exact calculations taking into account more details confirm this
order of magnitude estimate. Numerical calculations are needed to determine the $O(1)$
factor and deal with $\ell _{\max}$ more carefully. Nevertheless this rough estimate
reveals several characteristics of local non-Gaussianity, namely: (i) the presence of the
$\ln (\ell _{\max}/\ell _{\min})$ factor indicates that the bulk of the signal arises
from the coupling of large angle modes and small angle modes (i.e., the modulation of the
smallest scale power, near the resolution limit of the survey, by the very large angle
modes (i.e., $\ell =2,3, \ldots $), and (ii) the cosmic variance of the estimator derives
from the inability to measure the small-scale power accurately, and thus decreases in
proportion to the effective number of resolution elements of the survey. Cosmic variance
on large scales is not an issue. As long as we know the particular realization on large
scales accurately (which is not subject to cosmic variance), we know what kind of
modulation to look for in the small-scale anisotropies.

Theoretical models for other shapes for the bispectral anisotropy
can be motivated by models for fundamental physics (see for
example Refs.~\cite{babich} and \cite{KomatsuSpergel} and
\hbox{references} therein) and have been searched for as well. The
Planck 2013 Results gave the following limits \cite{planck-fnl}{:}
$f_{\it NL}^{{\rm local}}=2.7\pm 5.8,f_{{\rm NL}}^{{\rm
equil}}=-42\pm 75,$ and $f_{{\rm NL}}^{{\rm ortho}}=-25\pm 39.$
This result rules out many of the models developed to explain a
result at low statistical significance from some analyses of the
WMAP data and was a profound disappointment to those who hoped
that Planck would turn up striking evidence against the simplest
inflationary models. One may anticipate that the results from
Planck 2014 will improve modestly on the 2013 results by including
more modes because of the use of polarization. If there is a
next-generation all-sky CMB polarization satellite having good
angular resolution so that S/N$\approx 1$ maps may be obtained up
to $\ell _{\max}\approx 3000,$ an improvement on the limits on
$f_{{\rm NL}}$ by a factor of a few may be envisaged. However,
beyond $\ell \approx 3000,$ the primary CMB anisotropies become a
sideshow. As one pushes upward in $\ell,$ other nonprimordial
sources of anisotropy take over, such as gravitational lensing,
the Sunyaev-Zeldovich effect, and point sources of various sorts.
These contaminants have angular power spectra that rise steeply
with $\ell $ while the CMB damping tail is falling rapidly, in
fact almost exponentially. For including more modes, polarization
may be helpful. On the one hand, polarization measurements present
formidable instrumental challenges because of their lower
amplitude compared to the temperature anisotropies. However, for
the polarization, the ratio of foregrounds to the primordial CMB
signal is favorable up to higher $\ell $ than for temperature. For
searching for non-Gaussianity, the primordial CMB is highly
linear, unlike other tracers of the primordial cosmological
perturbations. However given that the basic observables are
two-dimensional maps, the small number of modes compared to
three-dimensional tracers of the primordial perturbations is a
handicap.

\section[B modes: a new probe of inflation]{B Modes: A New Probe of Inflation}

As discussed above, some limits on the possible contribution from
tensor modes, or primordial gravitational waves, can be obtained
from analyzing the $C^{TT}$ power spectrum. However, given that
$r$ is not large, these limits become highly model dependent, and
because of cosmic variance as well as uncertainty as to the
correct parametric model, cannot be improved to any substantial
extent. This difficulty is illustrated by the results from the
Planck 2013 analysis \cite{planck-inflation}, which did not
include the data on polarized anisotropies collected by Planck but
only the TT power spectrum. If one assumes the minimal
six-parameter model plus $r$ added as an extra parameter, a limit
of $r<0.11$ (95\% C.L.) is obtained. However, if the primordial
power spectrum in this model (where a form $P(k)\sim k^{n_s-1}$ is
assumed) is generalized to $P(k)\sim (k/k^*)^{n_s-1} \exp (\alpha
(\ln (k/k^*))^2),$ the limits on $r$ loosen by about a factor of
two to become $r<0.2$ (95\% confidence level). This lack of
robustness to the assumptions of the parametric model highlights
the fragility of this approach and demonstrates that the
statistical error bars are not to be trusted unless one has strong
reason to trust the underlying parametric model. However, given
our current understanding of inflationary potentials and dynamics,
we have no reason to trust any of the parametric models, which
serve more as fitting functions for summarizing the current state
of the observations.

Searching for tensor modes using B modes, on the other hand, allows us to probe much
lower values of $r$ in a manner that is substantially model independent because in the
linear theory ``scalar'' perturbations cannot generate any B modes, and this conclusion
is independent of the parametric model assumed.

\subsection{Suborbital searches for primordial B modes}
\label{SubortialBModeSearches}

In March 2014, the BICEP2 Collaboration \cite{bicep2a,bicep2b} claimed a detection of
primordial gravitational waves based primarily on observations at a single frequency
(150\,GHz) over a small patch of the sky (1\% of the sky) that was believed to be of
particularly low polarized dust emission. Using a telescope based at the South Pole, the
BICEP2 team created a map at 150\,GHz of the polarized sky emission from which the B mode
contribution was extracted. The BICEP2 analysis included many null tests to estimate and
exclude systematic errors as well as to estimate the statistical noise in the
measurement. However, in order to claim a detection of primordial gravitational waves,
the BICEP2 team had to exclude a nonprimordial origin for the observed B mode signal, for
which the main suspect would be polarized thermal dust emission. An independent analysis
\cite{flauger} called into question the BICEP2 estimate of the likely dust contribution
in their field, claiming that all the observed signal could be attributed to dust. A
subsequent Planck analysis estimating the dust contribution in the BICEP2 field by
extrapolating polarized dust maps at higher frequencies where dust is dominant down to
150\,GHz confirmed this finding \cite{planckXXX}. While this analysis does not
necessarily exclude a nonzero primordial contribution to the B modes observed by BICEP2
at 150\,GHz, the argument crucial to the BICEP2 claim that a dominant contribution from
dust can be excluded collapses in light of this new finding from Planck concerning the
expected contribution from thermal dust emission. A cross-correlation study is now
underway as a joint project of the two teams that will combine the BICEP2 map at 150\,GHz
with the Planck polarization maps higher frequencies (which are dominated by dust). It
remains to be seen whether this effort can establish a limit on $r$ better than what is
currently available or whether this combined analysis might even result in a detection of
primordial B modes albeit at a lower level of $r$.

In any case, a number of experiments are now underway involving observations at multiple
frequencies that will push down the limits on $r$ using B modes if not result in a first
detection if $r$ is not too small. The BICEP2 team has a series of upgrades to their
experiment expanding their frequency coverage as well as massively increasing the number
of detectors and hence their sensitivity. Other competing efforts include POLARBEAR
\cite{polarBear}, SPIDER \cite{spider} and QUBIC \cite{qubic}, as well as ACTPol and
SPTPol.

Moreover, farther into the future, a more ambitious upgrade to
ground based efforts is contemplated through the US Stage 4 (S4)
\cite{stageFour,stageFourBis} CMB experiment currently under
consideration by the US DOE. S4 contemplates deploying
approximately a total of $2\times 10^5$ detectors from the ground
in order to achieve a massive improvement in raw sensitivity.
Although it is claimed that an improvement in the control of
systematic errors to a comparable level can be achieved, it
remains to be seen whether S4 can realize its projected
performance.

\subsection{Space based searches for primordial B modes}

Several groups around the world have proposed space missions specifically
dedicated to mapping the microwave polarization over the full sky with a
sensitivity that would permit the near ultimate measurement of the CMB B modes. In
Europe, three ESA missions had been proposed: BPol in 2007 \cite{bpol}, COrE
\cite{coreWhitePaper} in 2010, both of which were medium-class, and the
large-class mission PRISM \cite{prism} in 2013. None of these missions was
selected. A mission proposal called COrE+ is currently in the process of being
submitted. In Japan a mission called LiteBird \cite{LiteBird} has been developed,
but final approval is still pending. In the US, CMBPol/EPIC \cite{CMBPol-epic} has
proposed a number of options for a dedicated CMB polarization space
mission,\footnote{See http:/$\!$/cmbpol.uchicago.edu/papers.php for a list of
papers on this effort.} but none of these has been successful in securing funding.
All the above concepts involve a focal plane including several thousand
single-mode detectors in order to achieve the required sensitivity. Another
proposal endeavors to achieve the needed sensitivity by another means: multi-moded
detectors. In the Pixie \cite{pixie} proposal, a Martin-Pupplet Fourier
spectrometer comparing the sky signal to that of an artificial blackbody is
proposed. The setup is similar to the COBE FIRAS instrument, except that
polarization sensitive bolometers and a more modern technology are used, which
will allow an improvement in the measurement of the absolute spectrum about two
orders of magnitude better than FIRAS.

\section[CMB anomalies]{CMB Anomalies}
\label{AnomaliesSection}

The story recounted so far has emphasized the agreement of the observations
accumulated over the years with the six-parameter so-called ``concordance'' model.
But this would not be a fair and accurate account without reporting a few wrinkles
to this remarkable success story. These wrinkles may either be taken as hints of
new physics, or discounted as statistical flukes, perhaps using the pejorative
term {\it a posteriori} statistics. On the one hand, one can argue that if one
looks at enough models that are in some sense independent, one is bound to turn up
something at high statistical significance, and by most standards the significance
of these anomalies is not high. Which interpretation is preferred is presently
under debate. Ultimately, despite all the fancy statistical terminology used in
this discourse, what one concludes inevitably relies on a subjective judgment of
theoretical plausibility.

One anomaly explored over the years, known under several names including ``hemispherical
asymmetry,'' ``dipolar modulation,'' and ``bipolar disorder,'' asks the question whether
the local angular power spectrum is identical when \hbox{compared} between opposite
directions on the celestial sphere. There is no unique way to pose this question
precisely. The statistical significance, amplitude, and direction of the asymmetry depend
somewhat on the precise formulation chosen. However, the WMAP finding of a dipolar
modulation with an amplitude of $\approx 7\% $ and a significance ranging within about
$2-4\sigma,$ and pointing approximately toward $(l, b) = (237^\circ, 20^\circ)$ in
galactic coordinates is broadly confirmed by the Planck 2013 Results. This confirmation
by Planck, owing to the additional high frequency coverage, renders less plausible a
nonprimordial explanation based on foreground \hbox{residuals}. The sensitivity to
dipolar modulation on these angular scales is primarily limited by \hbox{cosmic} variance
for both WMAP and Planck, so future experiments cannot hope to improve substantially on
the significance of these results. Planck, however, because of its superior angular
resolution, was able to probe for dipolar modulation pushing to smaller scales, where
cosmic variance is less of a problem and a possible modulation can be constrained more
tightly. [The measurement uncertainty in the \hbox{scale-dependent} dipole modulation of
the amplitude of the perturbations
 $\Delta
A(\ell )$ (assuming broad binning) scales as $1/\ell $ as long as
cosmic variance is the limiting factor.] The Planck 2013 data find
no evidence of dipolar modulation at the same amplitude extending
to small angular scales, suggesting that if the dipolar modulation
observed by WMAP is not a statistical fluke, a scale-dependent
theoretical mechanism for dipolar modulation is needed. Moreover,
Planck does not see \hbox{evidence} of modulation associated with
higher multipoles (e.g. quadrupolar and higher order disorders).
This last point is important for constraining theoretical models
producing statistical anisotropy by means of extra fields
disordered during inflation.

Another anomaly is the so-called ``cold spot.'' Rather than limiting ourselves to
two-point statistics, or to three-point statistics, searching for a bispectral signal
using theoretically motivated templates, as described in Sec.~\ref{bispectralNG}, we may
ask whether the most extreme values of an appropriately filtered pure CMB map lie within
the range that may be expected assuming an isotropic Gaussian stochastic process, or
whether their $p$-values (i.e., probability to exceed) render the Gaussian explanation
implausible. There are many ways to formulate questions of this sort, a fact that renders
the assessment of statistical significance difficult. One approach that has been applied
to the data is to filter the sky maps with a spherical Mexican hat wavelet (SMHW) filter,
which on the sphere would have the profile of the Laplacian operator applied to a
Gaussian kernel of a given width $\sigma,$ giving a broad [i.e., $(\Delta \ell )/\ell
=O(1)$] two-dimensional spatial bandpass filter. (See Ref.~\cite{vielva2011} for more
details concerning the methodology and Ref.~\cite{vielva2004} for the original claim
of a cold spot detection. See also Ref.~\cite{wmap2011d} for an assessment by the WMAP
team as well as Ref.~\cite{planck-iso-stats} for the analysis from Planck 2013.)

The final anomaly concerns possible alignments of the low-$\ell $
multipoles, imagi\-natively named the ``axis of evil'' by Land and
Magueijo \cite{axisOfEvil}. A Gaussian isotropic theory for the
temperature anisotropy\footnote{The discussion can readily be
generalized to polarization, but we shall stick to the temperature
anisotropy alone in our discussion.} predicts that the multipole
vectors $\mathbf{a}_\ell =\{ A_{\ell m}\} _{m=-\ell}^{+\ell
}$ are each Gaussian and independently distributed. If we add
three or more such vectors together according to the tensor
product of representations of $SO(3)$
\begin{eqnarray}
L_{1}\otimes L_{2}\otimes \cdots \otimes L_{N},
\end{eqnarray}
which may be decomposed into a direct sum of irreducible representations. Provided that
the triangle inequality is satisfied, we may extract one or more scalars (transforming
according to $L=0$) from the above tensor product. The mathematics is simply that of the
usual addition of angular momentum. For each of these scalars, we may ask whether the
observed value lies within the range expected from cosmic variance in the framework of a
Gaussian theory taking into account the uncertainty in the determination of the
underlying power spectrum. The approach just described was not the approach of Land and
Magueijo, who maximized $\max _{m\in [-\ell, +\ell ]}|a_{\ell m}(\hat {\mathbf{n}})|$
as a function of
$\hat{\mathbf{n}}$ where $a_{\ell m}(\hat {\mathbf{n}})$ is the expansion coefficient in
the coordinate system with $\hat {\mathbf{n}}$ pointing to the north pole. In this way,
for each generic value of the multipole number $\ell $ a unique double headed vector, or
axis, $\pm \hat {\mathbf{n}}$ may be extracted, and the alignments between
these axes may be assessed. Land and Magueijo report an alignment of the $\ell =3,$
$\ell =4,$ and $\ell =5$ axes with a $p$-value less than $10^{-3}$ in the Gaussian
theory. Accurate $p$ values may be obtained for each of these invariant quantities by
resorting to MC simulations to account for the practicalities of dealing with a cut sky,
etc. However, there still remains a subjective element to assessing statistical
significance.  One may ask how many invariants were tried before arriving at a reportable
statistically significant anomaly.

\section[Sunyaev-Zeldovich effects]{Sunyaev-Zeldovich Effects}

\begin{wrapfigure}{L}{0.6\textwidth }
\begin{center}
\includegraphics[width=10cm]{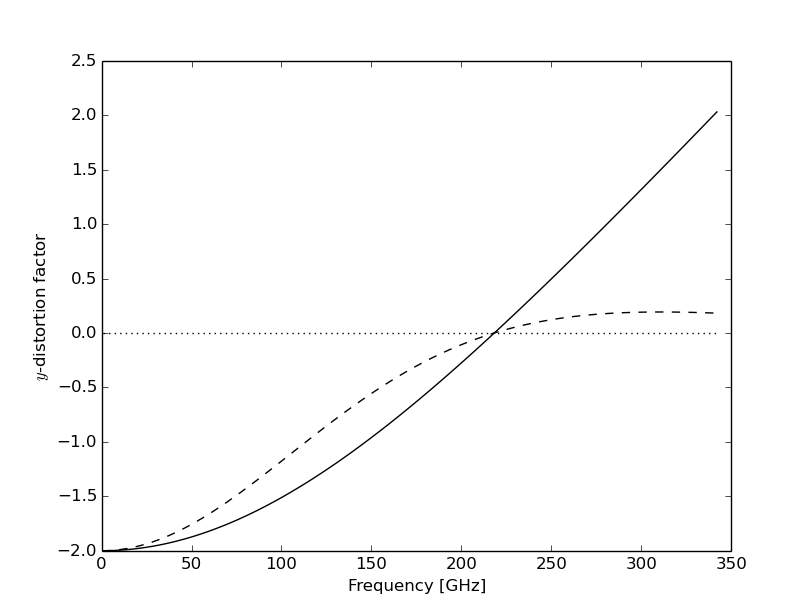}
\end{center}
\caption{{\bf Frequency dependence of thermal Sunyaev-Zeldovich spectral distortions.} The
solid and dashed curves show the fraction spectral distortion in terms of the CMB
thermodynamic and R-J temperatures, respectively. For low frequencies ($\nu <217$\,GHz),
there is a decrement in temperature, corresponding to the fact that low frequency photons
are on the average Doppler shifted to higher frequencies by the hot electrons. On the
other hand, for $\nu >217$\,GHz the net effect is to increase the photon phase space
density.
\label{SZSpectrum}}
\end{wrapfigure}

In the simplified discussion of reionization presented in Sec.~\ref{ReionizationSection},
it was assumed that the electrons responsible for the rescattering are at rest with
respect to the cosmic rest frame. This approximation treats the reionized gas as a cold
plasma having a vanishing peculiar velocity field. As pointed out by Sunyaev and
Zeldovich \cite{sz-orig}, corrections to this approximation arise in two ways: (i) In the
so-called thermal Sunyaev-Zeldovich effect (tSZ), the random thermal motions of the free
electrons alter the spectrum of the rescattered CMB photons through the Doppler effect
and Compton recoil. Because $\langle\mathbf{v}\rangle =0,$ this effect is second order in
the velocity, or linear in the electron temperature, and results in a spectral distortion
not respecting the frequency dependence of a blackbody spectrum with a perturbed
temperature. (See Fig.~\ref{SZSpectrum}.) (ii) In the kinetic Sunyaev-Zeldovich (kSZ)
effect the peculiar velocity of the gas results in a shift in temperature proportional to
the component of the peculiar velocity along the line of sight. The kSZ effect has the
same frequency dependence as the underlying primary CMB perturbations, thus making it
hard to detect given its small magnitude. There is also a much smaller effect where the
transverse peculiar velocity imparts a linear polarization to the scattered photons for
which the polarization tensor $P_{ij}$ is proportional to $(\mathbf{v}_{\perp }\otimes
\mathbf{v}_{\perp }-\frac{1}{2}{v_\perp }^2\boldsymbol{\delta }_\perp )$, where
$\mathbf{v}_{\perp }$ is the peculiar velocity perpendicular to the line of sight and
$\boldsymbol{\delta }_\perp $ is the Kronecker delta function in the plane perpendicular
to the line of sight.

Galaxy clusters are filled with hot gas [$T_{{\rm gas}}=O(10\,{\rm KeV})$] that
emits primarily at X-ray frequencies. The hot electrons of this fully ionized gas,
or plasma, scatter CMB photons by Thomson, or Compton, scattering, shifting their
frequency by a factor of approximately $(1+\beta \cos \theta )$ where $\beta
=k_{\rm B}T/m_ec^2$ and we have ignored higher order corrections in $\beta.$ The
$y$-distortion parameter along a line of sight is given by the integral
 \begin{eqnarray}
y=\int d\tau \frac{k_{\rm B}T_e}{mc^2}= \int d\ell \sigma _T n_e \frac{k_{\rm
B}T_e}{mc^2}, \label{yDistDef}
\end{eqnarray}
and the fractional perturbation in the CMB thermodynamic temperature in the
nonrelativistic approximation is given by
 \begin{eqnarray}
\frac{\delta T_{{\rm CMB}}(\nu, \hat \Omega )}{\bar T_{{\rm CMB}}}=\left(
\frac{x(e^x+1)}{e^x-1}-4\right) y(\hat \Omega ),
\end{eqnarray}
where $x=h\nu /k_{\rm B}T_{{\rm CMB}}=\nu /(57\,{\rm GHz})$ and $y(\hat \Omega )$
is the $y$-distortion map as defined by the line of sight integral in
Eq.~(\ref{yDistDef}).

On the one hand, the thermal Sunyaev-Zeldovich effect provides a
powerful probe of the dynamics of galaxy clusters, which can be
used to discover new clusters and to probe the structure of known
clusters away from their central core. On the other hand, for
observing the primordial CMB, the thermal Sunyaev-Zeldovich
effect constitutes a contaminant that must either be removed or
modelled. The review by Birkinshaw \cite{MBirkinshaw} recounts the
early history of tSZ measurements. See also the reviews by
Carlstrom {\it et~al.} \cite{CarlstromSZReview} and by Rephaeli
\cite{rephaeli}. For the observation of the kSZ, see
Ref.~\cite{kSZobs}. For
 more recent measurements, see the SZ survey papers from the ACT \cite{act-sz},
SPT\cite{spt-sz}, and Planck \cite{planck-cluster-counts,planck-all-sky-compton-map}
collaborations. See Ref.~\cite{prism} for a discussion of what might ultimately be
possible from space in the future.

\section[Experimental aspects of CMB observations]%
{Experimental Aspects of CMB Observations}

Like astronomical observations at other wavelengths, most modern
microwave experiments consist of a telescope and a number of
detectors situated at its focal plane. The COBE experiment was an
exception, as it used single-moded feed horns pointed directly at
the sky to define its beam, which was relatively broad ($7^\circ $
FWHM). However for measurements at higher angular resolution, it
is difficult to construct a feed horn forming a sufficiently
narrow beam on the sky without the help of intermediate optics,
unless one resorts to interferometry as for example in the DASI
experiment.

Observations of the sky at microwave frequencies present a number of challenges unique to
this frequency range. Unlike at other frequencies, the microwave sky is remarkably
isotropic. The CMB monopole moment, and to a much lesser extent the dipole moment,
dominates. In the bands where other contaminants contribute least---that is, roughly in
the range 70--150\,GHz---virtually all the photons collected result from the isotropic
2.725\,K background. The brightest feature superimposed on this uniform background is the
CMB dipole, resulting from our proper motion with respect to the cosmic rest frame, at
the level of 0.1\%. When this dipole component is removed, the dominant residual is the
primordial CMB anisotropy, whose amplitude on large angular scales is about $35\,\mu$K,
or roughly $10^{-5}$. This situation is to be contrasted for example with X-ray astronomy
where only some 50 photons need to be collected to make an adequate galaxy cluster
detection. For the CMB about $10^{10}$ photons must be collected in each pixel just to
obtain the temperature anisotropy at a signal-to-noise ratio of one, and many more
photons are needed for measuring polarization at the same marginal signal-to-noise. The
challenge is to measure minute differences in temperature between different points on the
celestial sphere without introducing spurious effects. These can arise from a variety of
sources: drifts in the zero point of the detector, detector noise, hot objects in the far
sidelobes of the beam, and diffraction around the edges of the mirrors, to name some of
the most common problems.

\begin{figure}
\begin{center}
\includegraphics[width=15cm]{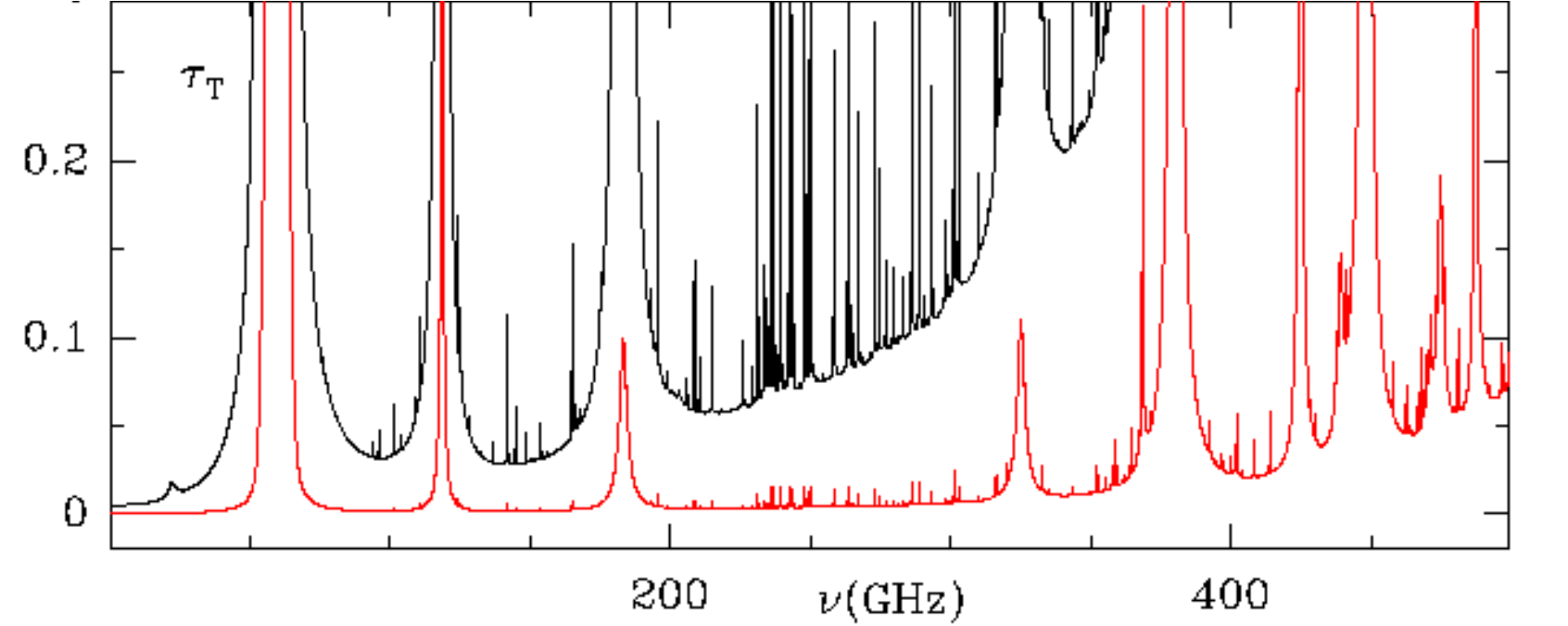}
\end{center}
 \caption{\textbf{Atmospheric optical depth as a function of frequency for 1\,mm of
precipitable water vapor.} Water vapor is an important but not the only source of atmospheric
contamination. See Ref.~\protect\cite{PardoCernicharo} for more details. A column density of
0.5--1\,mm of precipitable water vapor corresponds to good conditions in the Atacama deserts. Conditions
at the South Pole are often better by a factor of a few. The black curve shows the opacity, or optical
depth, and the red curve shows the opacity multiplied by 1/20 in order to highlight the structure around
the resonances. The limited transmission $0<t<1$ [where $t=\exp (-\tau )],$ on the one hand, attenuates
the desired sky signal. But more serious is the superimposed additive noise, having a brightness
temperature $(1-t)T_{{\rm atm}}.$ Here $T_{{\rm atm}}$ is the effective temperature of the atmosphere.
({\it Credit: J. Cernicharo})
\label{atmosphere}}
\end{figure}

For this reason, almost all measurements of the microwave sky are in some sense
differential. Drifts in the zero point of the detector (known as $1/f$ noise) can
be mitigated by rapidly switching either between the sky and an artificial cold
load with a long time constant or between different parts of the sky. This can be
accomplished by moving the beam rapidly across the sky, so that in one way or the
other the basic observable becomes differences in intensity of pixels whose
measurements are closely separated in time. The point to take away is that one
absolutely avoids making direct rather than differential measurements of the sky
temperature.

Another specificity of astronomical observations at microwave
frequencies is that extremely stringent requirements must be
imposed limiting the magnitude of the far sidelobes of the beam.
Given the smallness of the differences in temperature of interest,
contamination from the ground, the earth, the sun, and moon, and
also from the galaxy near the galactic plane, can easily introduce
spurious signals if the beam does not fall sufficiently rapidly
away from its central peak. Ground pickup (from a spherical angle
of $2\pi $) is particularly challenging to shield, and this is why
the WMAP and Planck satellites were situated at L2 (the second
Lagrange point, where the earth and the moon have a tiny angular
diameter) rather than in a low-earth orbit, which would equally
well avoid atmospheric emission, which is the other major
motivation for going to space. The second earth-sun Lagrange point
(L2) is $1.5\times 10^6$\,km from the earth---that is, about 4
times the earth-moon distance. From L2 the earth subtends an
angle of only 16 arcminutes. More importantly from L2, the sun,
earth, and moon all appear at approximately the same position in
the sky making it easier to keep away from the far sidelobes of
the beam.

A formidable challenge for suborbital observations is emission from the atmosphere,
primarily but not exclusively from water vapor. The emission from the atmosphere has a
complicated frequency dependence, as shown in Fig.~\ref{atmosphere}, where the optical
depth as a function of frequency is shown. This optical depth must be multiplied by the
atmospheric temperature (i.e., $\approx 270$\,K) to obtain the sky brightness. The
structure in frequency of the regions of high opacity implies that bands must be
carefully chosen to avoid emission lines and bands, giving less flexibility for wide
frequency coverage than from space. The problems arising from atmospheric emission can be
mitigated to some extent by choosing a site on the ground with a particularly small water
column density, such as the South Pole or the Atacama desert in Chile, or by observing
from a stratospheric balloon. Atmospheric emission is problematic in two respects. First,
if the atmospheric emission were stable in time, it would merely introduce additional
thermal loading on the detectors, which could be mitigated simply by deploying more
detectors or by observing over a longer time. If the contribution from the sky is
$T_{{\rm sky}},$ then the fluctuations in intensity at the detector (assumed perfect for
the moment) would increase by a factor of $(T_{{\rm sky}}+T_{{\rm CMB}})/T_{{\rm CMB}},$
meaning that we could measure the sky temperature with the same error by increasing
either the number of detectors or the total observation time by a factor of $((T_{{\rm
sky}}+T_{{\rm CMB}})/T_{{\rm CMB}})^2.$ At 150\,GHz from the South Pole, for example, to
choose a band where the sky brightness temperature is not very high, one has $T_{{\rm
sky}}\approx 16$\,K. This means that under the idealized assumptions made here, almost 40
times more detectors would be required than for a space-based experiment. But the
situation is worse than this, because the atmosphere contamination is not stable in time
but rather fluctuates in a complicated way characterized by a wide range of time scales
(owing to the underlying turbulence). Moreover, the atmospheric load varies with zenith
angle, roughly according to a secant law. Of course, observations from the ground are
attractive because of their low cost, the fact that telescopes can be improved from
season to season to address problems as they arise, and larger telescopes with many more
detectors can be deployed. As experiments increase in sensitivity, requirements for
mitigating these problems will become more stringent. At present it is not known what is
the limit on the quality of measurements that can be achieved from the ground or from
balloons.

\subsection{Intrinsic photon counting noise: ideal detector behavior}

In radio astronomy (where the frequency of observations is typically lower than
for CMB observations) the following expression, known as the radio astronomers'
equation \cite{dickeRSI,KrausBook}, gives the fractional error of an observation
under slightly idealized assumptions:
\begin{eqnarray}
\frac{\delta I}{I}= \left[ \frac{T_{{\rm sky}}+T_{{\rm sys}}}{T_{{\rm sky}}}\right]^2
\frac{1}{\sqrt{(\Delta B)t_{{\rm obs}}}}. \label{RadioAstrFormula}
\end{eqnarray}
Here $t_{{\rm obs}}$ is the time of observation of a pixel and $(\Delta B)$ is the
bandwidth. $T_{{\rm sky}}$ is the sky brightness temperature averaged over the pixel and
$T_{{\rm sys}}$ represents the additional noise introduced within the receiver. This
expression can be understood by considering the source in the sky as having thermal
statistics---that is, fluctuating in intensity as a Gaussian random field. In radio
astronomy, detection is coherent and each detector picks up only a single transverse mode
of the electromagnetic field incident from the sky, which is converted by an antenna or a
feed horn into an electrical signal, which depends only on time. This signal is
coherently \hbox{amplified} and nowadays digitized. Mathematically, in the absence of
additional detector noise, the sky signal can be thought of as a signal whose complex
amplitude fluctuates according to a Gaussian distribution with a variance proportional to
$T_{{\rm sky}}.$ The coherence time is $1/(\Delta B),$ so in a time interval $t_{{\rm
obs}},N_{{\rm samp}}=2(\Delta B)t_{{\rm obs}}$ independent realizations of this Gaussian
random process whose variance we are trying to measure are collected, leading to a
fractional error in the determination of the variance, or $T_{{\rm sky}},$ of
$1/\sqrt{N_{{\rm samp}}}.$ For a more realistic measurement noise is also introduced, and
in the above formula the noise is idealized as an independent additive Gaussian random
signal, characterized in terms of a system temperature $T_{{\rm sys}},$ to be added to
$T_{{\rm sky}}$ to obtain the variance actually measured. The additive noise lumped
together and known as the ``system'' temperature includes thermal emission from the
atmosphere, Johnson noise from dielectric losses in the feedlines, noise in the detector
itself---in other words, everything other than the sky temperature that one would
measure from space with an ideal measuring device.

The above formula was derived treating the incoming electromagnetic field as entirely
classical, and for a thermal source this treatment is valid in the R-J part of the
spectrum---that is, when $T_{{\rm source}}\gg h\nu /k_{\rm B}.$ For the CMB, with
$T_{{\rm CMB}}=2.725$\,K, this requires that $\nu \ltorder \nu _{{\rm CMB}}=k_{\rm
B}T_{{\rm CMB}}/h=57$\,GHz, meaning that quantum effects introduce significant
corrections to the above formula, but one is still far from the regime where photons
arrive in a nearly uncorrelated way, obeying Poissonian counting statistics, which is
what happens for observations in the extreme Wien tail of the blackbody distribution.
Before introducing the quantum corrections to Eq.~(\ref{RadioAstrFormula}), let us finish
our discussion of this result based on classical theory extracting all the lessons to be
learnt from this result. In the R-J regime, where $N\gg 1,$ $N$ being the photon
occupation number, roughly speaking, photons arrive in bunches of $N$ photons, so the
discreteness of the photons is not an issue. As long as a fraction greater than about
$1/N$ of the photons is captured, little additional noise is introduced. The intrinsic
noise of the incoming electromagnetic field is almost entirely classical in origin and
can be modeled faithfully using stochastic classical electromagnetic formalism. (For a
more detailed discussion of these issues, see Ref.~\cite{mandelWolf}.) From the above
formula, we learn that noise can be reduced by the following measures: (i) choosing as
wide a bandwidth as possible, (ii) increasing the observation time, and (iii) increasing
the number of detectors. We also learn that although lowering the internal noise of the
detection system increases the accuracy of the measurements, once the system temperature
is approximately equal to the brightness temperature of the source, lowering the system
temperature further leads only to marginal improvement in sensitivity.

We now derive the quantum corrected version of the radio astronomers equation,
using a Planck distribution instead of the classical description based on a
deterministic stochastic Gaussian field. For the Planck distribution, where
$p_N=x^N/(1-x)$ with $x=\exp (-h\nu /k_{\rm B}T),$
\begin{eqnarray}
\langle (\delta N)^2 \rangle = \langle N \rangle ^2 + \langle N \rangle = {\bar
N}^2+\bar N.
\end{eqnarray}
In the R-J regime, where $\bar N\gg 1,$ the first term dominates, reproducing the result
obtained using a classical random field as described above. But as the frequency is
increased and one starts to pass toward the Wien tail of the distribution, the second
term increases in importance. In the extreme Wien tail, where the second term dominates,
we observe the fluctuations characterized by Poisson statistics, where the (rare)
arrivals of photons are completely uncorrelated. This means that the fractional error for
the number of photons counted must be increased by a factor~of
\begin{eqnarray}
\left( 1 + \frac{1}{\bar N}\right) = \exp\!\left( \frac{h\nu }{2k_{\rm B}(T_{{\rm
sky}}+T_{{\rm sys}})} \right)\!,
\end{eqnarray}
and the quantum-corrected version of Eq.~(\ref{RadioAstrFormula}) becomes
\cite{lamarreBolo}
\begin{eqnarray}
\frac{\delta I}{I}= \left( \frac{T_{{\rm sky}}+T_{{\rm sys}}}{T_{{\rm sky}}}\right)^2
\frac{1}{\sqrt{( \Delta B)t_{{\rm obs}}}} \exp\!\left( \frac{h\nu }{2k_{\rm B}(T_{{\rm
sky}}+T_{{\rm sys}})} \right)\!. \label{RadioAstrFormulaBis}
 \end{eqnarray}

\subsection{CMB detector technology}

Having analyzed the performance of an ideal detector limited only
by the intrinsic fluctuations of the incoming electromagnetic
field, we now review the state of the art of existing detector
technologies, which may be divided into two broad classes: (i)
coherent detectors, and (ii) incoherent detectors.

\looseness=-1 Coherent detectors transfer the signal from a feed horn or antenna onto a transmission line
and then coherently amplify the signal using a low-noise amplifier, generally using high electron
mobility transistors (HEMTs) and switching with a cold load or between different points on the sky in
order to mitigate $1/f$ noise \cite{PospieszalskiWollack}. Coherent detection is an older technology.
Coherent detectors have the advantage that they can operate at much higher temperatures than the
competing incoherent bolometric detectors. Avoiding active cooling through a cryogenic system is a
definite plus, especially in space, where reducing risk of failure is important. However, unlike their
incoherent competitors, at the high frequencies of interest for CMB observations, coherent detectors have
noise levels far above the quantum noise limit, and until now coherent detectors have not been able to
reach frequencies above circa 100\,GHz, although some believe that this situation may improve. It is not
presently known how coherent amplification would function in the Wien part of the blackbody spectrum.
Coherent detectors were used for the COBE DMR experiment, for WMAP (including the frequencies 20, 30, 40,
60, and 90\,GHz), and the Planck LFI (low frequency instrument), which included the frequencies 30, 45,
and 70\,GHz. One advantage of coherent detectors is their insensitivity to cosmic rays.

Another property of coherent detectors is that for a single mode
measurement, all four Stokes parameters (i.e., I, Q, U, V) can be
measured simultaneously. For incoherent detectors one can measure
only two Stokes parameters simultaneously. However for
measurements outside the R-J part of the spectrum, this advantage
of coherent detectors disappears, because in the Wien part of the
spectrum, the photons are not bunched to any appreciable degree.
On the other hand, for coherent detectors, it is not possible to
construct multi-moded detectors, where the number of detectors can
be greatly reduced by sacrificing angular resolution, but not at
the cost of less sensitivity on large angular scales. Coherent
detectors are always single-moded, sampling at the diffraction
limit.

Incoherent detectors, unlike their coherent counterparts, do not
attempt to amplify the incident electromagnetic wave. Rather the
incident electromagnetic wave, or equivalently the stream of
photons, is directly converted into heat, changing the temperature
of a small part of the detector of small heat capacity, whose
temperature is monitored using a thermistor of some sort.
Typically the detector is cooled to an average temperature much
lower than that of the incident radiation. The bolometers of the
Planck high frequency instrument (HFI) were cooled to 0.1\,K.
Consequently each CMB photon produces many phonons in the
detector, and therefore the counting statistics of the phonons do
not to any appreciable extent limit the measurement. Heat flows
from the sky onto the detector and is then conducted to a base
plate of high heat capacity and thermal stability. A thermal
circuitry (which can be modelled as a sort of RC circuit with as
many loops as time constants) is chosen with a carefully optimized
time response.

Several technologies are used for the thermistor. One of the most
popular nowadays is transition edge sensors (TES) \cite{lee-tes},
where a superconducting thin film is kept at the edge of the
superconducting transition, so that minute changes in temperature
give rise to large changes in resistivity. A feedback circuitry
with heating is used to maintain the assembly near a fixed
location on the normal-superconducting transition edge. One
disadvantage of TES detectors is their low electrical impedance,
which requires that SQUIDs (superconducting quantum interference
devices), which need careful magnetic shielding, be used for the
readout.

Several new promising technologies are being developed and perfected for the next
generation of bolometric detectors. One such technology is kinetic inductance detectors
(also known as KIDs) \cite{dayArticle,kids}. For these detectors, a resonant electric
circuit is constructed by depositing a pattern of superconducting film on a thick
dielectric.  Photons incident on the superconducting film break Cooper pairs causing its
surface impedance to vary with the incident photon flux, which in turn alters the
resonant frequency. The shifts in frequency of the resonator allow the KID to be read out
in a simple way without using SQUIDs. The surface impedance of the superconducting film
is largely determined by the inertia of the superconducting electrons, hence the name
``kinetic inductance.''

Another possibly promising technology is the cold electron bolometer
\cite{kuzmin-ceb}, involving a small metal filament connected to a superconductor
on both sides by means of a junction with a thin insulator in between. A bias
current that passes through the assembly acts to cool the electrons in the thin
wire through a Peltier-like effect to a temperature far below that of the
substrate. CEBs are hoped to be more resistant to cosmic ray spikes than other
detector technologies.

Cosmic rays have proved to be more problematic than anticipated in the Planck experiment,
and substantial effort was needed to remove the contamination from cosmic ray events in
the raw time stream \cite{planck-hfi-cosmic-rays}. Some of the data was vetoed and
corrections were applied to the long-time tails of the larger magnitude events. While
experiments from the ground benefit from substantial shielding of cosmic rays by the
Earth's atmosphere, how best to minimize interference from cosmic rays will be a major
challenge for the next generation of experiments from space, which target sensitivities
more than two orders of magnitude beyond the Planck HFI. Another challenge will be
multiplexing, which is necessary to reduce the cooling requirements when the number of
detectors is greatly expanded.

\begin{figure}[t]
\begin{center}
\includegraphics[width=12cm]{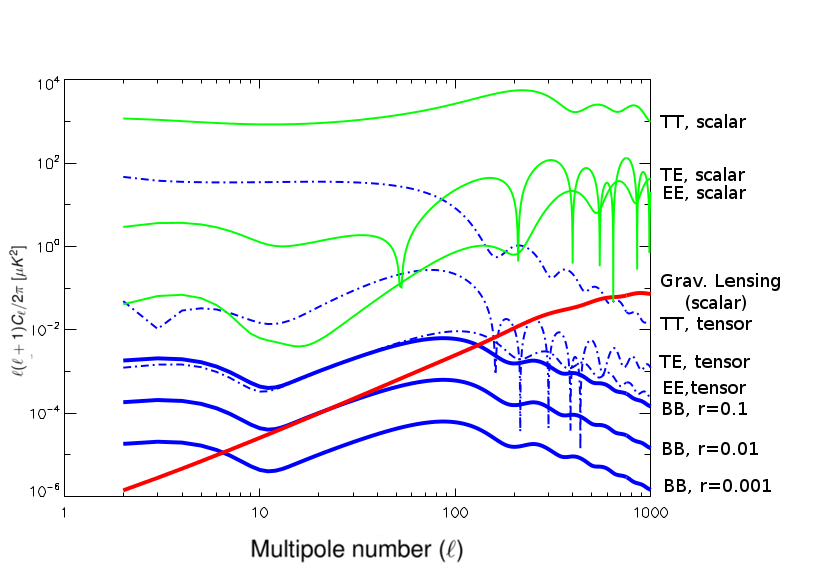}
\end{center}
\caption{{\bf Summary of T,E, and B anisotropy power spectra from
scalar and tensor modes.} The green curves (from top to bottom)
represent the TT, TE and EE CMB power spectra for the ``scalar''
mode, while the blue curves represent the CMB isotropies for the
tensor mode. The~broken blue curves (from top to bottom) and the
top solid blue curve represent the TT, TE, EE and BB anisotropies
assuming a value of the tensor-to-scalar ratio of $r=0.1.$ The
lower two solid curves represent the predicted BB anisotropy for
$r=0.01$ and $r=0.001.$ The red curve shows the BB power spectrum
arising nonlinearly from the scalar mode as a result of
gravitational lensing. 
({\it Credit: M.  Bucher})
\label{AllSpectraPlot}}
\end{figure}

\subsection{Special techniques for polarization}

If it were not for the fact that the polarized CMB anisotropy
(i.e., the $Q$ and $U$ Stokes parameters) is much smaller than the
anisotropy in the $I$ Stokes parameter, measuring polarization
would not pose any particular problems, because most detectors are
polarization sensitive or can easily be made so, and there would
be no need to discuss any techniques particular to polarization
measurements. An inspection of the relative amplitudes of the
spectra in Fig.~\ref{AllSpectraPlot} highlights the problems
encountered. In particular if one wants to search for $B$ modes at
the level of $r\approx 10^{-3}$ or better, the requirements for
preventing leakages of various types become exceedingly stringent,
as we now describe.

Let us characterize the problem for the most difficult polarization observation:
measuring the primordial B modes. We may classify the leakages to be avoided by
ordering them from most to least severe: (i) leakage of the CMB temperature
monopole to the $B$ mode, (ii) leakage of the $T$ anisotropies to the $B$ mode,
and (iii) leakage of the $E$ mode to the $B$ mode.

\begin{figure}[h] 
\begin{center}
\includegraphics[width=6cm]{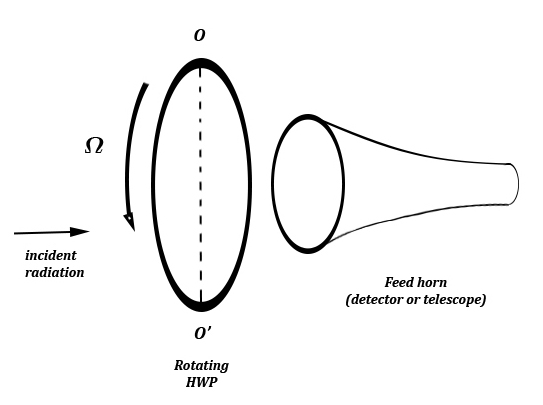}
\end{center}
 \caption{{\bf Polarization modulation by a rotating half-wave plate.} We show
a transmissive rotating half-wave plate (HWP), here placed in front of a microwave feed
horn pointing directly at the sky. However for modern CMB experiments, the horn would
generally be replaced with a telescope having intermediate optical elements and horns on
the focal plane. The rotating HWP is birefringent, with its fast polarization axis
represented by the line $OO'$. As explained in the main text, by measuring only that part
of the signal varying at an angular frequency $4\Omega $ (where $\Omega $ is the angular
velocity of the rotating HWP), one can prevent leakage from the much larger Stokes
parameter $I$ into the linearly polarized components $Q$ and $U.$ The presence of the
HWP, while complicating the instrument design somewhat, allows many hardware requirements
to be relaxed substantially compared to what would be needed without polarization
modulation.
\label{hwp:fig}}
\end{figure}

One of the solutions to many of these problems is polarization modulation. The idea is to
place an element such as a rotating half-wave plate, whose orientation we shall denote by
the angle $\theta,$ between the sky and the detector, preferably as the first element of
the optical chain, as sketched in Fig.~\ref{hwp:fig}. Half-wave plates can be constructed
using anisotropic crystals such as sapphire or using a mirror with a layer of wires a
certain distance above it so that one linear polarization reflects off the wires while
the other reflects off the mirror, resulting in a difference in path length.
Mathematically, we have
\begin{eqnarray}
\begin{split}
 \left(\begin{array}{@{}c@{}}
E_{x}^{{\rm (out)}}\\[3pt]
E_{y}^{{\rm (out)}}\\
\end{array}\right)&=
 \left(\begin{array}{@{}c@{\quad}c@{}}
 +\cos (\theta) & -\!\sin (\theta) \\[3pt]
 +\sin (\theta) & +\cos (\theta)\\
\end{array}\right)
 \left(\begin{array}{@{}c@{\quad}c@{}}
1& 0\\[3pt]
0 & -1\\
\end{array}\right)
 \left(\begin{array}{@{}c@{\quad}c@{}}
+\cos (\theta) & +\sin (\theta) \\[3pt]
-\!\sin (\theta) & +\cos (\theta)\\
\end{array}\right)
 \left(\begin{array}{@{}c@{}}
E_{x}^{{\rm (in)}}\\[3pt]
E_{y}^{{\rm (in)}}\\
\end{array}\right)\\
&=
 \left(\begin{array}{@{}c@{\quad}c@{}}
+\cos (2\theta) & +\sin (2\theta) \\[3pt]
-\!\sin (2\theta) & -\!\cos (2\theta)\\
\end{array}\right)
 \left(\begin{array}{@{}c@{}}
E_{x}^{{\rm (in)}}\\[3pt]
E_{y}^{{\rm (in)}}\\
\end{array}\right)\!,
\end{split}
 \end{eqnarray}
so that in terms of the Stokes parameters, expressed below as the matrix-valued
expectation value $\langle {\cal E}_i < {\cal E}_j^*\rangle,$ so that
\begin{eqnarray}
\begin{split}
& \left(\begin{array}{@{}c@{\quad}c@{}}
I+Q & U+iV\\[4pt]
U-iV & I-Q\\
\end{array}\right)_{{\rm det}}\\
&\qquad=
 \left(\begin{array}{@{}c@{\quad}c@{}}
\cos 2\Omega t & \sin 2\Omega t \\
\sin 2\Omega t & -\!\cos 2\Omega t\\
\end{array}\right)
 \left(\begin{array}{@{}c@{\quad}c@{}}
I+Q & U+iV\\[4pt]
U-iV & I-Q\\
\end{array}\right)_{{\rm sky}}
 \left(\begin{array}{@{}c@{\quad}c@{}}
\cos 2\Omega t & \sin 2\Omega t \\
\sin 2\Omega t & -\!\cos 2\Omega t\\
\end{array}\right)\\
&\qquad=
 \left(\begin{array}{@{}c@{\quad}c@{}}
I +\cos (4\Omega t)Q +\sin (4\Omega t)U & -\!\cos (4\Omega t)U +\sin (4\Omega t)Q-iV \\[4pt]
-\!\cos (4\Omega t)U +\sin (4\Omega t)Q+iV & I -\cos (4\Omega t)Q -\sin (4\Omega
t)U\\
\end{array}\right)\!.\\
\end{split}
\end{eqnarray}
By measuring only the component that varies with an angular frequency $4\Omega,$
we can measure the polarization without making two independent measurements of
large numbers that are then subtracted from each other. The much larger Stokes
intensity $I$ does not mix because it remains constant in time.

To see how polarization modulation helps, let us suppose that the vector
$\mathbf{P}_{{\rm det}}^{{\rm (ideal)}}= ( I_{{\rm det}}^{{\rm (ideal)}}, Q_{{\rm
det}}^{{\rm (ideal)}}, U_{{\rm det}}^{{\rm (ideal)}}, V_{{\rm det}}^{{\rm
(ideal)}} ) $ is replaced by the vector $\mathbf{P}_{{\rm det}}^{({\rm actual})},$
with $\mathbf{P}_{{\rm det}}^{({\rm actual})}= ( I_{{\rm det}}^{({\rm actual})},$
$Q_{{\rm det}}^{({\rm actual})}, U_{{\rm det}}^{({\rm actual})}, V_{{\rm
det}}^{({\rm actual})} ) = ( \mathbf{I}+ \boldsymbol{\epsilon }) \mathbf{P}_{{\rm
det}}^{{\rm (ideal)}}$, where $\mathbf{I}$ is the identity matrix and
$\boldsymbol{\epsilon }$ represents a hopefully small but unknown uncorrected
residual error in the actual linear response of the detector pair.

\begin{figure}[t]
\begin{center}
\includegraphics[width=8cm]{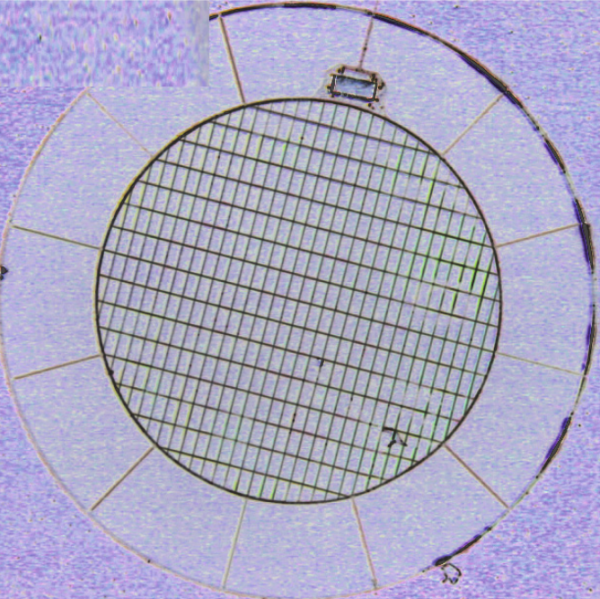}
\end{center}
 \caption{{\bf Planck polarization sensitive bolometers.} Shown above is a
photograph of one of the Planck polarization sensitive bolometers consisting of two grids
of parallel wires oriented orthogonally and placed on top of each other but separated by
a small distance. Each grid has its own thermistor, allowing the two selected components
of the linear polarizations to be read out separately. 
({\it Credit: ESA/Planck Collaboration})
\label{PlanckPSB:Fig}}
\end{figure}

We discuss the setup for measuring polarization used in Planck as
described above (see Fig.~\ref{PlanckPSB:Fig}). We describe some
of the possible systematic effects and how polarization modulation
by means of a rotating half-wave plate would remove these
systematic effects. Ideally the two orthogonal grids of wires
would correspond, but for the direction of the linear
polarization, to exactly the same circular beam on the sky and be
perfectly calibrated, with no offset between the beam centers, nor
any time-dependent drift of the differential gain. But in practice
the beams are not identical and the gain and offset of the
electronics drift in time. The centers of the beams do not
coincide, and without corrections this would cause a local
gradient of the $T$ to be mistaken for $E$ or $B$ polarization.
Differential ellipticity would cause the second derivative of $T$
to leak into $E$ and $B.$ Moreover, the actual beams are more
complicated and require three functions [i.e., maps of $I(\hat
\Omega ),$ $Q(\hat \Omega ),$ and $U(\hat \Omega )$] for a
complete description. All the above effects disappear, at least
ideally, with the polarization modulation scheme described above.
Polarization modulation can resolve or mitigate a lot of problems
but is not a panacea. For example, polarization modulation cannot
prevent $E\to B$ leakages. If there are any errors in the
calibration of the angles of the direction of the linear
polarization in the sky, these errors act to rotate $E$ into $B$
and vice versa. Moreover polarization modulation has been
difficult to realize in practice for a number of reasons such as
spurious signals due to microphonic coupling. Both continuously
rotating and discretely stepped half-wave plates are options for
polarization modulation.

\clearpage
\newpage 

\section[CMB statistics revisited: dealing with realistic observations]%
{CMB Statistics Revisited: Dealing with Realistic Observations}

The analysis of a CMB experiment going from the raw time stream of
data taken, to sky maps at a given frequency, then to a clean sky
map containing only the CMB, and finally to a likelihood function,
whose input is a complete theoretical power spectrum, involves
many steps and details. Space does not permit a complete
discussion. The following Planck papers and references therein
describe this analysis for the Planck experiment from the Planck
2013 Results series. See in particular the following papers:
Overview of products and results \cite{planck-mission-paper}, low
frequency instrument data processing
\cite{planck-lfi-data-processing}, high frequency instrument data
processing \cite{planck-hfi-data-processing}, component separation
\cite{planck-component-separation}, and HFI energetic particle
effects \cite{planck-hfi-cosmic-rays}. Similar descriptions can be
found for other experiments. In this Section, we limit ourselves
to sketching a few issues and to describing an idealized
statistical analysis.

One of the main steps in CMB analysis is constructing a likelihood
for the data given a particular theoretical model. Relative
likelihood lies at the heart of almost all statistical inference,
whether one uses frequentist methods, Bayesian methods, or an
ecumenical approach borrowing from both doctrines. Here we shall
discuss only how to formulate the likelihood, as how to exploit
the likelihood is less specific to the CMB and details can be
found in standard treatises on modern statistics. (See for example
Ref.~\cite{luptonBook} for a brief overview and
Ref.~\cite{kendall} for an authoritative treatment emphasizing
the frequentist approach. For Bayesian sampling as applied to the
CMB, see Refs.~\cite{cosmomc} and \cite{DunkleyMCMC} and
references therein.)

Constructing a likelihood is straightforward for an idealized survey for which the noise
in the sky maps is Gaussian and isotropic. For the simplest likelihood function, the
input argument is a model for the CMB sky signal, or more precisely the parameters
defining this model. This theoretical model defines an isotropic Gaussian stochastic
process whose parameters one is trying to infer. In order to introduce some of the
complications that arise, we start by writing down the likelihood ${\cal L}$ for the data
given the model, or rather the variable $-2\ln [{\cal L}],$ which in some ways resembles
$\chi ^2$ in the Gaussian approximation. We have
\begin{eqnarray}
-\!2\ln ({\cal L})=\sum _{\ell =2}^{\infty } (2\ell +1) \left\{ \ln\!\left( \frac{ B_\ell
C^{{\rm TT},(th)}_\ell +N_\ell } { C^{{\rm TT,(sky)}}_\ell +N_\ell } \right) +
\left(\frac{ C^{{\rm TT,(sky)}}_\ell +N_\ell } { B_\ell C^{{\rm
TT},(th)}_\ell +N_\ell } \right) -1 \right\}\!.
\label{IsotropicGaussianTemperatureLikelihood}
 \end{eqnarray}
Here the parameters $N_\ell $ represent the noise of the measurement. In the
approximation of white instrument noise--- that is, with no unequal time
correlations---we have $N_\ell =N_0.$ The factor $B_\ell $ represents the
smearing of the sky signal due to the finite beam width. For the beam profile
approximated as having a Gaussian shape, $B_\ell =\exp [-\sigma ^2\ell ^2]$ where
the sky intensity map is convolved with the Gaussian function $K(\theta
)=(1/\sqrt{2\pi \sigma ^2})\exp [-\theta ^2/(2\sigma ^2)].$\footnote{Note that in
the literature, beam widths are generally specified in terms of FWHM rather than
using the Gaussian definition above.} Note that we ignore constant offsets in the
log likelihood because for most statistical analysis only differences in the log
likelihood are relevant. The presence of the normalization of the Gaussian imply
that a quadratic representation of the log likelihood is not exact and higher
order corrections are needed to avoid bias, especially at low $\ell.$

To generalize the above expression to include polarization, we define the matrix
\begin{eqnarray}
{\bm{\mathcal{C}}}_\ell =
\left(\begin{array}{@{}c@{\quad}c@{}}
C^{{\rm TT}}_\ell +{B_\ell }^{-1}N_\ell^{{\rm TT}} & C^{{\rm TE}}_\ell \\[4pt]
C^{\rm TE}_\ell & C^{{\rm EE}}_\ell +{B_\ell }^{-1}N_\ell^{{\rm EE}}\\
\end{array}\right)\!,
\end{eqnarray}
so that
\begin{eqnarray}
-\!2\ln ({\cal L})=
\sum _{\ell =2}^\infty (2\ell +1)\{\ln [ {\bm{\mathcal{C}}}_\ell ^{(th)}( 
{{\bm{\mathcal{C}}}_\ell ^{{\rm (obs)}})}^{-1}] +\textrm{tr} [
{\bm{\mathcal{C}}}_\ell ^{{\rm (obs)}}({\bm{\mathcal{C}}}_\ell ^{(th)}) ^{-1}]-2\}. 
\label{IsotropicGaussianTP}
\end{eqnarray}

In order to generalize to less idealized (i.e., more realistic) situations, it is useful
to rewrite the above results in a more abstract matrix notation, in which the expressions
can be interpreted simultaneously as expressions in real (i.e., angular) space and as
expressions in harmonic space.
\begin{eqnarray}
-\!2\ln [{\cal L}]= {\rm
det}(\mathbf{C}_{th}+\mathbf{N})+\mathbf{t}^T(\mathbf{C}_{th}+\mathbf{N})^{-1}\mathbf{t}.
\label{GeneralGaussianLikelihood}
\end{eqnarray}
Here the vector $\mathbf{t}$ represents the observed sky map,
$\mathbf{C}_{th}$ is the covariance matrix of the underlying sky
signal (in general depending on a number of parameters whose
values one is trying to infer), and $\mathbf{N}$ represents the
covariance matrix of the instrument (and other) noise. In the
idealized case where everything is isotropic, this is simply a
rewriting of Eq.~(\ref{IsotropicGaussianTemperatureLikelihood}),
but in less idealized cases the likelihood in
Eq.~(\ref{GeneralGaussianLikelihood}) is correct whereas
Eqs.~(\ref{IsotropicGaussianTemperatureLikelihood}) and
(\ref{IsotropicGaussianTP}) are not applicable to the more general
case.

Let us now consider the effect of partial sky coverage, which may result from a survey
that does not cover the whole sky or from masking portions of the sky such as the
galaxy and bright point sources where contamination of the primordial
signal cannot be corrected for in a reliable way. If the vectors and matrices in
Eq.~(\ref{GeneralGaussianLikelihood}) are understood as representations in pixel space,
at least formally there is no difficulty in evaluating this expression in the subspace of
pixel space representing the truncated sky. If the dimensionality of the pixelized maps
were small, there would be no difficulty in simply evaluating this expression by brute
force. Another complication arises from inhomogeneous noise. Most surveys do not cover
the sky uniformly because requirements such as avoiding the sun, the earth, and the
planets as well as instrumental considerations lead to nonuniform sky coverage. This
leads to a noise matrix that has a simple representation in pixel space (under the
assumption of white noise), but not in harmonic space.

Unfortunately evaluating likelihoods such as in
Eq.~(\ref{GeneralGaussianLikelihood}) exactly is not possible for
full-sky surveys at high angular resolution such as the Planck
survey for which pixelized maps include up to about $N_{{\rm
pix}}=5\times 10^7$ pixels. While $\mathbf{C}_{th}$ is simple
(i.e., sparse) in harmonic space, it is dense in pixel space, and
the opposite holds for $\mathbf{N}$ (ignoring correlated noise),
so no representation can be found to simplify the calculation in
which all the matrices are sparse. Since even enumerating the
matrices involves $O({N_{{\rm pix}}}^2)$ operations, and inverting
these matrices or taking their determinant requires $O({N_{{\rm
pix}}}^3)$ operations, it is immediately apparent that brute force
will not work, and more clever, approximate techniques are
required. Pixel based likelihoods are however used at low-$\ell,$
where other approximations have difficulty representing the
likelihood, in large part because the cosmic variance is
non-Gaussian.

While the likelihoods above assumed an isotropic Gaussian stochastic process for
the underlying theoretical model generating the anisotropy pattern in the sky,
these likelihoods can be generalized to include nonlinear effects, such as
non-Gaussianity through $f_{\it NL},$ gravitational lensing, and certain models of
weak statistical anisotropy, to name just a few examples. Using the likelihood as
a starting point, possibly subject to some approximations, has the virtue that one
is guaranteed that an optimal statistical analysis will result without any special
ingenuity.

In the above simplified discussion, we have not touched on a number of very important
issues, of which we list just a few examples: asymmetric beams, far side-lobe
corrections, bandpass mismatch, errors from component separation, estimating noise from
the data, and null tests (also known as ``jackknives''). The reader is referred to the
references at the beginning of this Section for a more complete discussion and references
to relevant papers.

\section[Galactic synchrotron emission]{Galactic Synchrotron Emission}
\label{GalacticSynchrotronEmission}

An inspection of the single frequency temperature maps from the Planck space mission (see
Fig.~\ref{Planck-Monofrequency}), shown here in galactic coordinates so that the equator
corresponds to the galactic plane, suggests that the 70\,GHz frequency map is the
cleanest, as the excess emission around the galactic plane is the narrowest at this
frequency. But at lower frequencies, the region where the galactic contamination
dominates over the primordial CMB temperature anisotropies widens. This contamination at
low frequencies is primarily due to galactic synchrotron emission, resulting from
ultrarelativistic electrons spiralling in the galactic magnetic field, whose strength is
of order a few microgauss. (See for example Ref.~\cite{HanEtAl}.) This emission is
generally described as nonthermal because the energy spectrum of charged particles is
non-Maxwellian, especially at high energies. The electrons in question reside in the high
energy tail of the electron energy spectrum, described empirically as a falling power law
spectrum. Models of cosmic ray acceleration, for example arising about a shock from an
expanding supernova remnant, are able to explain spectra having such a power law form
\cite{BlandfordEichler}. If the radiating charged particles were nonrelativistic, their
radiation would be emitted at the cyclotron frequency $\omega _{c}=eB/mc,$ with very
little emission in higher harmonics, and given the measured values of the galactic
magnetic field, one could not thus explain the observed high frequency emission. However,
if the charged particles are highly relativistic, due to beaming effects the bulk of the
radiation is emitted in the very high order harmonics, allowing the observed emission to
be explained for reasonable values of the galactic magnetic field \cite{ginzburgReview}.
Although the data concerning the three-dimensional structure of the galactic magnetic
field is at present in a very rudimentary state, and little is known observationally
about the cosmic ray spectrum outside our immediate solar neighborhood, the physical
mechanism underlying galactic synchrotron emission is well understood physics. One
important point is that no matter what the details of the cosmic ray energy spectrum may
be, the kernel of the integral transform relating this distribution to the frequency
spectrum of the observed synchrotron emission applies considerable smoothing. Thus we can
assume with a high degree of confidence that the synchrotron spectrum is smooth in
frequency. Theory also predicts that this synchrotron emission will be highly polarized,
at least to the extent that the component of the galactic magnetic field coherent over
large scales is not negligible. This expectation is borne out by observations. It is
customary to describe the synchrotron spectrum using a power law, so that its R-J
brightness temperature may be fit by an Ansatz of the form
\begin{eqnarray}
T_{R-J}(\hat \Omega, \nu )=T_{R-J}(\hat \Omega, \nu _0)\left( \frac{\nu }{\nu
_0}\right)^\alpha,
 \end{eqnarray}
where empirically it has been established that $\alpha \approx 2.7$--3.1. The same holds
for the polarization---that is, the Stokes $Q$ and $U$ (and also $V$) parameters. In
the ultrarelativistic approximation this emission is primarily linearly polarized, but
there is also a smaller circularly polarized component suppressed by a factor of
${1}/{\gamma}.$ Variations in the spectral index depending on position in the sky have
been observed \cite{Fuskeland}.

A description of the WMAP full-sky observations of synchrotron emission appears in the
first-year WMAP foregrounds paper \cite{wmap2003d} and the polarization observations are
described in the three-year WMAP paper on foreground polarization \cite{wmap2007a}.
A~greater lever arm, in particular for studying the synchrotron spectral index, can be
obtained by including the 408\,MHz Haslam map covering almost the entire sky
\cite{Haslam}.

\section[Free-free emission]{Free-Free Emission}

Another source of low-frequency galactic contamination is free-free emission
arising from Bremsstrahlung photons emitted from electron-electron collisions in
the interstellar medium, and to a lesser extent from electron-ion collisions. Like
galactic synchrotron emission, the free-free emission brightness temperature
falls with increasing frequency, but the fall-off is slower than that of the
galactic synchrotron emission.

Unlike synchrotron emission, free-free emission is not polarized. H$\alpha $ ($n=3\to
n=2$) emission resulting from recombination of ionized atomic hydrogen can be used as a
template for removing the free-free component, because the H$\alpha $ emission likewise
is proportional to the square of the electron density. However, because the $\Gamma
_{H\alpha }(T)$ and $\Gamma _{{\rm free}\text{--}{\rm free}}(T,\nu )$ do not have an
identical temperature dependence, free-free removal using an H$\alpha$ template has some
intrinsic error. $\Gamma _{H\alpha}(T)$ is defined as the constant of proportionality in
the emissivity relation: $\epsilon _{H\alpha}= \Gamma _{H\alpha}(T) {n_e}^2$ where
$\epsilon _{H\alpha}$ is a bolometric emissivity (power per unit volume) and $n_e$ is the
density of free electrons. Likewise $\epsilon _{\operatorname{free--free}}(\nu )= \Gamma
_{\operatorname{free--free}}(T,\nu ) {n_e}^2$ where the emissivity has units of power per
unit frequency per unit volume per unit solid angle.

\section[Thermal dust emission]{Thermal Dust Emission}
\label{ThermalDust}

In Sec.~\ref{GalacticSynchrotronEmission} we discussed galactic synchrotron
emission, whose R-J brightness temperature rises with decreasing frequency,
making it exceedingly difficult to measure CMB anisotropies at frequencies below
$\approx 20$\,GHz, especially close to the galactic plane where the galactic
synchrotron emission is most intense. We also saw how the fact that the brightness
temperature increases with decreasing frequency can be understood as a consequence
of the synchrotron optical depth being much smaller than unity and increasing with
decreasing frequency. Eventually when the frequency is low enough so that the
optical depth is near one, the brightness temperature stops rising and approaches
a temperature related to the cosmic ray electron kinetic energy.

This Section considers another component contributing to the microwave sky: the thermal
emission from interstellar dust, whose brightness temperature increases in the opposite
way---that is, the brightness temperature increases with increasing frequency. This
behavior results because at microwave frequencies, the wavelength greatly exceeds the
typical size of a dust grain. At long wavelengths, the cross-section for absorbing or
elastically scattering electromagnetic radiation is much smaller than the geometric
cross-section $\sigma _{{\rm geom}}\approx a^2$ where $a$ is the effective grain radius.
The dust grains can be approximated as dipole radiators in this
frequency range because higher order multipoles are irrelevant. The dependence of the
cross-section on frequency can be qualitatively understood by modeling the electric
dipole moment of the grain as a damped harmonic oscillator where the resonant frequency
is much higher than the frequencies of interest. In this low frequency
approximation
\begin{eqnarray}
\sigma_{\rm abs}\sim \left( \frac{a}{\lambda }\right)^{\!2} a^2
 \end{eqnarray}
for absorption, and for the elastic, or Rayleigh, scattering component, which
becomes subdominant at very low frequencies,
\begin{eqnarray}
\sigma _{{\rm elastic}}\sim \left( \frac{a}{\lambda }\right)^{\!4} a^2.
 \end{eqnarray}
This is a simplistic approximation that we later shall see is not accurate, but it
does provide intuition about the qualitative behavior expected.

Our knowledge of interstellar dust derives from combining different observations
spanning a broad range of frequencies, from the radio to the UV and even X-ray
bands, in order to put together a consistent theoretical model able to account
simultaneously for all the observations. (For a nice recent overview, see for
example Refs.~\cite{draine2003} and \cite{Draine2004}.) In this Section, we
restrict ourselves to discussing the thermal microwave emission properties of the
dust.

An empirical law commonly used to model thermal dust emission at low frequencies
is a Planck blackbody spectrum modulated by a power law emissivity, or graybody
factor
\begin{eqnarray}
I_{{\rm dust}}(\nu ) = I_{{\rm dust}}(\nu_0) \left(\frac{\nu}{\nu
_0} \right)^\beta \frac{B(\nu , T_{{\rm dust}})}{B(\nu _0, T_{{\rm
dust} })}
\end{eqnarray}
where the Planck spectrum is given by
\begin{eqnarray}
B(\nu ;T)=\frac{2h\nu ^3}{c^2}\frac{1}{\left(\exp\!\left(\frac{h\nu}{k_{\rm
B}T}\right)-1\right)}.
\end{eqnarray}

The physical basis for this modification to the Planck spectrum
can be partially motivated by the following argument based on the
linear electric susceptibility of the dust grain, following a line
of reasoning first suggested by EM Purcell in 1969
\cite{purcell1969}. The simple argument, subject to a few caveats,
suggests that $\beta =2.$ However the observational data do not
bear out this prediction and instead are better fit by a power law
with $\beta \approx 1.4-1.6$ \cite{paradis,planckDust}. Let $\chi
(\omega )$ represent the linear susceptibility of a dust grain,
which as a consequence of the Kramers-Kronig dispersion relations
must include both nonvanishing real and imaginary parts. Here
$\mathbf{d}(\omega )=\chi (\omega ) \mathbf{E}(\omega ),$ where
$\mathbf{d}(\omega )$ and $\mathbf{E}(\omega )$ are the grain
electric dipole moment and the surrounding electric field,
respectively. As a consequence of causality, $\chi (\omega )$ is
analytic on the lower half-plane. Its poles in the upper
half-plane represent those decaying mode excitations of the grain
that are coupled to the electromagnetic field. Moreover, on the
real axis $\chi (-\omega )=[\chi (+\omega )]^*.$ If we assume
analyticity in a neighborhood of the origin, we may expand as a
power series about the origin, so that
\begin{eqnarray}
\chi (\omega )=\omega _0+i\chi _1\omega +\chi _2\omega ^2+\cdots,
\end{eqnarray}
\noindent and $\chi _0,$ $\chi _1,$ $\chi _2, \ldots $ are all real. It follows that the energy
dissipated is characterized by the absorptive cross-section
\begin{eqnarray}
\sigma _{{\rm abs}}(\omega )=\frac{4\pi \omega \textrm{Im}[\chi (\omega )]}{c},
 \end{eqnarray}
and at low frequencies
\begin{eqnarray}
\sigma_{\rm abs}(\omega )\approx \left(\frac{4\pi \chi ^1}{c}\right)\omega ^2,
 \end{eqnarray}
which corresponds to an emissivity index $\beta =2$ as $\omega \to 0.$ The perhaps
questionable assumption here is that there exists a circle of nonzero radius about
the origin of the $\omega $-plane where $\chi (\omega )$ has no singularities. If
we postulate that there is an infinite number of poles of suitably decreasing
strength whose accumulation point is $\omega =0,$ we may evade the conclusion that
asymptotically as $\omega \to 0,$ $\beta \to 2.$ For an insightful discussion of
how such low-frequency poles could arise, see C. Meny {\it et~al.} \cite{meny}.

\section[Dust polarization and grain alignment]{Dust Polarization and Grain Alignment}

The fact that interstellar dust grains are aligned was first seen through the
polarization of starlight at optical frequencies in 1949
\cite{hall,hiltner1949a,hiltner1949b}. Since this discovery an abundant literature
has been amassed elucidating numerous aspects of dust grain dynamics. The best bet
solution to the alignment problem has evolved over the years as additional
relevant physical effects were pointed out. It is unclear whether the current
understanding will remain the last word on this subject. The most naive solution
to the alignment problem would be some sort of compass needle type mechanism with
some dissipation. However, it was discovered early on that in view of the
plausible magnitudes for the galactic magnetic field, such an explanation cannot
work \cite{SpitzerTukey}. The theoretical explanation that has evolved involves
detailed modeling of the rotational dynamics of the grains, which as we shall see
below almost certainly rotate suprathermally.

For microwave observations of the primordial polarization of the CMB at ever greater sensitivity, the
removal of contamination from polarized thermal dust emission is of the utmost importance. This fact is
highlighted by the discussion of how to interpret the recent BICEP2 claim (discussed in
Sec.~\ref{SubortialBModeSearches}) of a detection of a primordial B mode signal based on essentially a
single frequency channel. This interpretation however seems unlikely given subsequent Planck estimates of
the likely dust contribution \cite{planckXXX}. The problem is that little data is
presently available on polarized dust emission at microwave frequencies. Current models of polarized dust
involve a large degree of extrapolation and are in large part guided by simplicity. From a theoretical
point of view, polarized dust could be much more complicated than the simplest empirical modeling
suggests. For example there is no {\it a priori} reason to believe that polarized emission from dust must
follow the same frequency spectrum as its emission in intensity. It could well be that the weighting over
grain types gives a different frequency dependence, or that as more precise dust modeling is called for,
the need to include several species of dust with spatially nonuniform distributions will become apparent.
Moreover, based on theoretical considerations, it is not clear that there should be a simple
proportionality between polarization and the strength of the magnetic field. If the BICEP2 B mode signal
is confirmed as a dust artefact, we could see ourselves in the coming years trying to detect values of
the tensor-to-scalar ratio around $r\approx 10^{-3},$ which would require exceedingly accurate polarized
dust removal from the individual frequency maps, and before this happens a wealth of new observational
data on polarized dust emission will become available. If, on the other hand, a detection of $B$ modes at
a large value of $r$ is made, as one tries to map out the precise primordial B mode spectrum, for example
to measure $n_T$ and thus check the consistency between ${\cal P}_S(k)$ and ${\cal P}_T(k),$ the need for
accurate dust removal will likewise be great, but for slightly different reasons.

Since the observational inputs are likely to change, here we
particularly emphasize the underlying physics of the alignment,
which may be updated, but is likely to change less. The
theoretical issues of dust alignment have so far not been widely
discussed among those not specialized in the ISM (interstellar
medium). Nevertheless, we believe that the questions raised will
become increasingly relevant for CMB observations in the future,
and this is why a thorough discussion is presented here.

We shall find that the alignment of interstellar dust grains is closely linked to
the rotational dynamics of a dust grain. We therefore start with a study of the
rotation of a dust grain. Mathematically, the simplest case to analyze is a
prolate (needle-like) or oblate (flattened spherical) dust grain with an axis of
azimuthal symmetry, which constitutes a symmetric top. In this case, the motions
can be solved in terms of elementary functions. For the more general case of an
asymmetric grain (where none of the three principal values of the moment of
inertia tensor are equal) integration of the equations of motion in the absence of
torques is more difficult, but can be achieved using Jacobian elliptic functions
\cite{Whittaker}.

In the noninertial frame rotating with the dust grain, which is convenient because in
this frame the moment of inertia tensor is constant in time, two types of inertial forces
arise: (i) a centrifugal force, described by the potential $V_{ctr}(\mathbf{x})/m=-\Omega
^2(t)\mathbf{x}^2 + (\boldsymbol{\Omega }(t)\cdot \mathbf{x})^2,$ and (ii) a Coriolis
force $\mathbf{F}_{{\rm cor}}/m=-2\boldsymbol{\Omega }(t)\times \mathbf{v},$ which for
our purposes is unimportant because the velocities are negligible in the co-rotating
frame. In order to understand the alignment process, we must model the solid as being
elastic with dissipation included.

\subsection{Why do dust grains spin?}

A flippant answer to the above question might be: Why not? If we can associate a
temperature with the three kinetic degrees of freedom of the dust grain, we would
conclude that
\begin{eqnarray}
\frac{1}{2}I\omega ^2\sim \frac{3}{2}k_{\rm B}T_{{\rm rot}}.
 \end{eqnarray}
Early discussions of rotating dust grains supposed that the rotational degrees of
freedom are subject to random torques, so that
\begin{eqnarray}
\langle {\bm \Gamma } \rangle =0.
\end{eqnarray}
Later it was realized that suprathermal rotation is plausible,
where the implied temperature of the rotation can greatly exceed
the effective temperature of any of the other relevant degrees of
freedom.

The question now arises what temperature one should use for $T_{{\rm rot}}.$ A
natural candidate might be the kinetic temperature of the ambient gas $T_{{\rm
kin}}.$ Indeed one expects molecules of the ambient gas to collide with the dust
grains, and such collisions do provide random torques, causing the angular
momentum to undergo a random walk accompanied by a dissipation mechanism, as must
be present to prevent the angular velocity from diverging with time.

\subsection{About which axis do dust grains spin?}

This question has two parts. First, we may ask what is the alignment of the
co-rotating internal (body) coordinates with respect to the angular momentum
vector. Second, we may ask how the angular momentum aligns itself with respect to
the inertial coordinates of the ambient space. In this Section, we consider only
the first question, postponing the second question to a later section.

But for exceptional situations where two or more eigenvalues of the moment of
inertia tensor coincide, the principal axes of moment of inertia tensor provide a
natural set of axes for the dust grain, and we may order the eigenvalues so that
$I_1<I_2<I_3.$ We argue that the dust grain tends to align itself so that its
$I_3$ axis is parallel to the angular momentum. This orientation of the grain
minimizes the rotational energy subject to the constraint that the total angular
momentum remains constant. If the grain starts in a state of random orientation,
the grain wobbles. While $\mathbf{L}$ remains constant,
$\boldsymbol{\Omega}(t)=\mathbf{I}^{-1}(t) \mathbf{L}$ (expressed in inertial
coordinates) does not, but rather fluctuates, exciting internal vibrational modes
that dissipate. As the excess rotational energy is dissipated, an alignment as
described above is achieved.\footnote{Much to the dismay of NASA engineers, this
is exactly what happened to the Explorer 1 rocket, the first US satellite launched
in 1958 following Sputnik, due primarily to dissipation from its whisker antennas.
The elongated streamlined cylinder was initially set in rotation about its long
axis, but subsequently, after an initial wobble phase, ended up spinning about an
axis at right angles in the body coordinates. Since then rotational stability,
also relevant in ballistics, has become a carefully considered design issue for
spacecraft.} The relevant question is how this time scale compares to other time
scales of the grain rotation dynamics. It turns out that this time scale, which we
denote $\tau _{I-L}$, is short. In the equations that follow, we exploit the
shortness of $\tau _{I-L}$ compared to the other relevant time scales. This
hierarchy of time scales allows us to simplify the equations by excluding the
degrees of freedom that describe the lack of alignment of the principal axis of
maximum moment of inertia from $\mathbf{L},$ leading to a considerable
simplification of the formalism.

\subsection{A stochastic differential equation for $\mathbf{L}(t)$}

Mathematically, we may describe the combined effects of the torque from random
collisions and its associated damping through the coherent torque using a
stochastic differential equation of the form
\begin{eqnarray}
\dot {\mathbf{L}}(t) =-\alpha \left({\mathbf{L}}(t) - \bar L \frac
{\mathbf{L}(t)}{| \mathbf{L}(t)|}\right) +\boldsymbol{\Gamma }_{{\rm ran}}(t).
\label{LangevinOne}
 \end{eqnarray}
In this equation, the torque described by the first term is coherent with respect
to the body coordinates of the dust grain. Later we shall also consider torques
that are coherent with respect to the ambient inertial coordinates, which must be
modeled in a different way. Here the stochastic Gaussian forcing term has the
expectation values
\begin{eqnarray}
\begin{split}
\langle\Gamma _{{\rm ran},i}(t) \rangle &=0, \\[4pt]
\langle \Gamma _{{\rm ran},i}(t) \Gamma _{{\rm ran},j}(t') \rangle &= \mu _L \delta _{ij} \delta (t-t'),
\end{split}
 \end{eqnarray}
where the angular momentum diffusion coefficient has units $[\mu _L]=(\textrm{Torque})^2/(\textrm{Time}).$

Let us analyze the stationary ensemble defined by the above
equation---in other
words, the probability function $p(\mathbf{L})$ toward which the biased random
walk defined by the above equation evolves after a long period of time. In the
low-temperature limit, $| \mathbf{L}| \approx \bar L,$ with negligible fractional
fluctuations, and the direction of the angular momentum vector $\hat
{\mathbf{n}}_\mathbf{L}={\mathbf{L}(t)}/{| \mathbf{L}(t)| }$ follows a random walk
on the surface of the sphere with diffusion coefficient $\mu _{\hat n}=\mu _L/\bar
L^2.$ In the opposite high temperature limit, the $\bar L$ term is a sideshow or
small correction, and the main effect is the competition between the first and
second terms leading to a thermal ensemble, which for an isotropic moment of
inertial tensor would take the form
\begin{eqnarray}
p(\mathbf{L})= \left( \frac{\beta }{2\pi } \right) ^{3/2} \exp[ -\beta
\mathbf{L}^2], \label{thermEnsemble}
 \end{eqnarray}
where $\beta =\alpha /\mu _L.$ Without other sources of dissipation or random
torques, we expect that $\beta =1/(2k_{\rm B}T_{{\rm kin}}I_{\max})$ where
$I_{\max}$ is the largest eigenvalue of the moment of inertia tensor of the
particular grain being modelled and $T_{{\rm kin}}$ is the kinetic temperature of
the ambient gas. It is straightforward to generalize Eq.~(\ref{thermEnsemble}) to
the intermediate case. It was however later noted that a number of mechanisms
exist for which $\langle \boldsymbol{\Gamma } \rangle \ne 0,$ which can give rise
to what is known as ``suprathermal rotation'' \cite{purcellSupra}---that is,
rotation at an angular velocity larger than any temperature associated with the
other degrees of freedom interacting with the dust grain.

\begin{figure}
\begin{center}
\includegraphics[width=7cm]{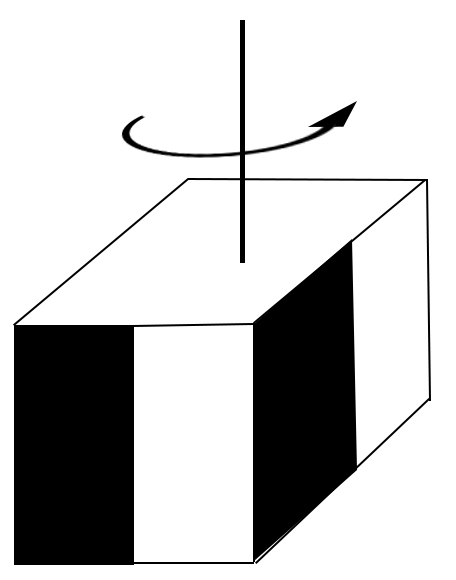}
\end{center}
\caption{{\bf Coherent torques from grain radiation pressure.} \label{GrainRadiometer}}
\end{figure}

\subsection{Suprathermal rotation}
\label{SupraThermalSection}

To see how such {\it suprathermal rotation} might come about,
consider the caricature of a dust grain as sketched in
Fig.~\ref{GrainRadiometer}. Let us assume that the dust grain
emits radiation isotropically in the infrared, but that in the
visible and UV bands radiation is absorbed or reflected by the
surfaces indicated in the figure. Let us further assume an
isotropic visible or UV illumination. The blackened (perfectly
absorbing) faces suffer a radiation pressure equal to half that of
the mirrored surfaces, giving a coherent torque in the direction
indicated in the figure. The presence of two distinct
temperatures, reminiscent of a thermal engine, is crucial to avoid
thermal equilibrium for the rotational degree of freedom. As in
the thermodynamics of an engine, there is a heat flow from a
hotter temperature to a colder temperature with some of the energy
being siphoned off in the form of work exerted on the three
rotational degrees of freedom of the dust grain. Other
possibilities for coherent torques result from photoelectron
emission at privileged sites and the catalysis of molecular
hydrogen on favored sites on the grain surface. In the former
case, the incident UV photon is almost pure energy due to its
masslessness, but the emitted electron carries substantial
momentum for which there is a recoil. Molecular hydrogen forms
almost exclusively through catalysis on a grain surface. Unlike in
ordinary chemistry where truly two-body collisions are an
exception and activation barriers prevent reactions from
establishing equilibrium, for molecular hydrogen formation in the
cold interstellar medium, the bottleneck is the lack of three-body
collisions and the lack of an efficient mechanism for carrying
away the excess energy, so that the hydrogen atoms combine rather
than simply bouncing off each other. In principle, the energy can
be lost radiatively, but the rate for this is too slow. Given that
4.2\,eV is liberated, the recoil at selected sites can produce a
substantial coherent torque on the grain.

\subsection{Dust grain dynamics and the galactic magnetic field}

Much of the dust grain rotational dynamics does not single out any preferred
direction. To explain the polarization of starlight and of the thermal dust
emission at microwave frequencies, some mechanism is needed that causes the dust
grains to align themselves locally at least along a common direction such as that
of the magnetic field.

\subsubsection{Origin of a magnetic moment along $\mathbf{L}$}
\label{MagMoment}
Two effects generate a magnetic moment along the axis of $\mathbf{L}$:

\medskip
\noindent (i) {\bf Inhomogeneous electric charge distribution in
the dust grain.} The second moment of the electric charge density
about the principal axis of the largest eigenvalue of the moment
of inertia tensor (denoted $z$) induces a magnetic dipole moment
as the grain rotates about this axis, according to
 \begin{eqnarray}
\boldsymbol{\mu }=\mathbf{\hat z} {\Omega _z}\int d^3r \rho _e(\mathbf{r})(x^2+y^2),
 \end{eqnarray}
where $\rho _e(\mathbf{r})$ is the electrostatic charge density of the dust grain.
In general, and particularly when the grain structure is amorphous, the net charge
density is nonvanishing, so the integral above does not vanish. This effect is
known as the {\it Rowland effect}.

\medskip
\noindent (ii) {\bf The Barnett effect.} Minimizing the free
energy tends to align the free spins of a rotating body with the
direction of rotation. In nonferromagnetic substances this effect
is usually negligible, but rotating dust grains present an extreme
environment because of their minute size and isolation from their
environment. Interstellar dust grains are among the fastest
rotating macroscopic objects known.

To understand the origin of the Barnett effect, consider an isolated rotating
body, whose conserved total angular momentum may be decomposed into two parts
 \begin{eqnarray}
\mathbf{L}_{{\rm tot}}=\mathbf{L}_{{\rm bulk}}+\mathbf{S}_{{\rm free}}.
 \end{eqnarray}
We consider the energetics of the exchange of angular moment between the bulk and
spin degrees of freedom. We assume the spins to be free neglecting any
interactions between them. In this case, the Hamiltonian for the system is
 \begin{eqnarray}
H=\frac{\mathbf{L}_{{\rm bulk}}^2}{2I}= \frac{(\mathbf{L}_{{\rm
tot}}-\mathbf{S})^2}{2I},
\end{eqnarray}
and retaining only the term linear in the spin, we obtain
 \begin{eqnarray}
H=- \boldsymbol{\Omega }\cdot \mathbf{S},
\end{eqnarray}
where $\boldsymbol{\Omega }(t)=\mathbf{I}^{-1}(t) \mathbf{L}$ is the angular
rotation velocity. The spins want to align themselves with the rotation to lower
the rotational energy \cite{landau}. So far we have not included any terms
opposing complete alignment, so at $T=0$ complete alignment would result with
$\mathbf{S}=S_{{\rm available}} {\mathbf{\Omega }}/{| \mathbf{\Omega }| }. $ But
at finite temperature entropic considerations oppose such alignment, and it is the
free energy $F=H-TS(\mathbf{S})$ or
 \begin{eqnarray}
F(\mathbf{S};\boldsymbol{\Omega })=-\boldsymbol{\Omega }\cdot
\mathbf{S}+aT\mathbf{S}^2
\end{eqnarray}
\noindent that is minimized. Far from saturation, when the degree of alignment is small, this quadratic
term provides a good approximation to the entropy of the spin system.

\subsection{Magnetic precession}\label{magPrec}
Having established that a spinning dust grain has a magnetic moment parallel to
its angular momentum vector, which is aligned with the principal axis of the
largest eigenvalue of the moment of inertia tensor, we now consider the effect of
the torque resulting from the interaction with the ambient magnetic field. The
interaction Hamiltonian is
 \begin{eqnarray}
H=-\boldsymbol{\mu }\cdot \mathbf{B},
 \end{eqnarray}
from which it follows that
 \begin{eqnarray}
\dot {\mathbf{L}} =\boldsymbol{\mu }\times \mathbf{B} =\gamma _m\mathbf{L}\times
\mathbf{B}.
 \end{eqnarray}
In other words, the angular momentum vector precesses at a frequency $\omega
_{{\rm prec}}=\gamma _m| \mathbf{B}| $ around the magnetic field. Under this
precession, the angle between $\mathbf{L}$ and $\boldsymbol{\mu }$ remains
constant, meaning that this effect neither aligns nor disaligns the grain with
respect to the local magnetic field direction.

We estimate the time scale of this precession. Let
$\boldsymbol{\mathcal {S}}$ denote the spin polarization density
(angular momentum per unit volume), which under saturation (i.e.,
all free spins aligned) would have a magnitude $\mathcal{S}_{{\rm
sat}},$ whose order of magnitude can be estimated by assuming
approximately Avogadro's number $N_A$ of spins having a magnitude
$\hbar$ in a cubic centimeter volume, giving $\mathcal{S}_{{\rm
sat}}\approx O(1) (4\times 10^{-4}) \textrm{erg} \,{\cdot}\,
\textrm{s} \,{\cdot}\, \textrm{cm}^{-3}.$ We obtain the following
free energy density (per unit volume):
\begin{eqnarray}
F_0(\mathbf{S};T)= \frac{1}{2} (k_{\rm B}T)\mathcal{N}_{{\rm spin}} \left(
\frac{\boldsymbol{\mathcal {S}}}{\mathcal {S}_{{\rm sat}}} \right) ^2.
 \end{eqnarray}
Here for a paramagnetic substance the order of magnitude for
$\mathcal{N}_{{\rm spin}}$ might be around
$10^{24}\,\textrm{cm}^{-3}$ but for a ferromagnetic substance
could be much smaller, corresponding to the density of domains.
Minimizing the free energy with respect to $\boldsymbol{\mathcal
{S}}$ gives the spin polarization density
 \begin{eqnarray}
\boldsymbol{\mathcal{S}}= \frac{\mathcal{S}_{{\rm sat}}^2} { (k_{\rm
B}T)\mathcal{N}_{{\rm spin}} } =I_S \boldsymbol{\Omega }.
\end{eqnarray}
We express the constant of proportionality as if it were a moment of inertia
because it has the same units.
 \begin{eqnarray}
\frac{I_S}{I_L}&\approx &\frac{\mathcal{N}_{{\rm free}}\hbar ^2}{k_{\rm B}T\rho
a^2}\cr
&\approx&(3\times 10^{-5}) \left( \frac{N_{{\rm free}}}{10^{24}\,{\rm cm}^{-3}} \right)
\left(\frac{100\,{\rm K}}{T} \right) \left(\frac{3g\,{\rm cm}^{-3}}{\rho} \right)
\left(\frac{10\,{\rm nm}}{a} \right) ^2.
\end{eqnarray}

\subsubsection{Barnett dissipation}

The coupling in the Barnett effect between the bulk and spin angular momenta does
not act instantaneously. Rather as the instantaneous angular velocity
$\boldsymbol{\Omega }(t)$ changes, there is a lag in the response of
$\mathbf{S}(t).$ In his landmark 1979 paper \cite {purcellSupra}, Purcell pointed
out that this lag would help align the body coordinates so that the axis of
maximum moment of inertia is parallel to the angular momentum vector. Purcell
coined the name {\it Barnett dissipation} for this effect, whose magnitude is
greater than the mechanical dissipation mentioned above.

Two types of free spins participate in the Barnett effect: the electronic spins and the
nuclear spins. The electronic spins are much more relevant for establishing a magnetic
moment along $\mathbf{L}$ as discussed above in Sec.~\ref{MagMoment} because electronic
magnetic moments are larger than nuclear magnetic moments by a factor of about
$(m_{p}/m_{e}).$ But for dissipation, this strong coupling is a liability. The nuclear
spins contribute predominantly to Barnett dissipation because of their longer lag time.

Mathematically, the coupled equations
 \begin{eqnarray}
\begin{split}
\dot {{\bf S}}(t)&= -\gamma ( \mathbf{S}(t) -\alpha
\boldsymbol{\Omega
}(t)),\\[4pt]
\dot {\mathbf{L}}(t)&= +\gamma( \mathbf{S}(t) -\alpha \boldsymbol{\Omega }(t)),
\end{split}
\end{eqnarray}
describe this effect. We may integrate out ${\bf S}(t)$ to obtain
the following equation in inertial coordinates for the torque on
$\mathbf{L}$ resulting from the exchange of angular momentum
between the bulk and spin degrees of freedom
 \begin{eqnarray}
\dot {\mathbf{L}}(t)&=& -\alpha \gamma \boldsymbol{\Omega }(t) +\alpha \gamma
^2\int _0^\infty d\tau \exp [-\gamma \tau ]\boldsymbol{\Omega }(t-\tau ).
\end{eqnarray}

To calculate how fast this exchange of angular momentum aligns the axis of maximum
moment of inertia with the angular momentum vector, we calculate the averaged
rotational energy loss rate of the bulk degrees of freedom retaining only secular
contributions. This calculation is simpler than keeping track of torques. The
degree of alignment of $\boldsymbol{\Omega }(t)$ with the axis of maximum moment
of inertia is a function of the rotational energy in the bulk rotational degrees
of freedom. The instantaneous power transferred from the bulk rotation to the spin
degrees of freedom is given by
 \begin{eqnarray}
P_{{\rm loss}}(t)&=& \boldsymbol{\Omega }(t)\cdot \dot {\mathbf{L}}(t)=
-\boldsymbol{\Omega }(t)\cdot \dot {\mathbf{S}}(t) \cr 
&=&-\gamma I_s \boldsymbol{\Omega }(t) -\gamma \left\{ \boldsymbol{\Omega }(t)- \gamma
 \int _0^\infty dt' \exp [-\gamma t'] \boldsymbol{\Omega
 }(t-t')\right\}\!,
\end{eqnarray}
and to obtain the averaged or secular contribution, we decompose
$\boldsymbol{\Omega }(t)$ into harmonic components of amplitude
$\boldsymbol{\Omega }_a$ and frequency $\omega _a,$ so that
 \begin{eqnarray}
\left\langle \boldsymbol{\Omega }(t)\cdot \left\{ \boldsymbol{\Omega }(t) -\gamma
\int _0^\infty dt' \exp [-\gamma t'] \boldsymbol{\Omega }(t-t')\right\}
\right\rangle =\frac{1}{2}\sum _a \boldsymbol{\Omega }_a^2 \frac{{\omega _a}^2}
{{\gamma }^2+{\omega _a}^2}.\qquad
\end{eqnarray}
\noindent It follows that
 \begin{eqnarray}
\langle P_{{\rm loss}}\rangle =-\frac{1}{2}I_S\sum _a \boldsymbol{\Omega }_a^2
\frac{{\omega _a}^2} {{\gamma }^2+{\omega _a}^2}.
 \end{eqnarray}
We see that the stationary component (with $\omega _a=0$) does not contribute to
the loss but the precessing components dissipate.

We obtain an order of magnitude for the alignment rate by assuming that $(\delta
I)/I\approx O(1)$ and that except for the stationary component $\omega _a\approx
\omega _{\max}=L/I_{\max},$ so that the excess energy is approximately
\begin{eqnarray}
E_{{\rm excess}}\approx O(1) I_{\max}{\omega _{\max}}^2 \sin ^2 \theta
\end{eqnarray}
where $\theta $ is the disalignment angle, and the rate of dissipation is
approximately
 \begin{eqnarray}
\bar P_{{\rm loss}} \approx O(1) \gamma I_S \omega _{\max}^2 \frac{{\omega
_{\max}}^2}{\gamma ^2+{\omega _{\max}}^2} \sin ^2 \theta,
 \end{eqnarray}
so that
 \begin{eqnarray}
\gamma _{{\rm align}}\approx \gamma \frac{I_S}{I_3} \frac{{\omega
_{\max}}^2}{\gamma ^2+{\omega _{\max}}^2}.
 \end{eqnarray}

The action of the imperfect coupling of the spins to the bulk angular momentum and
bulk rotational velocity is somewhat analogous to that of a harmonic balancer on a
rotating shaft (such as a crankshaft), a common device in mechanical engineering.
Here however the weakly and dissipatively coupled flywheel is replaced by the
internal spin degrees of freedom.

\subsection{Davis-Greenstein magnetic dissipation}

If a grain rotates with an angular velocity along the direction of the magnetic
field, in the frame co-rotating with the grain, the magnetic field is time
independent. However the component of the angular velocity in the plane normal to
the magnetic field $\boldsymbol{\Omega }_\perp $ gives rise to an oscillating
magnetic field of angular frequency $\Omega _\perp.$ In general, the magnetic
susceptibility at nonzero frequency has a nonzero imaginary part $\chi _{m,i},$
where
\begin{eqnarray}
\chi _{m}(\omega )= \chi _{m,r}(\omega )+i \chi _{m,i}(\omega ).
\end{eqnarray}
The lag in the response of the magnetization to the applied driving field gives
rise to dissipation accompanied by a torque that tends to align (or antialign) the
spin with the magnetic field. This effect is known as the Davis-Greenstein
mechanism \cite{davisGreenstein}.

The alignment tendency of the Davis-Greenstein mechanism may be described by the
pair of equations
 \begin{eqnarray}
\dot {\mathbf{L}}_\perp =-\frac{1}{\tau _{DG}}{\mathbf{L}}_\perp, \quad \dot
{\mathbf{L}}_\parallel =0,
 \end{eqnarray}
where the decay time $\tau _{DG}$ indicates the time scale of the
Davis-Greenstein relaxation process. In deriving this simple form, we have
assumed that $\tau _{{\rm prec}}\ll \tau _{DG},$ so~that we may average over many
precessions. We also assume that the alignment with the axis of maximum moment of
inertia likewise is very fast. If these assumptions do not apply, an averaged
equation of the simple form above does not hold, and these effects need to be
analyzed simultaneously.

A key issue in evaluating the viability of the Davis-Greenstein mechanism is the
frequency of reversals of the direction of the coherent torque, termed ``cross-over''
events in the literature. If we consider for example coherent torques from the formation
of molecular hydrogen at a small number of catalysis sites on the grain surface, the
orientation of this torque depends on the structure of the outermost surface atomic
monolayer, and its structure is likely to change as new layers are deposited or as old
layers are eroded away. Consequently, the long term coherence time of the ``coherent''
torque is uncertain. Under the assumption that the alignment of the body coordinates with
the angular momentum is fast, it is the evolution of the coherent torque projected onto
the axis of maximum moment of inertia that matters. While the grain rotates
suprathermally, its direction is relatively stable against random perturbing torques, but
when its rotation velocity is in the process of changing sign and of the same order as
the would-be ``thermal'' rotation velocity, the grain is particularly susceptible to
changing orientation as a result of random torques as discussed in detail in
Refs.~\cite{WW1} and~\cite{WW2}.

In deciding whether $\tau _{DG}$ is too long for producing a significant degree of
alignment, we must consider competing effects that tend to disalign the grains, which
modify the above equation to become a stochastic differential equation of the form
\begin{equation}
\mathbf{\dot L} = -\frac{1}{\tau _{DG}} \left(\mathbf{L} -\mathbf{\hat e}_B (\mathbf{\hat
e}_B \cdot \mathbf{L}) \right) -\frac{1}{\tau _{\rm rel}} \left(\mathbf{L} -\bar L
\mathbf{\hat e}_L \right) +\frac{\bar L}{\tau _{\rm rand}^{1/2}}\mathbf{n}(t)
\end{equation}
where $\mathbf{\hat e}_B=\mathbf{B}/\vert \mathbf{B}\vert ,$ $\mathbf{\hat
e}_L=\mathbf{L}/\vert \mathbf{L}\vert ,$ and $\mathbf{n}(t)$ is a normalized white noise
source whose expectation value is given by
\begin{equation}
\left< n_i(t)~ n_j(t') \right> =\delta (t-t').
\end{equation}
If $\tau _{\rm rel}\ltorder \tau _{\rm rand},$ in the absence of a magnetic field, $\mathbf{L}$
undergoes a random walk on the surface of a sphere of radius $\bar L.$ If $\tau _{DG}\ll
\tau _{\rm rand}$ the magnetic dissipation is strong enough to bring about almost complete
alignment. However if, on the other hand, $\tau _{DG}\gg \tau _{\rm rand},$ the magnetic
dissipation leads to a negligible amount of alignment.

\subsection{Alignment along $\mathbf{B}$ without Davis-Greenstein dissipation}

Because of the magnetic precession discussed in
Sec.~\ref{magPrec}, an average alignment with the magnetic field
direction (or more precisely with a direction very close to the
magnetic field direction) can be obtained if there is a tendency
toward alignment with some other direction characterized by a time
scale $t_{{\rm align}}$ where $t_{{\rm align}}\gg t_{{\rm prec}}.$
The effect of precession about an axis $\boldsymbol{\Omega }$
(presumably along $\pm \mathbf{B}$) in competition with alignment
about another direction $\mathbf{\hat n}$ is described by the
ordinary differential equation
 \begin{eqnarray}
\dot {\mathbf{L}}= \boldsymbol{\Omega }\times {\mathbf{L}} -\gamma
({\mathbf{L}}-{\mathbf{\hat n}}\bar L ). \label{PrecEqn}
\end{eqnarray}
This equation has two time scales: $\Omega ^{-1}$ and $\gamma
^{-1}.$ We are primarily interested in fast precession (i.e.,
$\Omega \gg \gamma ).$ Equation~(\ref{PrecEqn}) is a linear
equation whose solution decomposes into a stationary part and
transient spiralling in to approach the stationary attractor. We
solve
 \begin{eqnarray}
\left(\begin{array}{@{}c@{\quad}c@{\quad}c@{}}
-\gamma & +\Omega & 0\\[3pt]
-\Omega & -\gamma & 0\\[3pt]
 0& 0& -\gamma\\
\end{array}\right)
\left(\begin{array}{@{}c@{}}
L_{\perp 1}\\[3pt]
L_{\perp 2}\\[3pt]
L_{\parallel}\\
\end{array}\right)= -\gamma \bar L
\left(\begin{array}{@{}c@{}}
\hat n_{\perp 1}\\[3pt]
\hat n_{\perp 2}\\[3pt]
\hat n_{\parallel}\\
\end{array}
\right)\!,
\end{eqnarray}
so that
\begin{eqnarray}
\begin{split}
L_{\perp 1} &= \frac{\gamma ^2}{\gamma ^2+\Omega ^2} \bar L \hat
n_{\perp 2}
-\frac{\gamma \Omega }{\gamma ^2+\Omega ^2} \bar L\hat n_{\perp 1},
\\[4pt]
L_{\perp 2} &= \frac{\gamma ^2}{\gamma ^2+\Omega ^2} \bar L \hat
n_{\perp 1}
+\frac{\gamma \Omega }{\gamma ^2+\Omega ^2} \bar L\hat n_{\perp 2},
\\[2pt]
L_{\parallel} &=\bar L \hat n_{\parallel }.\phantom{\int }
\end{split}
 \end{eqnarray}
In the fast precession regime of most interest to us, $\mathbf{L}= \bar L
\boldsymbol{\Omega }(\boldsymbol{\Omega }\cdot \mathbf{\hat n})/\Omega ^2,$ with
corrections suppressed by a factor $(\gamma /\Omega ).$ The precession has no
alignment tendency of its own, but if the precession period is shorter than the
time scale of the alternative alignment mechanism, the precession acts to suppress
alignment in the plane normal to $\boldsymbol{\Omega }.$

If it turns out that radiative torques combined with the fast
precession described above is the solution to the alignment
problem, a number of interesting observational consequences
follow. Blind data analysis might suggest that one should be able
to predict polarization from the thermal dust emission intensity
combined with a map of the galactic magnetic field, perhaps
deduced from rotation measures and synchrotron emission. However,
if the above mechanism is correct, this procedure will not work,
even with a three-dimensional model of the unpolarized dust
emission and galactic magnetic field. In the above equations,
provided that the galactic magnetic field is strong enough, its
direction serves to define the direction of the grain alignment
but its precise magnitude has little relevance. The above
mechanism can also accommodate a large amount of small-scale power
in the polarized dust signal, even if both the galactic magnetic
field and the dust density are smooth. One might expect the field
characterizing the radiative alignment tendency $\bar
L(\mathbf{x})\mathbf{\hat n}(\mathbf{x})$ to include an
appreciable component varying on small scales.

\subsection{Radiative torques}

Electromagnetic radiation scattered or absorbed by the dust grain
in general exerts a torque on the dust grain, as already modeled
in Sec.~\ref{SupraThermalSection} for isotropic illumination,
which is simpler because the torque direction is constant in the
co-rotating body coordinates. Here we generalize to anisotropic
illumination.

Let $\boldsymbol{\Gamma }(\nu, \boldsymbol{\hat n})$ expressed in
body coordinates be the torque exerted on a dust grain by
monochromatic electromagnetic radiation of frequency $\nu $ and
unit spectral radiance propagating in the $\boldsymbol{\hat n}$
direction. For an illumination described by the spectral radiance
$I(\nu, \boldsymbol{\hat n}),$ the torque exerted on the dust
grain in inertial coordinates is given~by
\begin{eqnarray}
\boldsymbol{\Gamma }(t) =\int _0^\infty d\nu \int _{S^2}d \boldsymbol{\hat n}
\boldsymbol{\Gamma }(\nu, {\cal R}(t) \boldsymbol{\hat n}) I(\nu, \boldsymbol{\hat
n}).
\end{eqnarray}
The complication here arises from the need to include both inertial and
co-rotating body coordinates, related by the transformation $ \mathbf{x}_{{\rm
iner}} = {\cal R}(t) \mathbf{x}_{{\rm body}} $.

Assuming that the time scale for radiative torques is long compared to the moment
of inertia alignment time scale $\tau_{I-L},$ we may take
$\boldsymbol{\Omega }$ to be aligned with $\mathbf{L}$ and define an averaged
kernel $\bar {\boldsymbol{\Gamma }}(\nu, \boldsymbol{\hat n}_I; \boldsymbol{\hat
n}_L ),$ defined as
 \begin{eqnarray}
\bar {\boldsymbol{\Gamma }}(\nu, \boldsymbol{\hat n}_I; \boldsymbol{\hat n}_L )
=\frac{1}{2\pi}\int _0^{2\pi }d\chi \boldsymbol{\Gamma }(\nu, \mathbf{R}_{(\chi,
\boldsymbol{\hat n}_L)} \boldsymbol{\hat n}_I),
\end{eqnarray}

\noindent where $\mathbf{R}_{(\chi, \boldsymbol{\hat n}_L)}$ represents a rotation by $\chi$ about the
$\boldsymbol{\hat n}_L$ axis. Apart from a few general observations, little can be said about the form
and properties of the radiative torque kernel, which differ from grain to grain. At very long
wavelengths---that is, long compared to the size of the dust grain---it is a good approximation to
treat the grain as a dipole (as one does for Rayleigh scattering). In the dipole approximation there are
no torques. Therefore, $(a/\lambda )$ serves as an expansion parameter for which the torque approaches
zero as this parameter tends toward zero. Likewise a spherical grain gives a vanishing torque at any
frequency as the result of symmetry arguments.

Convolving the incident radiation intensity $I(\nu, \boldsymbol{\hat n_I})$ with
this kernel, we obtain the angle-averaged torque $\boldsymbol{\Gamma }_{{\rm
tot,rad}}( \boldsymbol{\hat n_L}).$ If there were no magnetic field, we could
incorporate the torque above into the previously derived stochastic differential
equation for the evolution of $\mathbf{L}$ [Eq.~(\ref{LangevinOne})] to obtain
 \begin{eqnarray}
\dot {\mathbf{L}}(t)= - \beta {\mathbf{L}}(t) + \beta \bar L
\frac{{\mathbf{L}}(t)}{| {\mathbf{L}}(t)| } +\boldsymbol{\Gamma }_{{\rm tot,rad}}
\!\left( \frac{{\mathbf{L}}(t)}{| {\mathbf{L}}(t)| } \right) +\mathbf{n}_{{\rm
ran}}(t),
 \end{eqnarray}
where a coherent torque in body coordinates has been included as well. The above
equation describes a random walk $\mathbf{L}(t)$ in three-dimensional angular
momentum space. Owing to the stochastic random torque noise term, the rotational
degree of freedom loses memory of its initial state after some characteristic
decay time $t_{{\rm ran}},$ after which its state can be described by a
probability distribution $p(\mathbf{L})$ calculable based on the above equation.
The stochastic equation above has the general form
\begin{eqnarray}
\dot {\mathbf{L}}= \mathbf{v}(\mathbf{L})-\mathbf{n}(t),
\end{eqnarray}
\noindent for which the associated Fokker-Planck equation describing the evolution of the probability
density is
 \begin{eqnarray}
\frac{\partial p (\mathbf{L}) }{\partial t}=\nabla ^2p (\mathbf{L}) +\nabla \cdot
( p (\mathbf{L}) \mathbf{v} (\mathbf{L})).
\end{eqnarray}
With the magnetic field included, we add the magnetic precession term to obtain
 \begin{eqnarray}
\dot {\mathbf{L}}(t)= - \beta {\mathbf{L}}(t) + \beta \bar L
\frac{{\mathbf{L}}(t)}{| {\mathbf{L}}(t)| } +\boldsymbol{\Gamma }_{{\rm tot,rad}}
\left( \frac{{\mathbf{L}}(t)}{| {\mathbf{L}}(t)| } \right) +\gamma _{{\rm
mag}}\mathbf{L} \times \mathbf{B} +\mathbf{n}_{{\rm ran}}(t),\qquad
 \end{eqnarray}
where the constant $\gamma _{{\rm mag}}$ relates the magnetic moment to the
angular momentum according to $\boldsymbol{\mu }_B= \gamma _{{\rm
mag}}\mathbf{L}.$

\clearpage

\begin{figure}
\begin{center}
\includegraphics[width=8cm]{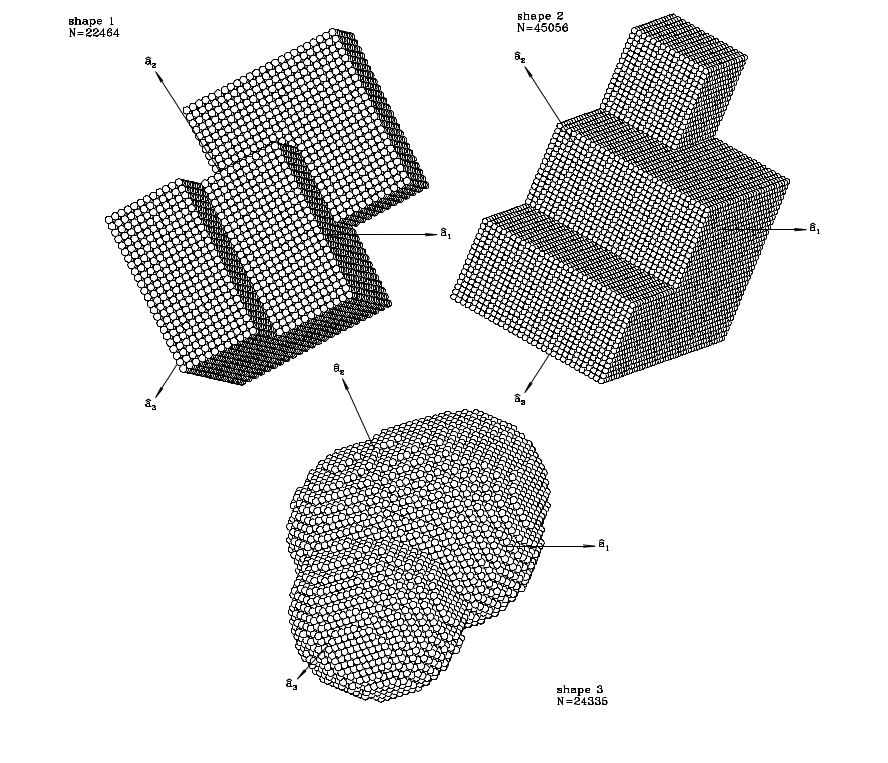}
\end{center}
\caption{{\bf Selection of grain shapes for studying radiative torques.} In
Ref.~\cite{draine1996}, Draine and Weingartner calculated the radiative torque on a dust grain
as a function of frequency $\nu $ and incident photon direction $\hat {\mathbf{n}}$ for the three grain
shape illustrated above. ({\it Credit: B. Draine and J. Weingartner})
\label{GrainShapeFig}}
\end{figure}

When $t_{{\rm ran}}, t_{{\rm rad}}\gg t_{{\rm prec}},$ it is possible to reduce
the dimension of the stochastic differential equation to two dimensions, namely
the angular velocity $\omega $ and the angle $\theta _{BL}$ between the vectors
$\mathbf{L}$ and $\mathbf{B},$ obtaining
 \begin{eqnarray}
\dot \theta _{BL}=F(\theta _{BL}, \omega )+n_\theta (t),\quad \dot
\omega =G(\theta _{BL}, \omega )+n_\omega (t).
\label{planeAutoStochastic}
 \end{eqnarray}
Unfortunately, very little can be said about the form of the
functions $F(\theta _{{\it BL}},\omega )$ and $G(\theta _{{\it
BL}},\omega ),$ which depends on the detailed grain shapes and
other properties. Draine and Weingartner
\cite{draine1996,draine1997,weingartner2003} studied three
possible irregular grain shapes (shown in
Fig.~\ref{GrainShapeFig}), calculating numerically the radiative
torque as a function of illumination direction and frequency. Then
assuming a certain anisotropic illumination, they calculated the
functions $F$ and $G$ in Eq.~(\ref{planeAutoStochastic}) and
studied the solutions to this equation numerically establishing
the orbits in the absence of noise. Their simulations conclude
that radiative torques can provide a plausible alignment
mechanism.

The proposal that radiative torques resulting from anisotropic illumination align the interstellar dust
grains with the magnetic field offers a promising alternative to the Davis-Greenstein magnetic
dissipation mechanism whose characteristic time scale may be too long to provide the
needed degree of alignment. However, its viability is hard to evaluate given our ignorance concerning the
shapes of the dust grains. Without such shape information, the best case that can be put forward in favor
of this scenario based on radiative torques and anisotropic illumination is to show that some candidate
shapes exist that give the right orders of magnitude combined with the argument that the
Davis-Greenstein scenario cannot account for the data. It is discouraging that it is not possible to
formulate precise predictions for the radiative torque scenario. The origin of grain alignment seems to
remain an open question.

\subsection[Small dust grains and anomalous microwave emission (AME)]
{Small dust grains and anomalous microwave emission (AME)}

The thermal emission of interstellar dust grains discussed above was based on the assumption that its
thermal state may be characterized by a mean temperature (around 20\,K) and that fluctuations about this
temperature subsequent to the absorption of an UV photon constitute a negligible correction. If a dust
grain is sufficiently large, this is a good approximation, but for small molecules an FUV (far
ultraviolet) photon can raise the grain temperature momentarily by an order of magnitude or more,
allowing emission in bands where there would be almost no emission for a grain at its average
temperature. The bulk of the energy absorbed from the UV photon is then rapidly re-emitted through IR
fluorescence as the molecule rapidly cools back down to its mean temperature. Hints of such emission
requiring a temperature significantly above the mean temperature were noted through the observation of
aromatic infrared emission features consistent with C--H bonds on the edge of an aromatic 
molecule---that is, around 3.3 and 11.3 microns \cite{duleyAndWilliams,legerPuget1984}. Later, analysis of the IRAS
dust maps showed that a significant fraction of the UV photons absorbed through extinction were
re-emitted in bands of wavelength shorter than $60\,\mu,$ where under the assumption of a reasonable
average temperature and a modified blackbody law there should be no emission \cite{PugetLegerReview}. As
a consequence, the dust models were revised to extend the size distribution to include small grains
thought to be PAHs as a result of the infrared emission features described above. PAHs (polycyclic
aromatic hydrocarbons) are essentially small monolayer planar sheets of graphite with hydrogen frills on
their edges.

A template for the small grain population can be constructed using maps of
infrared emission at $\lambda \ltorder 60\mu $ where the thermal emission from
large grains does not contribute. Since the short wavelength emission from the
dust grains is proportional to the local FUV flux density, these templates can in
principle be improved by dividing the local FUV flux density to obtain a better
map of the projected small dust grain column density, which in turn can be
correlated with the AME (anomalous microwave emission) at low frequencies.

It was also pointed out that such a population of very small dust
grains would be expected to rotate extremely rapidly, giving rise
to electric dipole emission at its rotation frequency and possibly
at higher harmonics, since $P_{e}\sim \mu _E\omega ^4$ and $\omega
^2\approx (k_{\rm B}T_{{\rm rot}})/ma^2$ in the absence of
suprathermal rotation \cite{EDSpinning,diffuseSpinning,Haimoud}.
The possibility of magnetic dipole emission from small rotating
grains has also been investigated \cite{magDipole}.

Observationally, emission at low frequencies, where diffuse
thermal dust emission should be minuscule, was found correlated
with high frequency dust
maps \cite{angelika,kogutHighGalactic,Leitch}.
Such emission would be consistent with dipole emission from small
rapidly rotating dust grains, as described above. As an
alternative explanation for such a correlation, it was proposed
that regions of free-free emission could be correlated with
regions of thermal dust emission. However, this low frequency
excess was found not to be correlated with $H\alpha $ line
emission, which serves as a tracer of free-free emission. One
possibility to weaken such a correlation would be to postulate an
extremely high temperature, so that $\Gamma _{{\rm
free}\text{--}{\rm free}}(T)/\Gamma _{H\alpha }$ is large.
However, it was found that such a high temperature would require
an implausible energy injection rate to prevent this phase from
cooling.

\begin{figure}[t]
\begin{center}
\includegraphics[width=15cm]{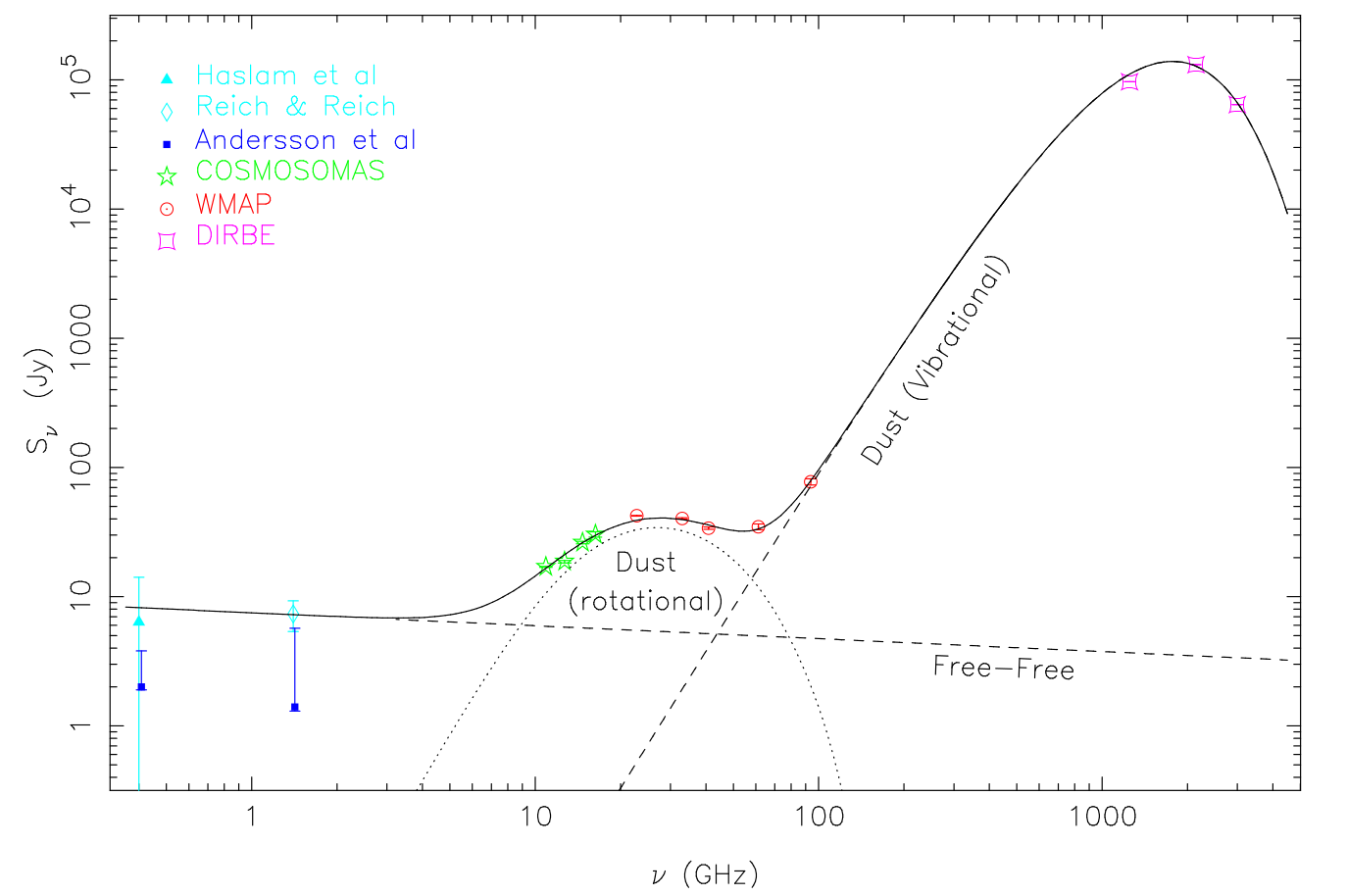}
\end{center}
\caption{{\bf Consistency of anomalous microwave emission with spinning dust.} The multi-frequency observations
shown here were taken for a molecular cloud believed to have strong anomalous dust emission. We observe
that given the shape a linear combination of thermal dust and free-free emission cannot explain the
observed shape. However, if an anomalous dust component with the spectral shape predicted by Draine and
Lazarian for electric dipole emission from spinning dust is included (dotted curve), a reasonable fit may
be obtained. For more details, see Ref.~\protect\cite{WatsonEtAl}. 
({\it Credit: COSMOMAS Collaboration}) 
\label{anomalousMicrowaveEmissionFigure}}
\end{figure}

The spectrum of anomalous microwave emission, presumably from spinning dust, has been measured by taking
spectra in targeted regions having a large contribution from AME and subtracting the expected
synchrotron, free-free, thermal dust, and primordial CMB signals \cite{WatsonEtAl,HildebrandtEtAl}.
Figure~\ref{anomalousMicrowaveEmissionFigure} shows the spectrum thus obtained. The observed fall-off at
low $\nu $ disfavors the hypothesis of free-free \hbox{emission} correlated with thermal dust emission
as the origin of the anomalous emission, because in that case, the spectrum of the residual should rise
as $\nu $ is lowered.

Most theoretical studies have assumed that the small grains do not become aligned
with the galactic magnetic field, but alignment mechanisms for this grain
population have not yet been investigated in detail. Observational limits on the
anomalous microwave emission constrain the degree of polarization to less than a
few per cent \cite{RubinoMartin}.

\section[Compact sources]{Compact Sources}

The above processes pertain primarily to diffuse galactic
emission, which contaminates the primordial CMB signal most
severely at low multipole number $\ell.$ On smaller scales,
however, compact sources become more of a concern and eventually
become the dominant contribution at all frequencies. We use the
term ``compact'' to denote both unresolved point sources and
partially resolved, localized sources.

The angular power spectrum of point sources distributed according
to a Poissonian distribution would scale with multipole number as
$\ell ^0.$ If ignoring the structure of the acoustic oscillations,
one approximates the primordial CMB spectrum as having the shape
$\ell ^{-2},$ one would conclude that compact sources rapidly take
over at high $\ell.$ Because of Silk damping combined with the
smoothing from the finite width of the last scattering surface,
the fall-off of the CMB at large $\ell $ is even more rapid. The
fact that point sources are clustered rather than distributed in
Poissonian manner causes the point source power spectrum to be
clustered, leading to an excess of power at small $\ell $ compared
to an $\ell ^0$ power law.

\subsection{Radio galaxies}

Section~\ref{GalacticSynchrotronEmission} described the physical
process of synchrotron emission, emphasizing its contribution to
the diffuse emission from our own galaxy. However the same process
is also at work in galaxies other than our own, and in particular
in the rare, but highly luminous radio galaxies at high redshift,
extreme processes fueled by a central black hole give rise to
intense emission in many bands, including synchrotron emission in
the radio and microwave bands. For our purposes, the most salient
and relevant feature of these sources is their inverted spectrum.
For a detailed survey of radio sources, see
Ref.~\cite{RadioSourceRef}. Since each source differs slightly
in its spectral properties, the strategy for dealing with these
sources is to mask out the brightest sources and then model the
residual arising from the remaining unmasked sources.

\subsection{Infrared galaxies}
In Sec.~\ref{ThermalDust}, we discussed at length the contamination from thermal
emission from the interstellar dust in our galaxy. Thermal dust emission, however,
is not restricted to our galaxy. All galaxies contribute some thermal dust
emission. However, the galaxies at high redshift in which star formation is
occurring at a rapid rate contribute predominantly to the extragalactic thermal
dust emission in the microwave bands of interest to us. In fact, most of the
energy of the radiation from these galaxies in which intense star formation is
taking place is emitted in the infrared, because the light and UV emitted is
absorbed by dust, which becomes heated and in turn re-emits in the infrared.

Some of these infrared galaxies are resolved by current
observation. Others remain unresolved by present day observations
and these are attributed to the diffuse infrared background. These
infrared galaxies do not all emit with the same frequency
spectrum, in part because they are spread over a wide range of
redshifts. For the early discovery papers on the CIB (cosmic
infrared background) from COBE see
Refs.~\cite{pugetCFIRB,hauserCFIRB} and \cite{fixsenCFIRB}.
For more recent observations from Planck, see
Ref.~\cite{planck-cib-ps}.

\section[Other effects]{Other Effects}

\subsection{Patchy reionization}

In Sec.~\ref{ReionizationSection}, we idealized the reionized hot electron gas extending from
intermediate redshifts to today as being isotropically distributed so that the free
electron density depends only on redshift. Such a reionization scenario is implausible but provides a
good approximation for the effect of reionization on the CMB power spectrum at low $\ell.$ In a more
realistic reionization scenario, the universe becomes reionized in an inhomogeneous manner. As the first
stars and quasars form, their UV radiation creates around them spheres of almost completely ionized HII
that grow and eventually coalesce. These spheres have sharp edges that act to transform large-angle power
from the primordial CMB into small-scale power. For some estimates of the impact due to patchy
reionization of the angular CMB spectrum, see Ref.~\cite{zahn}.

\subsection{Molecular lines}

The sources of diffuse galactic emission discussed so far all have
a broad continuum emission spectrum. But there also exist
molecular lines, mostly from rotational transitions of molecules,
that emit in the microwave bands of interest here. The brightest
of these are the CO lines \cite{dameCO}, of which the lowest is
situated at 115.3\,GHz. Since CO is a linear molecule, there are
also transitions at integer multiples of this frequency. Removing
CO emission from the Planck data has proven to be a challenge
\cite{planck-galactic-co}. For more sensitive future experiments,
it is likely that other lines will also present concerns.

\subsection{Zodiacal emission}
Besides the planets and asteroids, our solar system is filled with large dust grains, known as zodiacal
dust, which on the average are much larger than the interstellar dust grains described above. These
grains manifest themselves by a variety of means. In the visible they scatter light from the sun, and
this background must be subtracted when analyzing diffuse emission in the optical and near infrared
bands. The zodiacal dust is concentrated around the ecliptic plane. It is a good but not perfect
approximation to model the distribution of zodiacal dust assuming azimuthal symmetry. The temperature of
the zodiacal dust is much above that of the interstellar dust and roughly proportional to the inverse
square of the distance from the sun. Subtracting zodiacal emission is challenging because its
contribution to a given point on the celestial sphere depends on the time of year when the observation is
made, because observations at different times of the year look through different columns of dust, even
when azimuthal symmetry is assumed. The ideal observing program for characterizing the zodiacal dust
would cover the whole sky many times over the year in order to accumulate as many lines of sight to the
same point on the celestial sphere as possible and to carry out a sort of tomography. For a discussion of
zodiacal emission at long wavelengths as measured by Planck, see Ref.~\cite{planck-zodi}. For previous
work on zodiacal emission in the far infrared, see Ref.~\cite{cobe-zodi}.

\section[Extracting the primordial CMB anisotropies]%
{Extracting the Primordial CMB Anisotropies}

Most of the contaminants described above, with the exception of gravitational lensing and
the kinetic SZ effect, have a frequency dependence that differs from that of the primary
CMB anisotropies. This property can be exploited to remove these nonprimordial
contaminants and lies at the heart of all component separation techniques. If there are
$N$ distinct components, each of which is characterized by a fixed frequency spectrum, in
the absence of measurement noise, a perfect separation of these components is possible
using sky maps in $N$ frequency bands. With more bands, the separation is overdetermined
allowing validation of the model. The component separation strategy sketched above, which
may be described as a linear method, is based on a linear model, but this is not the only
way to clean maps. More sophisticated nonlinear models may be formulated, often based on
or inspired by Bayesian statistics that exploit other properties such as the positivity
of certain components, spatial variation of spectral indices, and so on. Some methods are
characterized as being ``blind,'' meaning that the number of components is indicated by
data itself, while other methods are more physically based.

Because of limitations of space, we do not go into the details or practicalities
of component separation, instead referring the reader to a few excellent reviews
for more details. For an early discussion of the component separation problem and
various approaches, see Refs.~\cite{bouchetGispert} and \cite{dcfPaper}.
Reference~\cite{leach} includes a comparison of methods used for the Planck
analysis. Reference~\cite{dunkleyCMBPOLfgr} describes foreground removal
challenges for future CMB B mode polarization experiments and includes extensive
references to the literature.

\section[Concluding remarks]{Concluding Remarks}

Prior to the discovery of the CMB anisotropy in 1992 by the COBE team, cosmology
was a field with a questionable reputation. Many from other fields described
cosmology as a field where theorists were free to speculate with almost no
observational or experimental constraints to contradict them. Much was made of the
supposed distinction between an ``observation'' and an ``experiment.''\footnote{It
is not clear whether proton-antiproton colliders would be classified as
constituting an ``experiment'' under this criterion, because the experimenter has
virtually no control over parameters of the hard parton-parton collisions of
interest due to the broad spread in $x$ of the parton distribution functions.}
However, between 1992 and today cosmology has transformed into a precision science
where it has become increasingly difficult to propose new models without
contradicting present observation. Ironically, much science having only a tenuous
connection to cosmology is now being explained to the public as serving to probe
the big bang.

In this review, we have tried to sketch the most important aspects of the physics of the
CMB in a self-contained way. Other reviews with a different emphasis include
Refs.~\cite{huDodelson}, \cite{PJEPee1987} and \cite{DSamSStaBWin2007}. We have
seen how the CMB provides a snapshot of the state of the universe at around $z\approx
1100,$ when the primordial perturbations were still very nearly linear, and how the
precise angular pattern on the sky is also sensitive to a number of other cosmological
parameters, which can be determined in many cases to the percent level or better. We have
also explored tests of non-Gaussianity. At present no non-Gaussianity of a primordial
origin has been observed, ruling out many nonminimal models of inflation, in particular
multi-field models developed to explain hints of non-Gaussianity in the WMAP data, which
subsequently were excluded by Planck.

We also saw how better measurements of the CMB absolute spectrum could reveal
deviations from a perfect blackbody spectrum, some of which are to be expected and
others of which would constitute signs of novel physics, such as energy injection
at high-redshift. The absolute spectrum remains a promising frontier of CMB
science.

The other promising area of CMB research is the characterization of the polarized sky, in particular to
search for primordial B modes from inflation, which constitutes the most unique prediction of inflation.
Maps of B modes on smaller angular scales would also allow characterization of the gravitational lensing
field at much greater precision than is possible from observations of the temperature anisotropy. Such
data would provide an important cross check of weak lensing surveys using the ellipticities of galaxies
and provide a unique probe of absolute neutrino masses, perhaps allowing one to distinguish the normal
from the inverted mass hierarchy.

Since this volume is dedicated to Albert Einstein and the
hundredth anniversary of his celebrated paper presenting the
general theory of relativity, we close with a few remarks about
Einstein, and in particular about his relation to the subject of
this chapter. Einstein is remembered by today's generation of
physicists mostly for his role in the development of the special
and general theories of relativity. This part of Einstein's work
sometimes overshadows Einstein's numerous achievements in other
areas and the extreme breadth of his scientific interests.
Einstein is often viewed today as the paragon of the abstruse,
highly mathematical theorist. Many physicists today would be
surprised to learn that Einstein patented a design for an
ammonia-butane refrigerator \cite{einsteinPatent}. Others might
be surprised to learn that Einstein is the author of one
experimental paper for work after which the Einstein-de Haas
effect is named. This effect may be thought of as the converse of
the Barnett effect, which we saw in Sec.~\ref{MagMoment} plays a
crucial role in the alignment of interstellar dust grains.
Moreover, the stochastic differential equations in that section
bear a close relation to Einstein's work on Brownian motion.
General relativity was once regarded as an abstruse, highly
mathematical subfield of physics. Within astrophysics, relativity
was first introduced through the emergence of a subfield known as
``relativistic astrophysics.'' But nowadays general relativity
pervades almost all areas of astrophysics and has become one of
the elements of the underlying fundamental physics that every
astrophysicist must know. One can no longer neatly separate
``relativistic astrophysics'' from the rest of astrophysics.

This chapter told the story of the big bang cosmology from a point of view emphasizing
the observations of the relic blackbody radiation. Einstein's theory of general
relativity culminating in his celebrated 1915 formulation of the gravitational field
equations\footnote{A.~Einstein, ``Die Feldgleichungen der Gravitation," Sitzungberichte
der K\"onigliche Preussiche Akademie der Wissenschaften, p. 844 (25 November,
1915).}$^{,}$\footnote{For an authoritative account from a scientific perspective of the
path involving over a dozen papers and several reversals that finally led Einstein to the
gravitational field equations as we know them today (i.e., $G_{\mu \nu }=R_{\mu \nu
}-\frac{1}{2}g_{\mu \nu }R=(8\pi G_N)T_{\mu \nu }$), see Abraham Pais, {\it Subtle is the
Lord} (New York, Oxford University Press, 1982).} enters as a key ingredient to this
story in two ways. First there is unperturbed spacetime solution now known as the FLRW
solutions. This family of spatially homogeneous and isotropic solutions to the Einstein
equations was first found by Alexandre Friedmann in 1922 and independently by Georges
Lema\^itre in 1927. At that time, the solutions were surprising because of their
prediction of a non-eternal dynamical universe whose natural state was either expansion
or contraction, an idea that was before its time. It was not until 1929 that Hubble
discovered that the recessional velocities of distant galaxies except for some noise vary
approximately linearly with distance (although in its first version the data was not
statistically compelling). Einstein's general theory was deduced by drawing mainly on
considerations of mathematical simplicity and elegance and on the equivalence principle,
but remarkably applied to the expanding universe through the FLRW solution and its
linearized perturbations, general relativity remains the basis of cosmology despite the
great improvement in the data that has taken place between the 1920s and now. Today there
is much talk about ``modified gravity,'' largely motivated by the discovery of the
so-called ``dark energy,'' which no one yet really understands. It can be argued whether
the dark energy belongs to the left hand side of the Einstein field equation, in which
case it would constitute a revision to Einstein's theory, or on the right-hand side, in
which case it would simply constitute a new form of exotic stress-energy, leaving the
gravitational sector unchanged. But so far Einstein's theory of relativity as a classical
field theory has survived without change. To be sure, we believe that gravitational
theory as formulated by Einstein is an incomplete story. Where quantum effects become
relevant, we know that there must be a more complete theory of which classical relativity
is but a limiting case. But we do not yet know what that larger theory is. In the
classical regime, Einstein's theory has survived many stringent tests, and perhaps more
importantly has become an indispensable foundation for our current understanding of the
universe and its origins.

\section*{Acknowledgments}
MB would like to thank Ken Ganga, Kavilan Moodley, Heather Prince, Doris Rojas, and Andrea Tartari for
many invaluable comments, corrections, and suggestions, and Jean-Luc Robert for making several of the
figures. MB thanks the University of KwaZulu-Natal for its hospitality where a large part of this review
was written.

\pagebreak


\addcontentsline{toc}{section}{References}

\begin{thebibliography}{000}
\bibitem{stageFour}
K. N. Abazajian {\it et~al.}, Neutrino physics from the cosmic microwave
background and large scale structure, arXiv:astro-ph/1309.5383.

\bibitem{stageFourBis}
K. N. Abazajian {\it et~al.}, Inflation physics from the cosmic microwave
background and large scale structure, arXiv:astro-ph/1309.5381.

\bibitem{abbottWise}
L. F. Abbott and M. B. Wise, Constraints on generalized inflationary cosmologies,
{\it Nucl. Phys. B} {\bf 244} (1984) 541.

\bibitem{aquaviva}
V. Acquaviva, N. Bartolo, S. Matarrese and A. Riotto, Gauge-invariant second-order
perturbations and non-Gaussianity from inflation, {\it Nucl. Phys. B} {\bf 667}
(2002) 119, arXiv:astro-ph/0209156.

\bibitem{AlbrechtCausality}
A. Albrecht, D. Coulson, P. Ferreira and J. Magueijo, Causality, randomness, and
the microwave background, {\it Phys. Rev. Lett.} {\bf 76} (1996) 1413.

\bibitem{Haimoud}
Y. Ali-Ha\"imoud, C. M. Hirata and C. Dickinson, A refined model for spinning dust
radiation, {\it Mon. Not. R. Astron. Soc.} {\bf 395} (2009) 1055.

\bibitem{AlpherBetheGamov}
R. A. Alpher, H. Bethe and G. Gamov, The origin of chemical elements, {\it Phys.
Rev.} {\bf 73} (1948) 803.

\bibitem{polarBear}
K. Arnold {\it et~al.}, The POLARBEAR CMB polarization experiment, {\it Proc.
SPIE} {\bf 7741} (2010) 77411E.

\bibitem{babich}
D. Babich, P. Creminelli and M. Zaldarriaga, The shape of non-Gaussianities, {\it
J. Cosmol. Astropart. Phys.} {\bf 08} (2004) 009, arXiv:astro-ph/0405356.

\bibitem{bardeenGIF}
J. M. Bardeen, Gauge-invariant cosmological perturbations, {\it Phys. Rev. D} {\bf
22} (1980) 1882.

\bibitem{Bardeen:1983qw}
J. M. Bardeen, P. J. Steinhardt and M. S. Turner, Spontaneous
creation of almost scale-free density perturbations in an
inflationary universe, {\it Phys.~Rev.~D} {\bf 28} (1983) 679.

\bibitem{BarkanaLoeb}
R. Barkana and A. Loeb, In the beginning: The first sources of light and the
reionization of the universe, {\it Phys. Rep.} {\bf 349} (2001) 125,
arXiv:astro-ph/0010468.

\bibitem{qubic}
E. Battistellie {\it et~al.}, QUBIC: The QU bolometric interferometer for
cosmology, {\it Astropart. Phys.} {\bf 34} (2011) 705, arXiv:astro-ph/1010.0645.

\bibitem{becker}
R. H. Becker {\it et~al.}, Evidence for reionization at $z\sim 6$: Detection of a
Gunn-Peterson trough in a $z=6.28$ quasar, {\it Astron. J.} {\bf 122} (2001)
2850, arXiv:astro-ph/0108097.

\bibitem{boomerang}
P. de Bernardis {\it et~al.}, A flat universe from high-resolution maps of the
cosmic microwave background radiation, {\it Nature} {\bf 404} (2000) 955,
arXiv:astro-ph/0004404.

\bibitem{cosmics}
E. Bertschinger, COSMICS: Cosmological initial conditions and microwave anisotropy
codes, arXiv:astro-ph/9506070.

\bibitem{bicep2a}
BICEP2 Collab. (P. A. R. Ade {\it et~al.}), Detection of B-mode polarization at
degree angular scales by BICEP2, {\it Phys. Rev. Lett.} {\bf 112} (2014) 241101.

\bibitem{bicep2b}
BICEP2 Collab. (P. A. R. Ade {\it et~al.}), BICEP2. II. Experiment and three-year
 data set, {\it Astrophys. J.} {\bf 792} (2014) 62, arXiv:astro-ph/1403.4302.

\bibitem{MBirkinshaw}
M. Birkinshaw, The Sunyaev-Zel'dovich effect, {\it Phys. Rep.} {\bf 310} (1999)
97, arXiv:astro-ph/9808050.

\bibitem{BlandfordEichler}
R. Blandford and D. Eichler, Particle acceleration at astrophysical shocks: A
theory of cosmic ray origin, {\it Phys.~Rep.} {\bf 154} (1987) 1.

\bibitem{cobe-noise}
N. W. Boggess {\it et~al.}, The COBE mission---Its design and performance two
years after launch, {\it Astrophys.~J.} {\bf 397} (1992) 420.

\bibitem{bondEfstathiou}
J. R. Bond and G. Efstathiou, Cosmic background radiation anisotropies in
universes dominated by nonbaryonic dark matterm, {\it Astrophys. J.} {\bf 285}
 (1984) L45.

\bibitem{BornAndWolf}
M. Born and E. Wolf, {\it Principles of Optics} (Pergamon, New York, 1970).

\bibitem{bouchetGispert}
F. R. Bouchet and R. Gispert, Foregrounds and CMB experiments: I. Semi-analytical
estimates of contamination, {\it New Astron.} {\bf 4} (1999) 443,
arXiv:astro-ph/9903176.

\bibitem{boughnCrittenden}
S. Boughn and R. Crittenden, A correlation between the cosmic microwave background
and large-scale structure in the universe, {\it Nature} {\bf 427} (2004) 45,
arXiv:astro-ph/0305001.

\bibitem{bgt}
M. Bucher, A. S. Goldhaber and N. Turok, An open universe from inflation, {\it
Phys.~Rev. D} {\bf 52} (1995) 3314, arXiv:hep-ph/9411206.

\bibitem{btArbMass}
M. Bucher and N. Turok, Open inflation with an arbitrary false vacuum mass, {\it
Phys. Rev. D} {\bf 52} (1995) 5538.

\bibitem{genCosmoPert}
M. Bucher, K. Moodley and N. Turok, General primordial cosmic perturbation, {\it
Phys. Rev. D} {\bf 62} (2000) 083508, arXiv:astro-ph/9904231.

\bibitem{CarlstromSZReview}
J. E. Carlstrom, G. P. Holder and E. D. Reese, Cosmology with the
Sunyaev-Zel'dovich effect, {\it Ann. Rev. Astron. Astrophys.} {\bf 40} (2002)
643, arXiv:astro-ph/0208192.

\bibitem{chlubaThomas}
J. Chluba and R. M. Thomas, Towards a complete treatment of the cosmological
recombination problem, {\it Mon. Not. R. Astron. Soc.} {\bf 412} (2011) 748,
arXiv:astro-ph/1010.3631.

\bibitem{energy-release}
J. Chluba and R. A. Sunyaev, Pre-recombinational energy release and narrow
features in the CMB spectrum, {\it Astron. Astrophys.} {\bf 501} (2009) 29,
arXiv:astro-ph/0803.3584.

\bibitem{contaldi}
C. R. Contaldi, J. Magueijo and L. Smolin, Anomalous CMB polarization and
gravitational chirality, {\it Phys. Rev. Lett.} {\bf 101} (2008) 141101, arXiv:
astro-ph/0806.3082.

\bibitem{coreWhitePaper}
COrE Collab. (F. Bouchet {\it et~al.}), COrE (Cosmic Origins
Explorer): A White
Paper, arXiv:1102.2181.


\bibitem{spider}
B. P. Crill {\it et~al.}, SPIDER: A balloon-borne large-scale CMB polarimeter,
 {\it Proc. SPIE 7010$,$ Space Telescopes and Instrumentation 2008\/$:$ Optical$,$ Infrared$,$ and
Millimeter$,$ 70102P} (July 12, 2008).

\bibitem{crittenden}
R. Crittenden, R. L. Davis and P. J. Steinhardt, Polarization of the microwave
background due to primordial gravitational waves, {\it Astrophys. J.} {\bf 417}
(1993) L13, arXiv:astro-ph/9306027.

\bibitem{TexturePeaks}
R. G. Crittenden and N. Turok, Doppler peaks from cosmic texture, {\it Phys. Rev.
Lett.} {\bf 75} (1995) 2642.

\bibitem{reesSciama}
R. G. Crittenden and N. Turok, Looking for a cosmological constant with the
Rees-Sciama effect, {\it Phys. Rev. Lett.} {\bf 76} (1996) 575,
arXiv:astro-ph/9510072.

\bibitem{dameCO}
T. M. Dame {\it et~al.}, The milky way in molecular clouds: A new complete CO
survey, {\it Astrophys. J.} {\bf 547} (2001) 792.

\bibitem{act-lensing}
S. Das {\it et~al.}, Detection of the power spectrum of cosmic microwave
background lensing by the atacama cosmology telescope, {\it Phys. Rev. Lett.} {\bf
107} (2011) 021301, arXiv:astro-ph/1103.2124.

\bibitem{davisGreenstein}
L. Davis and J. Greenstein, The polarization of starlight by aligned dust grains,
{\it Astrophys. J.} {\bf 114} (1951) 206.

\bibitem{dayArticle}
P. Day {\it et~al.}, A broadband superconducting detector suitable for use in
large arrays, {\it Nature} {\bf 425} (2003) 817.

\bibitem{angelika}
A. de Oliveira-Costa {\it et~al.}, Galactic microwave emission at degree angular
scales, {\it Astrophys. J.} {\bf 482} (1997) L17, arXiv:astro-ph/9702172.

\bibitem{bpol}
P. De Bernardis, M. Bucher, C. Burigana and L. Piccirillo, B-Pol: Detecting
primordial gravitational waves generated during inflation, {\it Exp. Astron.} {\bf
23} (2009) 5, arXiv:0808.1881.

\bibitem{dcfPaper}
J. Delabrouille, J.-F. Cardoso and G. Patanchon, Multi-detector multi-component
spectral matching and applications for CMB data analysis, {\it Mon. Not. R.
 Astron. Soc.} {\bf 346} (2003) 1089, arXiv:astro-ph/0211504.

\bibitem{dickeRSI}
R. H. Dicke, The measurement of thermal radiation at microwave frequencies, {\it
Rev. Sci. Instrum.} {\bf 17} (1946) 238.

\bibitem{DickePeeblesRollWilkinson}
R. H. Dicke, P. J. E. Peebles, P. G. Roll and D. T. Wilkinson, Cosmic black-body
radiation, {\it Astrophys. J.} {\bf 142} (1965) 414.

\bibitem{kids}
S. Doyle, P. Mauskopf, J. Naylon, A. Porch and C. Duncombe, Lumped element kinetic
inductance detectors, {\it J. Low Temp. Phys.} {\bf 151} (2008) 530.

\bibitem{draine2003}
B. T. Draine, Interstellar dust grains, {\it Ann. Rev. Astron. Astrophys.} {\bf
41} (2003) 241, arXiv:astro-ph/0304489.

\bibitem{Draine2004}
B. T. Draine, Astrophysics of dust in cold clouds, in {\it The Cold Universe$,$
Saas-Fee Advanced Course 32}, ed. D. Pfenniger (Springer-Verlag, Berlin, 2004),
pp.~213--303, arXiv:astro-ph/0304488.

\bibitem{EDSpinning}
B. T. Draine and A. Lazarian, Electric dipole radiation from spinning dust grains,
{\it Astrophys. J.} {\bf 508} (1998) 157, arXiv:astro-ph/9802239.

\bibitem{diffuseSpinning}
B. T. Draine and A. Lazarian, Diffuse galactic emission from spinning dust grains,
{\it Astrophys. J.} {\bf 494} (1998) L19.

\bibitem{magDipole}
B. T. Draine and A. Lazarian, Magnetic dipole microwave emission from dust grains,
{\it Astrophys. J.} {\bf 512} (1999) 740, arXiv:astro-ph/9807009.

\bibitem{draine1996}
B. T. Draine and J. C. Weingartner, Radiative torques on interstellar grains. I.
Superthermal spin-up, {\it Astrophys. J.} {\bf 470} (1996) 551.

\bibitem{draine1997}
B. T. Draine, J. C. Weingartner and C. Joseph, Radiative torques on interstellar
grains. II. Grain alignment, {\it Astrophys. J.} {\bf 480} (1997) 633.

\bibitem{duleyAndWilliams}
W. W. Duley and D. A. Williams, The infrared spectrum of interstellar 
dust---Surface functional groups on carbon, {\it Mon. Not. R. Astron. Soc.} {\bf 196}
(1981) 269.

\bibitem{dunkleyCMBPOLfgr}
J. Dunkley {\it et~al.}, CMBPol mission concept study: Prospects for polarized
foreground removal, {\it AIP Conf. Proc.} {\bf 1141} (2009) 222,
arXiv:astro-ph/0811.3915.

\bibitem{DunkleyMCMC}
J. Dunkley, M. Bucher, P. G. Ferreira and K. Moodley, C Skordis
``Fast and reliable MCMC for cosmological parameter estimation,
{\it Mon. Not. R. Astron. Soc.} {\bf 356} (2005) 925,
arXiv:astro-ph/0405462.

\bibitem{act-cosmo-params}
J. Dunkley {\it et~al.}, The Atacama cosmology telescope: Cosmological parameters
from the 2008 power spectrum, {\it Astrophys. J.} {\bf 739} (2011) 52,
arXiv:astro-ph/1009.0866.

\bibitem{einsteinPatent}
A. Einstein and L. Szil\'ard, Refrigeration (Appl: 16 December 1927; Priority:
Germany, 16 December 1926) U. S. Patent 1,781,541, 11 November 1930.

\bibitem{HairyBall}
M. Eisenberg and R. Guy, A proof of the hairy ball theorem, {\it Am. Math. Mon.}
{\bf 86} (1979) 571.

\bibitem{CMBPol-epic}
EPIC Collab. (J. Bock {\it et~al.}), Study of the experimental
probe of inflationary cosmology EPIC-Intemediate mission for
NASA's einstein inflation probe, arXiv:0906.1188.

\bibitem{RR1}
R. Fabbri and M. Pollock, The effect of primordially produced gravitons upon the
anisotropy of the cosmological microwave background radiation, {\it Phys. Lett. B} {\bf
125} (1983) 445.

\bibitem{ferte}
A. Ferte and J. Grain, Detecting chiral gravity with the pure pseudospectrum
reconstruction of the cosmic microwave background polarized anisotropies,
arXiv:astro-ph/1404.6660.

\bibitem{fixsenCFIRB}
D. J. Fixsen {\it et~al.}, The spectrum of the extragalactic far-infrared
background from the COBE FIRAS observations, {\it Astrophys. J.} {\bf 508} (1998)
123.

\bibitem{cobe-zodi}
D. J. Fixsen and E. Dwek, The zodiacal emission spectrum as determined by COBE and
its implications, {\it Astrophys. J.} {\bf 578} (2002) 1009.

\bibitem{COBEfiras}
D. J. Fixen {\it et~al.}, The cosmic microwave background spectrum
from the full COBE FIRAS data set, {\it Astrophys. J.} {\bf 473}
(1996) 576.

\bibitem{flauger}
R. Flauger, J. C. Hill and D. N. Spergel, Toward an understanding of foreground
emission in the BICEP2 region, arXiv:astro-ph/1405.7351.

\bibitem{Fuskeland}
U. Fuskeland, I. K. Wehus, H. K. Eriksen and S. K. Naess, Spatial variations in
the spectral index of polarized synchrotron emission in the 9 yr WMAP sky maps,
{\it Astrophys. J.} {\bf 790} (2014) 104, arXiv:astro-ph/1404.5323.

\bibitem{giardFirstDet}
M. Giard, G. Serra, E. Caux, F. Pajot and J. M. Lamarre, First detection of the
aromatic 3.3-micron feature in the diffuse emission of the Galactic disk, {\it
Astron. Astrophys.} {\bf 201} (1988) L1.

\bibitem{ginzburgReview}
V. L. Ginzburg and S. I. Syrovatsk, Developments in the theory of synchrotron
radiation and its reabsorption, {\it Ann. Rev. Astron. Astrophys.} {\bf 7} (1969)
375.

\bibitem{goldhaberOne}
A. S. Goldhaber and M. Nieto, Terrestrial and extraterrestrial limits on the
photon mass, {\it Rev. Mod. Phys.} {\bf 43} (1971) 277.

\bibitem{goldhaberTwo}
A. S. Goldhaber and M. Nieto, Photon and graviton mass limits, {\it Rev. Mod.
Phys.} {\bf 82} (2010) 939, arXiv:hep-ph/0809.1003.

\bibitem{gottA}
J. R. Gott III, Creation of open universes from de Sitter space, {\it Nature} {\bf
295} (1982) 304.

\bibitem{gottB}
J. Richard Gott III and T. S. Statler, Constraints on the formation of bubble
universes, {\it Phys. Lett. B} {\bf 136} (1984) 157.

\bibitem{GunnPeterson}
J. E. Gunn and B. A. Peterson, On the density of neutral hydrogen in intergalactic
space, {\it Astrophys. J.} {\bf 142} (1965) 1633.

\bibitem{Guth:1982ec}
A. H. Guth and S. Y. Pi, Fluctuations in the new inflationary universe, {\it Phys.
Rev. Lett.} {\bf 49} (1982) 1110.

\bibitem{hall}
J. S. Hall, Observations of the polarized light from stars, {\it Science} {\bf
109} (1949) 166.

\bibitem{HanEtAl}
J. L. Han {\it et~al.}, Pulsar rotation measures and the large-scale structure of
the galactic magnetic field, {\it Astrophys. J.} {\bf 642} (2006) 868,
arXiv:astro-ph/0601357.

\bibitem{maxima}
S. Hanany {\it et~al.}, MAXIMA-1: A measurement of the cosmic microwave background
anisotropy on angular scales of 10'-$5^\circ $, {\it Astrophys. J.} {\bf 545}
(2000) L5, arXiv:astro-ph/0005123.

\bibitem{kSZobs}
N. Hand {\it et~al.}, Evidence of galaxy cluster motions with the kinematic
Sunyaev-Zel'dovich effect, {\it Phys. Rev. Lett.} {\bf 109} (2012) 041101,
arXiv:astro-ph/1203.4219.

\bibitem{Haslam}
C. G. T. Haslam {\it et~al.}, A 408\,MHz all-sky continuum survey. I---
Observations at southern declinations and for the north polar region, {\it Astron.
Astrophys.} {\bf 100} (1981) 209.

\bibitem{hauserCFIRB}
M. G. Hauser {\it et~al.}, The COBE diffuse infrared background experiment search
for the cosmic infrared background. I. Limits and detections, {\it Astrophys. J.}
{\bf 508} (1998)~25.

\bibitem{Hawking:1982cz}
S. W. Hawking, The development of irregularities in a single bubble inflationary
universe, {\it Phys.~Lett.~B} {\bf 115} (1982) 295.

\bibitem{LiteBird}
M. Hazumi {\it et~al.}, LiteBIRD: A small satellite for the study of B-mode
polarization and inflation from cosmic background radiation
detection, {\it Proc. SPIE
8442$,$ Space Telescopes and Instrumentation 2012\/$:$ Optical$,$ Infrared$,$ and Millimeter
Wave$,$ 844219} (August 22, 2012).

\bibitem{hs2008}
C. M. Hirata and E. R. Switzer, Primordial helium recombination. II. Two-photon
processes, {\it Phys. Rev. D} {\bf 77} (2008) 083007, arXiv:astro-ph/0702144.

\bibitem{HildebrandtEtAl}
S. R. Hildebrandt, R. Rebolo, J. A. Rubi\~no-Martín, R. A. Watson, C. M.
Guti\'errez, R. J. Hoyland and E. S. Battistelli, COSMOSOMAS observations of the
cosmic microwave background and Galactic foregrounds at 11\,GHz: Evidence for
anomalous microwave emission at high Galactic latitude, {\it Mon. Not. R.
 Astron. Soc.} {\bf 382} (2007) 594, arXiv:0706.1873.

\bibitem{hiltner1949a}
W. A. Hiltner, Polarization of light from distant stars by interstellar medium,
{\it Science} {\bf 109} (1949) 165.

\bibitem{hiltner1949b}
W. A. Hiltner, On the presence of polarization in the continuous radiation of
stars. II., {\it Astrophys. J.} {\bf 109} (1949) 471.

\bibitem{psSD}
W. Hu, D. Scott and J. Silk, Power spectrum constraints from spectral distortions
in the cosmic microwave background, {\it Astrophys. J.} {\bf 430} (1994) L5,
arXiv:astro-ph/9402045.

\bibitem{peda-cosmo}
W. Hu and M. White, Acoustic signatures in the cosmic microwave background, {\it
Astrophys. J.} {\bf 471} (1996) 30, arXiv:astro-ph/9602019.

\bibitem{okamoto}
W. Hu and T. Okamoto, Mass reconstruction with CMB polarization, {\it Astrophys.
J.} {\bf 574} (2002) 566, arXiv:astro-ph/0111606.

\bibitem{huDodelson}
W. Hu and S. Dodelson, Cosmic microwave background anisotropies, {\it Ann. Rev.
Astron. Astrophys.} {\bf 40} (2002) 171, arXiv:astro-ph/0110414.

\bibitem{huHolder}
W. Hu and G. P. Holder, Model-independent reionization observables in the CMB,
{\it Phys. Rev. D} {\bf 68} (2003) 023001, arXiv:astro-ph/0303400.

\bibitem{huWhite}
W. Hu and M. White, A CMB polarization primer, {\it New Astron.} {\bf 2} (1997)
323, arXiv:astro-ph/9706147.

\bibitem{illarionov}
A. F. Illarionov and R. A. Siuniaev, Comptonization, the
background-radiation spectrum, and the thermal history of the
universe, {\it Sov. Astron.} {\bf 18} (1975)
691.

\bibitem{jarosik}
N. Jarosik {\it et~al.}, First year Wilkinson microwave anisotropy probe (WMAP)
observations: On-orbit radiometer characterization, {\it Astrophys. J.}
{\bf 148} (2003) 29, arXiv:astro-ph/0302224.

\bibitem{jungman}
G. Jungman, M. Kamionkowski, A. Kosowsky and D. N. Spergel, Cosmological parameter
determination with microwave background maps, {\it Phys. Rev. D} {\bf 54} (1996)
1332, arXiv:astro-ph/9512139.

\bibitem{MKamAKos1999}
M. Kamionkowski and A. Kosowsky, The cosmic microwave background
and particle physics, {\it Ann. Rev. Nucl. Part. Sci.} {\bf 49}
(1999) 77, arXiv:astro-ph/9904108.

\bibitem{stebbins}
M. Kamionkowski, A. Kosowsky and A. Stebbins, Statistics of cosmic microwave
background polarization, {\it Phys. Rev. D} {\bf 55} (1997) 7368,
arXiv:astro-ph/9611125.

\bibitem{kaplinghat}
M. Kaplinghat, M. Chu, Z. Haiman, G. Holder, L. Knox and C. Skordis, Probing the
reionization history of the universe using the cosmic microwave background
polarization, {\it Astrophys. J.} {\bf 583} (2003) 24, arXiv:astro-ph/0207591.

\bibitem{keating}
J. Kaufman, B. Keating and B. Johnson, Precision tests of parity violation over
cosmological distances, arXiv:astro-ph/1409.8242.

\bibitem{keisler}
R. Keisler {\it et~al.}, A measurement of the damping tail of the cosmic microwave
background power spectrum with the south pole telescope, {\it Astrophys. J.} {\bf
743} (2011) 28, arXiv:astro-ph/1105.3182.

\bibitem{kodamaSasaki}
H. Kodama and M. Sasaki, Cosmological perturbation theory, {\it Prog. Theor. Phys.
Suppl.} {\bf 78} (1984) 1.

\bibitem{cmb-dipole}
A. Kogut {\it et~al.}, Dipole anisotropy in the COBE DMR first-year sky maps, {\it
Astrophys. J.} {\bf 419} (1993) 1, arXiv:astro-ph/9312056.

\bibitem{pixie}
A. Kogut {\it et~al.}, The primordial inflation explorer (PIXIE): A nulling
polarimeter for cosmic microwave background observations, {\it J. Cosmol.
Astropart. Phys.} {\bf 1107} (2011) 025, arXiv:astro-ph/1105.2044.

\bibitem{kogutHighGalactic}
A. Kogut, A. J. Banday, C. L. Bennett, K. M. Gorski, G. Hinshaw and W. T. Reach,
High-latitude galactic emission in the COBE differential microwave radiometer 2
year sky maps, {\it Astrophys. J.} {\bf 460} (1996) 1.

\bibitem{KomatsuSpergel}
E. Komatsu and D. Spergel, Acoustic signatures of the primary microwave background
bispectrum, arXiv:astro-ph/000503.

\bibitem{kosowsky}
A. Kosowsky, Cosmic microwave background polarization, {\it Ann. Phys.} {\bf 246}
(1996) 49, arXiv:astro-ph/9501045.

\bibitem{act-instrument}
A. Kosowsky, The Atacama cosmology telescope, {\it New Astron.
Rev.} {\bf 47} (2003) 939, arXiv:astro-ph/0402234.

\bibitem{dasi}
J. Kovac {\it et~al.}, Detection of polarization in the cosmic microwave
background using DASI, {\it Nature} {\bf 420} (2002) 772,
arXiv:astro-ph/0209478.

\bibitem{KrausBook}
J. Kraus, {\it Radio Astronomy} (McGraw-Hill, New York, 1966).

\bibitem{kuzmin-ceb}
L. Kuzmin, Ultimate cold-electron bolometer with strong electrothermal feedback,
{\it Proc. SPIE Conf.$,$ Millimeter and Submillimeter Detectors for Astronomy II},
Vol.~5498 (2004), pp.~349.

\bibitem{lamarreBolo}
J. M. Lamarre, Photon noise in photometric instruments at far infrared and
submillimeter wavelengths, {\it Appl. Opt.} {\bf 25} (1986) 870.

\bibitem{axisOfEvil}
K. Land and J. Magueijo, Examination of evidence for a preferred axis in the
cosmic radiation anisotropy, {\it Phys. Rev. Lett.} {\bf 95} (2005) 071301,
arXiv:astro-ph/0502237.

\bibitem{landau}
L. D. Landau and E. M. Lifshitz, {\it Electrodynamics of a Continuous Media:
Course of Theoretical Physics}, Vol.~8 (First English edition) (Pergamon Press,
New York, 1960).

\bibitem{WW2}
A. Lazarian and B. Draine, Thermal flipping and thermal trapping: New elements in
grain dynamics, {\it Ap. J.} {\bf 516} (1999) 37.

\bibitem{leach}
S. M. Leach {\it et~al.}, Component separation methods for the Planck mission,
{\it Astron. Astrophys.} {\bf 491} (2008) 597, arXiv:astro-ph/0805.0269.

\bibitem{lee-tes}
A. T. Lee, P. L. Richards, S. W. Nam, B. Cabrera and K. D. Irwin, A
superconducting bolometer with strong electrothermal feedback, {\it Appl. Phys.
Lett.} {\bf 69} (1996) 1801.

\bibitem{legerPuget1984}
A. Leger and J. L. Puget, Identification of the ``unidentified'' IR emission
features of interstellar dust? {\it Astron. Astrophys.} {\bf 137} (1984) L5.

\bibitem{Leitch}
E. M. Leitch, A. C. S. Readhead, T. J. Pearson and S. T. Myers, An anomalous
component of galactic emission, {\it Astrophys. J.} {\bf 486} (1997) L23,
arXiv:astro-ph/9705241.

\bibitem{lesgourguesPastor}
J. Lesgourgues and S. Pastor, Massive neutrinos and cosmology, {\it Phys. Rep.}
{\bf 429} (2006) 307, arXiv:astro-ph/0603494.

\bibitem{lewisChallinorLasenby}
A. Lewis, A. Challinor and A. Lasenby, Efficient computation of cosmic microwave
background anisotropies in closed Friedmann-Robertson-Walker models, {\it
Astrophys. J.} {\bf 538} (2000) 473.

\bibitem{cosmomc}
A. Lewis and S. Bridle, Cosmological parameters from CMB and other data: A
Monte-Carlo approach, {\it Phys. Rev. D} {\bf 66} (2002) 103511,
arXiv:astro-ph/0205436.

\bibitem{lensingReview}
A. Lewis and A. Challinor, Weak gravitational lensing of the CMB,
{\it Phys. Rep.}\break {\bf 429} (2006) 1,
arXiv:astro-ph/0601594.

\bibitem{LiddleLythBook}
A. R. Liddle and D. H. Lyth, {\it Cosmological Inflation and Large-Scale
Structure} (Cambridge University Press, Cambridge, 2009).

\bibitem{LifshitzKhalatnikov}
E. M. Lifshitz and I. M. Khalatnikov, Investigations in relativistic cosmology,
{\it Adv. Phys.} {\bf 12} (1963) 185.

\bibitem{luptonBook}
R. Lupton, {\it Statistics in Theory and Practice} (Princeton University Press,
Princeton, 1993).

\bibitem{lyth-zeta}
D. Lyth, Large-scale energy-density perturbations and inflation, {\it Phys. Rev.
D} {\bf 31} (1985) 1792.

\bibitem{inflationReviews}
D. H. Lyth and A. Riotto, Particle physics models of inflation and the
cosmological density perturbation, {\it Phys. Rep.} {\bf 314} (1999) 1,
arXiv:hep-ph/9807278.

\bibitem{MaBert}
C.-P. Ma and E. Bertschinger, Cosmological perturbation theory in the synchronous
and conformal newtonian gauges, {\it Astrophys. J.} {\bf 455} (1995)
7,
arXiv:astro-ph/9506072.

\bibitem{maldacena}
J. Maldacena, Non-Gaussian features of primordial fluctuations in single field
inflationary models, {\it J. High Energy Phys.} {\bf 05} (2003) 013,
arXiv:astro-ph/0210603.

\bibitem{mandelWolf}
L. Mandel and E. Wolf, {\it Optical Coherence and Quantum Optics} (Cambridge
University Press, Cambridge, 1995).

\bibitem{act-sz}
T. A. Marriage {\it et~al.}, The Atacama cosmology telescope:
Sunyaev-Zel'dovich---Selected galaxy clusters at 148\,GHz in
the 2008 survey, {\it Astrophys. J.} {\bf 737} (2011) 61,
arXiv:astro-ph/1010.1065.

\bibitem{mather1990}
J. C. Mather {\it et~al.}, A preliminary measurement of the cosmic microwave
background spectrum by the cosmic background explorer (COBE) satellite, {\it
Astrophys. J.} {\bf 354} (1990) L37.

\bibitem{mather94}
J. C. Mather {\it et~al.}, Measurement of the cosmic microwave background spectrum
by the COBE FIRAS instrument, {\it Astrophys. J.} {\bf 420} (1994) 439.

\bibitem{firas-instrument-paper}
J. C. Mather, D. J. Fixsen and R. A. Shafer, Design for the COBE far-infrared
absolute spectrophotometer, {\it Proc. SPIE $2019,$ Infrared Spaceborne Remote
Sensing} (1993) 168.

\bibitem{meny}
C. Meny {\it et~al.}, Far-infrared to millimeter astrophysical dust emission I: A
model based on physical properties of amorphous solids, {\it Astron. Astrophys.}
{\bf 468} (2007) 171, arXiv:astro-ph/0701226.

\bibitem{millerToco}
A. D. Miller {\it et~al.}, A measurement of the angular power spectrum of the CMB
from $\ell = 100$ to 400, {\it Astrophys. J.} {\bf 524} (1999) L1,
arXiv:astro-ph/9906421.

\bibitem{Mukhanov:1985rz}
V. F. Mukhanov, Gravitational instability of the universe filled with a scalar
field, {\it JETP Lett.} {\bf 41} (1985) 493.

\bibitem{Mukhanov:1981xt}
V. F. Mukhanov and G. V. Chibisov, Quantum fluctuation and nonsingular universe
(in Russian), {\it JETP Lett.} {\bf 33} (1981) 532.

\bibitem{Mukhanov:1982nu}
V. F. Mukhanov and G. V. Chibisov, The vacuum energy and large scale structure of
the universe, {\it Sov.~Phys.~JETP} {\bf 56} (1982) 258.

\bibitem{MukhanovFeldmanBrandenberger}
V. F. Mukhanov, H. A. Feldman and R. H. Brandenberger, Theory of
cosmological perturbations, {\it Phys. Rep.} {\bf 215} (1992)
203.

\bibitem{boomerang-netterfield}
C. B. Netterfield {\it et~al.}, A measurement by BOOMERANG of multiple peaks in
the angular power spectrum of the cosmic microwave background, {\it Astrophys. J.}
{\bf 571} (2002) 604, arXiv:astro-ph/0104460.

\bibitem{niOne}
W.-T. Ni, Dilaton field and cosmic wave propagation, {\it Phys. Lett. A} {\bf 378}
(2014) 3413, arXiv:1410.0126.

\bibitem{niTwo}
W.-T. Ni, Searches for the role of spin and polarization in gravity, {\it Rep.
Prog. Phys.} {\bf 73} (2010) 056901,
arXiv:0912.5057.

\bibitem{okamotoHu}
T. Okamoto and W. Hu, CMB lensing reconstruction on the full sky, {\it Phys. Rev.
D} {\bf 67} (2003) 083002, arXiv:astro-ph/0301031.

\bibitem{paradis}
D. Paradis {\it et~al.}, Variations of the spectral index of dust emissivity from
Hi-GAL observations of the galactic plane, {\it Astron. Astrophys.} {\bf 520}
(2010) L8, arXiv:1009.2779.

\bibitem{PardoCernicharo}
J. R. Pardo, J. Cernicharo and E. Serabyn, Atmospheric transmission at microwaves
(ATM): An improved model for millimeter/submillimeter applications, {\it IEEE
 Trans. Antennas Propagation} {\bf 49} (2001) 1683.

\bibitem{PeacockBook}
J. A. Peacock, {\it Cosmological Physics} (Cambridge University Press, Cambridge,
1998).

\bibitem{partridgeBook}
P.J.E. Peebles, Lyman A. Page, Jr. and R. B. Partridge (eds.), {\it Finding
the Big Bang} (Cambridge University Press, Cambridge, England, 2009).

\bibitem{PeeblesLSSofUniverseBook}
P.J.E. Peebles, {\it The Large-Scale Structure of the Universe} (Princeton
University Press, Princeton, 1980).

\bibitem{peebles1968}
P.J.E. Peebles, Recombination of the primeval plasma, {\it Astrophys. J.} {\bf
153} (1968) 1.

\bibitem{PJEPee1987}
P.J.E. Peebles, Cosmic background temperature anisotropy in a
minimal isocurvature model for galaxy formation, {\it Astrophys.
J.} {\bf 315} (1987) L73.

\bibitem{peeblesYu}
P.J.E. Peebles and J. T. Yu, Primeval adiabatic perturbation in an expanding
universe, {\it Astrophys. J.} {\bf 162} (1970) 815.

\bibitem{PenSeljakTurok}
U.-L. Pen, U. Seljak and N. Turok, Power spectra in global defect theories of
cosmic structure formation, {\it Phys. Rev. Lett.} {\bf 79} (1997) 1611.

\bibitem{penSpergelTurok}
U.-L. Pen, D. N. Spergel and N. Turok, Cosmic structure formation and microwave
anisotropies from global field ordering, {\it Phys. Rev. D} {\bf 49} (1994) 692.

\bibitem{PenziasAndWilson1965}
A. A. Penzias and R. W. Wilson, A measurement of excess antenna temperature at
4080\,Mc/s, {\it Astrophys. J.} {\bf 142} (1965) 421.

\bibitem{supernovaA}
S. Perlmutter {\it et~al.}, Measurements of $\Omega $ and $\Lambda $ from 42
High-Redshift Supernovae, {\it Astrophys. J.} {\bf 517} (1999) 565,
arXiv:astro-ph/9812133.

\bibitem{planck-cib-ps}
Planck Collab. (P. A. R. Ade {\it et~al.}), Planck early results XVIII: The power
spectrum of cosmic infrared background anisotropies, {\it Astron. Astrophys.}
 {\bf 536} (2011) A18, arXiv:astro-ph/1101.2028.

\bibitem{newLight}
Planck Collab. (P. A. R. Ade {\it et~al.}), Planck early results. XX. New light on
anomalous microwave emission from spinning dust grains, {\it Astron. Astrophys.}
{\bf 536} (2011) A20, arXiv:1101.2031.

\bibitem{studyAnomalous}
Planck Collab. (P. A. R. Ade {\it et~al.}), Planck intermediate results. XV. A
study of anomalous microwave emission in Galactic clouds, {\it Astron. Astrophys.}
{\bf 565} (2014) A103, arXiv:1309.1357.

\bibitem{planckDust}
Planck Collab. (P. Ade {\it et~al.}), Planck intermediate results. XIV. Dust
emission at millimetre wavelengths in the Galactic plane, {\it Astron. Astrophys.}
{\bf 564} (2014) A45, arXiv:1307.6815.

\bibitem{planckXXX}
Planck Collab. (R. Adam {\it et~al.}), Planck intermediate
results. XXX. The angular power spectrum of polarized dust
emission at intermediate and high galactic latitudes,
arXiv:astro-ph/1409.5738.


\bibitem{planck-thermal-dust}
Planck Collab. (P. Ade {\it et~al.}), Planck 2013 results. XI.
All-sky model of thermal dust emission, {\it Astron. Astrophys.}
{\bf 571} (2014), arXiv:astro-ph/1312.1300.

\bibitem{planck-mission-paper}
Planck Collab. (P. Ade {\it et~al.}), Planck 2013 results: I.
Overview of products and results, {\it Astron. Astrophys.} {\bf
571} (2014) A1, arXiv:astro-ph/1303.5062.

\bibitem{planck-component-separation}
Planck Collab. (P. Ade {\it et~al.}), Planck 2013 results: XII.
Diffuse component separation, {\it Astron. Astrophys.} {\bf 571}
(2014) A12, arXiv:astro-ph/1303.5072.

\bibitem{planck-galactic-co}
Planck Collab. (P. Ade {\it et~al.}), Planck 2013 results: XIII.
Galactic CO emission, {\it Astron. Astrophys.} {\bf 571} (2014)
A13, arXiv:astro-ph/1303.5073.

\bibitem{planck-zodi}
Planck Collab. (P. Ade {\it et~al.}), Planck 2013 results: XIV.
Zodiacal emission, {\it Astron. Astrophys.} {\bf 571} (2014) A14,
arXiv:astro-ph/1303.5074.

\bibitem{planck-ps-likelihood}
Planck Collab. (P. Ade {\it et~al.}), Planck 2013 results: XV. CMB
power spectra and likelihood, {\it Astron. Astrophys.} {\bf 571}
(2014) A15, arXiv:astro-ph/1303.5075.

\bibitem{planck-parameters}
Planck Collab. (P. Ade {\it et~al.}), Planck 2013 results: XVI.
Cosmological parameters, {\it Astron. Astrophys.} {\bf 571} (2014)
A16, arXiv:astro-ph/1303.5076.

\bibitem{planck-lensing}
Planck Collab. (P. Ade {\it et~al.}), Planck 2013 results: XVII.
Gravitational lensing by large-scale structure, {\it Astron.
Astrophys.} {\bf 571} (2014) A17, arXiv:astro-ph/1303.5077.

\bibitem{planck-isw}
Planck Collab. (P. Ade {\it et~al.}), Planck 2013 results: XIX.
The integrated Sachs-Wolfe effect, {\it Astron. Astrophys.} {\bf
571} (2014) A19, arXiv:astro-ph/1303.5079.

\bibitem{planck-cluster-counts}
Planck Collab. (P. Ade {\it et~al.}), Planck 2013 results: XX.
Cosmology from Sunyaev-Zeldovich cluster counts, {\it Astron.
Astrophys.} {\bf 571} (2014) A20, arXiv:astro-ph/1303.5080.

\bibitem{planck-all-sky-compton-map}
Planck Collab. (P. Ade {\it et~al.}), Planck 2013 results: XXI.
All-sky Compton-parameter map and characterization, {\it Astron.
Astrophys.} {\bf 571} (2014) A21, arXiv:astro-ph/1303.5081.

\bibitem{planck-inflation}
Planck Collab. (P. Ade {\it et~al.}), Planck 2013 results: XXII.
Constraints on inflation, {\it Astron. Astrophys.} {\bf 571}
(2014) A22, arXiv:astro-ph/1303.5082.

\bibitem{planck-iso-stats}
Planck Collab. (P. Ade {\it et~al.}), Planck 2013 results: XXIII.
Isotropy and statistics of the CMB, {\it Astron. Astrophys.} {\bf
571} (2014) A23, arXiv:astro-ph/1303.5083.

\bibitem{planck-fnl}
Planck Collab. (P. Ade {\it et~al.}), Planck 2013 results: XXIV.
Constraints on primordial non-Gaussianity, {\it Astron.
Astrophys.} {\bf 571} (2014) A24, arXiv:astro-ph/1303.5084.

\bibitem{planck-defects}
Planck Collab. (P. Ade {\it et~al.}), Planck 2013 results: XXV.
Searches for cosmic strings and other topological defects, {\it
Astron. Astrophys.} {\bf 571} (2014) A25,
arXiv:astro-ph/1303.5085.

\bibitem{planck-geom-topo}
Planck Collab. (P. Ade {\it et~al.}), Planck 2013 results: XXVI.
Background geometry and topology of the Universe, {\it Astron.
Astrophys.} {\bf 571} (2014) A26, arXiv:astro-ph/1303.5086.

\bibitem{planck-eppur}
Planck Collab. (P. Ade {\it et~al.}), Planck 2013 results: XXVII.
Doppler boosting of the CMB: Eppur si muove, {\it Astron.
Astrophys.} {\bf 571} (2014) A27, arXiv:astro-ph/1303.5087.

\bibitem{planck-szc}
Planck Collab. (P. Ade {\it et~al.}), Planck 2013 results: XXIX.
The Planck catalogue of Sunyaev-Zeldovich sources, {\it Astron.
Astrophys.} {\bf 571} (2014) A29, arXiv:astro-ph/1303.5089.

\bibitem{planck-csc}
Planck Collab. (P. Ade {\it et~al.}), Planck 2013 results. XXVIII.
The Planck catalogue of compact sources, {\it Astron. Astrophys.}
{\bf 571} (2014) A28, arXiv:astro-ph/1303.5088.

\bibitem{planck-lfi-data-processing}
Planck Collab. (N. Aghanim {\it et~al.}), Planck 2013 results. II.
Low frequency instrument data processing, {\it Astron. Astrophys.}
{\bf 571} (2014) A2, arXiv:1303.5063.

\bibitem{planck-hfi-data-processing}
Planck Collab. (P. A. Ade {\it et~al.}), Planck 2013 results. VI.
High frequency instrument data processing, {\it Astron.
Astrophys.} {\bf 571} (2014) A6, arXiv:1303.5067.

\bibitem{planck-hfi-cosmic-rays}
Planck Collab. (P. A. Ade {\it et~al.}), Planck 2013 results. X.
HFI energetic particle effects, {\it Astron. Astrophys.} {\bf 571}
(2014) A10, arXiv:1303.5071.


\bibitem{polar-bear-bmode-lensing}
Polarbear Collab. (P. A. R. Ade {\it et~al.}), A measurement of
the cosmic microwave background B-mode polarization power spectrum
at sub-degree scales with POLARBEAR, {\it Astrophys. J.} {\bf 794}
(2014) 171, arXiv:1403.2369.

\bibitem{PospieszalskiWollack}
M. Pospieszalski and E. J. Wollack, Ultralow noise InP field effect transistor
radio astronomy receivers: State of the art, {\it 13 Int. Conf. Microwaves$,$
Radar and Wireless Communications}, MIKON-2000.

\bibitem{VMPreskill}
J. Preskill, Vortices and monopoles, {\it Architecture of Fundamental Interactions
at Short Distances}, ed. P. Ramond (North-Holland, Amsterdam, 1987).

\bibitem{prism}
PRISM (Polarized Radiation Imaging and Spectroscopy Mission): An Extended White Paper
PRISM Collab. (P. Andr\'e {\it et~al.}, {\it J. Cosmol. Astropart. Phys.} {\bf 02} (2014)
006, arXiv:1310.1554.

\bibitem{PugetLegerReview}
J. L. Puget and A. Leger, A new component of the interstellar matter: Small grains
and large aromatic molecules, {\it Ann. Rev. Astron. Astrophys.} {\bf 27} (1989)
161.

\bibitem{pugetCFIRB}
J.-L. Puget {\it et~al.}, Tentative detection of a cosmic far infrared background
with COBE, {\it Astron. Astrophys.} {\bf 308} (1996) L5.

\bibitem{purcell1969}
E. M. Purcell, On the absorption and emission of light by interstellar grains,
{\it Astrophys. J.} {\bf 158} (1969) 433.

\bibitem{purcellSupra}
E. M. Purcell, Suprathermal rotation of interstellar grains, {\it Astrophys. J.}
{\bf 231} (1979) 404.

\bibitem{spt-secondary}
C. L. Reichardt {\it et~al.}, A measurement of secondary cosmic microwave
background anisotropies with two years of South Pole Telescope Observations, {\it
Astrophys. J.} {\bf 755} (2012) 70, arXiv:astro-ph/1111.0932.

\bibitem{rephaeli}
Y. Rephaeli, Comptonization of the cosmic microwave background: The
Sunyaev-Zeldovich effect, {\it Ann. Rev. Astron. Astrophys.} {\bf 33} (1995) 541.

\bibitem{supernovaB}
A. G. Riess {\it et~al.}, Observational evidence from supernovae for an
accelerating universe and a cosmological constant, {\it Astron. J.} {\bf 116}
(1998) 1009, arXiv:astro-ph/9805201.

\bibitem{RR2}
V. Rubakov, M. Sazhin and A. Veryaskin, Graviton creation in the inflationary universe
and the grand unification scale, {\it Phys. Lett. B} {\bf 115} (1982) 189.

\bibitem{RubinoMartin}
J. A. Rubi\~no-Martín, C. H. L\'opez-Caraballo, R. G\'enova-Santos and R. Rebolo,
Observations of the polarization of the anomalous microwave emission: A review,
{\it Advances in Astronomy,} 2012, 351836 (2012).

\bibitem{spt-instrument}
J. Ruhl {\it et~al.}, The south pole telescope, {\it Proc. SPIE} {\bf 5498} (2004)
11, arXiv:astro-ph/0411122.

\bibitem{SW1967}
R. K. Sachs and A. M. Wolfe, Perturbations of a cosmological model and angular
variations of the microwave background, {\it Astrophys. J.} {\bf 147} (1967) 73.

\bibitem{DSamSStaBWin2007}
D. Samtleben, S. Staggs and B. Winstein, The cosmic microwave
background for pedestrians: A review for particle and nuclear
physicists, {\it Ann. Rev. Nucl. Part. Sci.} {\bf 57} (2007) 245,
arXiv:0803.0834 [astro-ph].

\bibitem{sss1999}
S. Seager, D, Sasselov and D. Scott, A new calculation of the recombination epoch,
{\it Astrophys. J.} {\bf 523} (1999) L1, arXiv:astro-ph/9909275.

\bibitem{sss2000}
S. Seager, D. Sasselov and D. Scott, How exactly did the universe become neutral?
 {\it Astrophys. J. Suppl.} {\bf 128} (2000) 407, arXiv:astro-ph/9912182.

\bibitem{Sellgren}
K. Sellgren, The near-infrared continuum emission of visual reflection nebulae,
{\it Astrophys. J.} {\bf 277} (1984) 623.

\bibitem{SeljakTwoFluid}
U. Seljak, A two-fluid approximation for calculating the cosmic microwave
background anisotropies, {\it Astrophys. J.} {\bf 435} (1994) L87.

\bibitem{lineOfSight}
U. Seljak and M. Zaldarriaga, A line-of-sight integration approach to cosmic
microwave background anisotropies, {\it Astrophys. J.} {\bf 469} (1996) 437,
arXiv:astro-ph/9603033.

\bibitem{silkDamping}
J. Silk, Cosmic black-body radiation and galaxy formation, {\it Astrophys. J.}
{\bf 151} (1968) 459.

\bibitem{SilkAndChluba}
J. Silk and J. Chluba, Next steps for cosmology, {\it Science} {\bf 344} (2014)
586.

\bibitem{lensing-cross}
K. M. Smith, O. Zahn and O. Dor\'e, Detection of gravitational
lensing in the cosmic microwave background, {\it Phys. Rev. D}
{\bf 76} (2007) 043510, arXiv:astro-ph/0705.3980.

\bibitem{DMR-instrument}
G. Smoot {\it et~al.}, COBE differential microwave radiometers:
Instrument design and implementation, {\it Astrophys. J.} {\bf
360} (1990) 685S.

\bibitem{COBEdmr}
G. F. Smoot {\it et~al.}, Structure in the COBE differential microwave radiometer
first-year maps, {\it Astrophys. J.} {\bf 396} (1992) L1.

\bibitem{sorbo}
L. Sorbo, Parity violation in the cosmic microwave background from
a pseudoscalar inflaton, {\it J. Cosmol. Astropart. Phys.} {\bf
1106} (2011) 003, arXiv:astro-ph/1101.\break1525.

\bibitem{WW1}
L. Spitzer Jr. and T. A. McGlynn, Disorientation of interstellar grains in suprathermal
rotation, {\it Ap. J.} {\bf 231} (1979) 417.

\bibitem{SpitzerTukey}
L. Spitzer, Jr. and J. W. Tukey, A theory of interstellar polarization, {\it
Astrophys. J.} {\bf 114} (1951) 187S.

\bibitem{sz-orig}
R. A. Sunyaev and Y. B. Zeldovich, Small-scale fluctuations of relic radiation,
{\it Astrophys. Space Sci.} {\bf 7} (1970) 3.

\bibitem{RR3}
A. Starobinsky, Spectrum of relict gravitational radiation and the early state of the
universe, {\it JETP. Lett.} {\bf 30} (1979) 682.

\bibitem{RR4}
A. Starobinsky, Cosmic background anisotropy induced by isotropic flat-spectrum
gravitational wave perturbations, {\it Sov. Astron. Lett.} {\bf 11} (1985) 133.

\bibitem{Starobinsky:1982ee}
A. A. Starobinsky, Dynamics of phase transition in the new inflationary universe
scenario and generation of perturbations, {\it Phys.~Lett. B} {\bf 117} (1982)
175.

\bibitem{kendall}
A. Stuart and K. Ord, {\it Kendall's Advanced Theory of Statistics$,$ Classical
Inference and Relationship}, Vol.~2 (Oxford University Press, New
York, 1991).

\bibitem{sh2008}
E. R. Switzer and C. M. Hirata, Primordial helium recombination. I. Feedback, line
transfer, and continuum opacity, {\it Phys. Rev. D} {\bf 77} (2008) 083006,
arXiv:astro-ph/0702143.

\bibitem{RadioSourceRef}
A. Taylor, K. Grainge, M. Jones, G. Pooley, R. Saunders and E. Waldram, The radio
source counts at 15\,GHz and their implications for cm-wave CMB imaging, {\it Mon.
Not. R. Astron. Soc.} {\bf 327} (2001) L1, arXiv:astro-ph/0102497.

\bibitem{CriticalDialogues}
N. Turok (ed.), {\it Critical Dialogues in Cosmology} (World Scientific
Publishing, Singapore, 1997).

\bibitem{spt-lensing}
A. van Engelen {\it et~al.}, A measurement of gravitational lensing of the
microwave background using south pole telescope data, {\it Astrophys. J.} {\bf
756} (2012) 142, arXiv:1202.0546.

\bibitem{vielva2011}
P. Vielva, E. Mart\'inez-Gonz\'alez, M. Cruz, R. B. Barreiro and M. Tucci, Cosmic
microwave background polarization as a probe of the anomalous nature of the cold
spot, {\it Mon. Not. R. Astron. Soc.} {\bf 410} (2011) 33.

\bibitem{vielva2004}
P. Vielva, E. Mart\'inez-Gonz\'alez, R. B. Barreiro, J. L. Sanz and L. Cay\'on,
Detection of non-Gaussianity in the Wilkinson microwave anisotropy probe
first-year data using spherical wavelets, {\it Astrophys. J.} {\bf 609} (2004) 22.

\bibitem{ShellardVilenkinBook}
A. Vilenkin and E. P. S. Shellard, {\it Cosmic Strings and Other Topological
Defects} (Cambridge University Press, Cambridge, 1994).

\bibitem{waldBook}
Wald, {\it General Relativity} (University of Chicago Press,
Chicago, 1984).

\bibitem{nucleo-review}
T. P. Walker, G. Steigman, H.-S. Kang, D. M. Schramm and K. A. Olive, Primordial
nucleosynthesis redux, {\it Astrophys. J.} {\bf 376} (1991) 51.

\bibitem{WatsonEtAl}
R. A. Watson {\it et~al.}, Detection of anomalous microwave emission in the
perseus molecular cloud with the COSMOSOMAS experiment, {\it Astrophys. J.} {\bf
624} (2005) L89, arXiv:astro-ph/0503714.

\bibitem{weinbergBook}
S. Weinberg, {\it Gravitation and Cosmology\/$:$ Principles and
Applications of the General Theory of Relativity} (J. Wiley~\&
Sons, New York, 1972).

\bibitem{weingartner2003}
J. C. Weingartner and B. T. Draine, Radiative torques on interstellar grains. III.
dynamics with thermal relaxation, {\it Astrophys. J.} {\bf 589} (2003) 289.

\bibitem{spt-sz}
R. Williamson {\it et~al.}, A Sunyaev-Zel'dovich-selected sample of the most
massive galaxy clusters in the 2500 deg2 south pole telescope survey, {\it
Astrophys. J.} {\bf 738} (2011) 139, arXiv:astro-ph/1101.1290.

\bibitem{whiteHuOneThird}
M. White and W. Hu, The Sachs-Wolfe effects, {\it Astron. Astrophys.} {\bf 321}
(1997) 8, arXiv:astro-ph/9609105.

\bibitem{Whittaker}
E. T. Whittaker, {\it A Treatise on the Analytical Dynamics of Particles and Rigid
Bodies} (Cambridge University Press, Cambridge, 1917).

\bibitem{wmap2003a}
WMAP Collab. (C. L. Bennett {\it et~al.}), First year Wilkinson microwave
anisotropy probe (WMAP) observations: Preliminary maps and basic results, {\it
Astrophys. J. Suppl.} {\bf 148} (2003) 1.

\bibitem{wmap2003b}
WMAP Collab. (G. Hinshaw {\it et~al.}), First year Wilkinson microwave anisotropy
probe (WMAP) observations: Data processing methods and systematic error limits,
{\it Astrophys. J. Suppl.} {\bf 148} (2003) 63.

\bibitem{wmap2003c}
WMAP Collab. (D. N. Spergel {\it et~al.}), First year Wilkinson microwave
anisotropy probe (WMAP) observations: Determination of cosmological parameters,
{\it Astrophys. J. Suppl.} {\bf 148} (2003) 175.

\bibitem{wmap2003d}
WMAP Collab. (C. L. Bennett {\it et~al.}), First year Wilkinson microwave
anisotropy probe (WMAP) observations: Foreground emission, {\it Astrophys. J.
Suppl.} {\bf 148} (2003)~97.

\bibitem{wmap2003e}
WMAP Collab. (H. V. Peiris {\it et~al.}), First year Wilkinson microwave
anisotropy probe (WMAP) observations: Implications for inflation, {\it Astrophys.
J. Suppl.} {\bf 148} (2003) 213.

\bibitem{wmap2003f}
WMAP Collab. (L. Page {\it et~al.}), First year Wilkinson microwave anisotropy
probe (WMAP) observations: Interpretation of the TT and TE angular power spectrum
peaks, {\it Astrophys. J. Suppl.} {\bf 148} (2003) 233.

\bibitem{wmap2003g}
WMAP Collab. (A. Kogut {\it et~al.}), First year Wilkinson microwave anisotropy
probe (WMAP) observations: Temperature-polarization correlation, {\it Astrophys.
J. Suppl.} {\bf 148} (2003) 161.

\bibitem{wmap2003h}
WMAP Collab. (E. Komatsu {\it et~al.}), First year Wilkinson microwave anisotropy
probe (WMAP) observations: Tests of Gaussianity, {\it Astrophys. J. Suppl.} {\bf
148} (2003) 119.

\bibitem{wmap2003i}
WMAP Collab. (G. Hinshaw {\it et~al.}), First year Wilkinson microwave anisotropy
probe (WMAP) observations: The angular power spectrum, {\it Astrophys. J. Suppl.}
{\bf 148} (2003) 135.

\bibitem{wmap2007a}
WMAP Collab. (A. Kogut {\it et~al.}), Three-year Wilkinson microwave anisotropy
probe (WMAP) observations: Foreground polarization, {\it Astrophys. J. Suppl.}
{\bf 665} (2007) 355.

\bibitem{wmap2007b}
WMAP Collab. (G. Hinshaw {\it et~al.}), Three-year Wilkinson microwave anisotropy
probe (WMAP) observations: Temperature analysis, {\it Astrophys. J. Suppl.} {\bf
170} (2007) 288.

\bibitem{wmap2007c}
WMAP Collab. (L. Page {\it et~al.}), Three-year Wilkinson microwave anisotropy
probe (WMAP) observations: Polarization analysis, {\it Astrophys. J. Suppl.} {\bf
170} (2007) 335.

\bibitem{wmap2007d}
WMAP Collab. (D. N. Spergel {\it et~al.}), Three-year Wilkinson microwave
anisotropy probe (WMAP) observations: Implications for cosmology, {\it Astrophys.
J. Suppl.} {\bf 170} (2007) 377.

\bibitem{wmap2009a}
WMAP Collab. (J. Dunkley {\it et~al.}), Five-year Wilkinson microwave anisotropy
probe (WMAP) observations: Bayesian estimation of CMB polarization maps, {\it
Astrophys. J. Suppl.} {\bf 701} (2009) 1804.

\bibitem{wmap2009b}
WMAP Collab. (G. Hinshaw {\it et~al.}), Five-year Wilkinson microwave anisotropy
probe (WMAP) observations: Data processing, sky maps, and basic results, {\it
Astrophys. J. Suppl.} {\bf 180} (2009) 225.

\bibitem{wmap2009c}
WMAP Collab. (B. Gold {\it et~al.}), Five-year Wilkinson microwave anisotropy
probe (WMAP) observations: Galactic foreground emission, {\it Astrophys. J. Suppl.}
{\bf 180} (2009) 265.

\bibitem{wmap2009d}
WMAP Collab. (M. Nolta {\it et~al.}), Five-year Wilkinson microwave anisotropy
probe (WMAP) observations: Angular power spectra, {\it Astrophys. J. Suppl.} {\bf
180} (2009)\break 296.

\bibitem{wmap2009e}
WMAP Collab. (J. Dunkley {\it et~al.}), Five-year Wilkinson microwave anisotropy
probe (WMAP) observations: Likelihoods and parameters from WMAP Data,
{\it
Astrophys. J. Suppl.} {\bf 180} (2009) 306.

\bibitem{wmap2009f}
WMAP Collab. (E. Komatsu {\it et~al.}), Five-year Wilkinson microwave anisotropy
probe (WMAP) observations: Cosmological interpretation, {\it Astrophys. J. Suppl.}
{\bf 180} (2009) 330.

\bibitem{wmap2011a}
WMAP Collab. (N. Jarosik {\it et~al.}), Seven-year Wilkinson microwave anisotropy
probe (WMAP) observations: Sky maps, systematic errors, and basic results, {\it
Astrophys. J. Suppl.} {\bf 192} (2011) 14J.

\bibitem{wmap2011b}
WMAP Collab. (B. Gold {\it et~al.}), Seven-year Wilkinson microwave anisotropy
probe (WMAP) observations: Galactic foreground emission, {\it Astrophys. J.
Suppl.} {\bf 192} (2011) 15G.

\bibitem{wmap2011c}
WMAP Collab. (D. Larson {\it et~al.}), Seven-year Wilkinson microwave anisotropy
probe (WMAP) observations: Power spectra and WMAP-derived parameters, {\it
Astrophys. J. Suppl.} {\bf 192} (2011) 16L.

\bibitem{wmap2011d}
WMAP Collab. (C. Bennett {\it et~al.}), Seven-year Wilkinson microwave anisotropy
probe (WMAP) observations: Are there cosmic microwave background anomalies? {\it
Astrophys. J. Suppl.} {\bf 192} (2011) 17.

\bibitem{wmap2011e}
WMAP Collab. (E. Komatsu {\it et~al.}), Seven-year Wilkinson microwave anisotropy
probe (WMAP) observations: Cosmological interpretation, {\it Astrophys. J. Suppl.}
{\bf 192} (20011) 18K.

\bibitem{wmap2013a}
WMAP Collab. (C. L. Bennett {\it et~al.}), Nine-year Wilkinson microwave
anisotropy probe (WMAP) observations: Final maps and results, {\it Astrophys. J.
Suppl.} {\bf 208} (2013) 20B.

\bibitem{wmap2013b}
WMAP Collab. (G. F. Hinshaw {\it et~al.}), Nine-year Wilkinson microwave
anisotropy probe (WMAP) observations: Cosmology results, {\it Astrophys. J.
Suppl.} {\bf 208} (2013) 19H.

\bibitem{sasakiLD}
K. Yamamoto, M. Sasaki and T. Tanaka, Large angle CMB anisotropy in an open
universe in the one-bubble inflationary scenario, {\it Astrophys. J.} {\bf 455}
(1995) 412, arXiv:astro-ph/9501109.

\bibitem{zahn}
O. Zahn, M. Zaldarriaga, L. Hernquist and M. McQuinn, The influence of nonuniform
reionization on the CMB, {\it Astrophys. J.} {\bf 630} (2005) 657,
arXiv:astro-ph/0503166.

\bibitem{allSkyPol}
M. Zaldarriaga and U. Seljak, All-sky analysis of polarization in the microwave
background, {\it Phys. Rev. D} {\bf 55} (1997) 1830, arXiv:astro-ph/9609170.

\end{thebibliography}
\end{document}